\documentclass[twocolumn]{aastex631}
\usepackage{amsmath}
\usepackage{graphicx}
\usepackage[toc,page]{appendix}
\usepackage{xspace}

\newcommand{\Lya}{\ifmmode{{\mathrm{Ly}\alpha}}\else Ly$\alpha$\xspace\fi}
\newcommand{\kms}{\,\ifmmode{\mathrm{km}\,\mathrm{s}^{-1}}\else km\,s${}^{-1}$\fi\xspace}

\shorttitle{Ly$\alpha$ Scattering in T Tauri Systems}
\shortauthors{Arulanantham et al.}

\begin{document}

\title{Ly$\alpha$ Scattering Models Trace Accretion and Outflow Kinematics in T Tauri Systems \thanks{Based on observations collected at the European Southern Observatory under ESO programme 106.20Z8}}

\author[0000-0003-2631-5265]{Nicole Arulanantham}
\affil{Space Telescope Science Institute, 3700 San Martin Drive, Baltimore, MD 21218, USA}
\author{Max Gronke}
\affil{Max Planck Institut fur Astrophysik, Karl-Schwarzschild-Stra{\ss}e 1, D-85748 Garching bei M\"{u}nchen, Germany}
\author{Eleonora Fiorellino}
\affil{Konkoly Observatory, Research Centre for Astronomy and Earth Sciences, E\"{o}tv\"{o}s Lor\'{a}nd Research Network (ELKH), Konkoly-Thege Mikl\'{o}s \'{u}t 15-17, 1121 Budapest, Hungary}
\affiliation{INAF-Osservatorio Astronomico di Capodimonte, via Moiariello 16, 80131 Napoli, Italy}
\author{Jorge Filipe Gameiro}
\affil{Instituto de Astrof\'{i}sica e Ci\^{e}ncias do Espa\c{c}o,  Universidade do Porto, CAUP, Rua das Estrelas, 4150-762 Porto, Portugal}
\affil{Departamento de F\'{\i}sica e Astronomia, Faculdade de Ci\^encias, Universidade do Porto, rua do Campo Alegre 687, 4169-007 Porto. Portugal}
\author{Antonio Frasca}
\affil{INAF - Osservatorio Astrofisico di Catania, via S. Sofia, 78, 95123 Catania, Italy}
\author{Joel Green}
\affil{Space Telescope Science Institute, 3700 San Martin Drive, Baltimore, MD 21218, USA}
\author{Seok-Jun Chang}
\affil{Max Planck Institut fur Astrophysik, Karl-Schwarzschild-Stra{\ss}e 1, D-85748 Garching bei M\"{u}nchen, Germany}
\author{Rik A. B. Claes}
\affil{European Southern Observatory, Karl-Schwarzschild-Strasse 2, 85748 Garching bei M\``unchen, Germany}
\author[0000-0001-9227-5949]{Catherine C. Espaillat}
\affil{Institute for Astrophysical Research, Department of Astronomy, Boston University, 725 Commonwealth Avenue, Boston, MA 02215, USA}
\author[0000-0002-1002-3674]{Kevin France}
\affil{Laboratory for Atmospheric and Space Physics, University of Colorado Boulder, Boulder, CO 80303, USA}
\author{Gregory J. Herczeg}
\affil{Department of Astronomy, Peking University, Yiheyuan Road 5, Haidian District, Beijing 100871, People's Republic of China}
\affil{Kavli Institute for Astronomy and Astrophysics, Peking University, Yiheyuan Road 5, Haidian District, Beijing 100871, People's Republic of China}
\author{Carlo F. Manara}
\affil{European Southern Observatory, Karl-Schwarzschild-Strasse 2, 85748 Garching bei M\``unchen, Germany}
\author[0000-0002-4115-0318]{Laura Venuti}
\affil{SETI Institute, 339 Bernardo Ave, Suite 200, Mountain View, CA 94043, USA}
\author[0000-0001-6015-646X]{P\'eter \'Abrah\'am}
\affiliation{Konkoly Observatory, Research Centre for Astronomy and Earth Sciences, E\"otv\"os Lor\'and Research Network, Konkoly-Thege Mikl\'os \'ut 15-17, 1121 Budapest, Hungary}
\affiliation{ELTE E\"otv\"os Lor\'and University, Institute of Physics, P\'azm\'any P\'eter S\'et\'any 1/A, 1117 Budapest, Hungary}
\affiliation{CSFK, MTA Centre of Excellence, Konkoly-Thege Mikl\'os \'ut 15-17, 1121 Budapest, Hungary}
\author[0000-0001-6410-2899]{Richard Alexander}
\affil{School of Physics and Astronomy, University of Leicester, Leicester, LE1 7RH, UK}
\author[0000-0002-7450-6712]{Jerome Bouvier}
\affil{Univ. Grenoble Alpes, CNRS, IPAG, 38000 Grenoble, France}
\author[0000-0002-3913-3746]{Justyn Campbell-White}
\affil{European Southern Observatory, Karl-Schwarzschild-Strasse 2, 85748 Garching bei M\``unchen, Germany}
\author[0000-0001-6496-0252]{Jochen Eislöffel}
\affil{Thüringer Landessternwarte, Sternwarte 5, D-07778 Tautenburg, Germany}
\author[0000-0002-3747-2496]{William J. Fischer}
\affil{Space Telescope Science Institute, 3700 San Martin Drive, Baltimore, MD 21218, USA}
\author[0000-0001-7157-6275]{\'Agnes K\'osp\'al}
\affiliation{Konkoly Observatory, Research Centre for Astronomy and Earth Sciences, E\"otv\"os Lor\'and Research Network, Konkoly-Thege Mikl\'os \'ut 15-17, 1121 Budapest, Hungary}
\affiliation{ELTE E\"otv\"os Lor\'and University, Institute of Physics, P\'azm\'any P\'eter S\'et\'any 1/A, 1117 Budapest, Hungary}
\affiliation{CSFK, MTA Centre of Excellence, Konkoly-Thege Mikl\'os \'ut 15-17, 1121 Budapest, Hungary}
\affil{Max Planck Institute for Astronomy, K\"onigstuhl 17, 69117 Heidelberg, Germany}
\author[0000-0002-4147-3846]{Miguel Vioque}
\affil{Joint ALMA Observatory, Alonso de C\'ordova 3107, Vitacura, Santiago 763-0355, Chile}
\affil{National Radio Astronomy Observatory, 520 Edgemont Road, Charlottesville, VA 22903, USA}

\begin{abstract}

T Tauri stars produce broad Ly$\alpha$ emission lines that contribute $\sim$88\% of the total UV flux incident on the inner circumstellar disks. Ly$\alpha$ photons are generated at the accretion shocks and in the protostellar chromospheres and must travel through accretion flows, winds and jets, the protoplanetary disks, and the interstellar medium before reaching the observer. This trajectory produces asymmetric, double-peaked features that carry kinematic and opacity signatures of the disk environments. To understand the link between the evolution of Ly$\alpha$ emission lines and the disks themselves, we model \emph{HST}-COS spectra from targets included in Data Release 3 of the \emph{Hubble} UV Legacy Library of Young Stars as Essential Standards (ULLYSES) program. We find that resonant scattering in a simple spherical expanding shell is able to reproduce the high velocity emission line wings, providing estimates of the average velocities within the bulk intervening H I. The model velocities are significantly correlated with the $K$ band veiling, indicating a turnover from Ly$\alpha$ profiles absorbed by outflowing winds to emission lines suppressed by accretion flows as the hot inner disk is depleted. Just 30\% of targets in our sample have profiles with red-shifted absorption from accretion flows, many of which have resolved dust gaps. At this stage, Ly$\alpha$ photons may no longer intersect with disk winds along the path to the observer. Our results point to a significant evolution of Ly$\alpha$ irradiation within the gas disks over time, which may lead to chemical differences that are observable with ALMA and \emph{JWST}.   
\end{abstract}


\section{Introduction}

Classical T Tauri stars (CTTSs) are readily identified by the infrared excess associated with their circumstellar disks and strong excess UV and optical emission generated at their accretion shocks, which become visible as the natal clouds disperse (see e.g., \citealt{Hartmann2016, Schneider2020} for reviews). UV photometry from a few thousand of these targets has been measured with the \emph{Galaxy Evolution Explorer} (\emph{GALEX}) satellite, showing that both the FUV (1400-1700 \AA) and NUV (1800-2750 \AA) broadband fluxes decrease with stellar age, as circumstellar material disperses and accretion slows (see e.g., \citealt{Findeisen2011, GomezdeCastro2015}). The disappearance of accretion signatures (and therefore the protoplanetary disks) roughly sets the upper limit on timescales for giant planet formation ($t < 10$ Myr; see e.g., \citealt{Hartmann1998, Lubow2006, Fedele2010, Manara2019}). Residual optically thick, gas-rich inner disks may linger, however, as evidenced by large accretion rates derived from H$\alpha$ emission, $U$ band excess \citep{Najita2007, Espaillat2012}, and UV-near-infrared continuum fits \citep{Manara2014} for many transitional disk systems with disk dust gaps or cavities that may be associated with protoplanet formation. Such reservoirs have subsequently been resolved around TW Hya \citep{Ercolano2017}, SR 24S and CIDA 1 \citep{Pinilla2019, Pinilla2021}, DM Tau \citep{Hashimoto2021, Francis2022}, and LkCa 15 \citep{Blakely2022}.

Physical-chemical models of circumstellar disks demonstrate that the UV fluxes from accreting CTTSs play a critical role in regulating the abundances and spatial distributions of volatile-bearing molecules (e.g., C$_2$H, C$_2$H$_2$, HCN, and CN; \citealt{vanZadelhoff2003, Walsh2015, Cazzoletti2018}). For example, larger input UV fluxes are expected to produce brighter, more radially extended rings of CN emission than weaker radiation fields \citep{Cazzoletti2018}, and species like H$_2$O, CO$_2$, and HCOOH are expected to have enhanced abundances near water snowlines due to larger desorption rates at these radii \citep{Ruaud2019}. This work is consistent with sub-mm observations of C$_2$H, which can only be reproduced in models with high UV radiation and C/O ratios greater than unity \citep{Bergin2016, Bergner2019, Miotello2019}. However, the observed ratios of submillimeter CN and HCN emission unexpectedly do not show any clear correlation with total UV luminosity across targets spanning two orders of magnitude in irradiation \citep{Bergner2021}, despite models predicting increased photodissociation and lower HCN abundances in strong UV fields \citep{Bergin2003}. 

One possible explanation for this discrepancy between model predictions and observations is that the UV radiation fields produced by accreting CTTSs are dominated by the Ly$\alpha$ transition of H I ($\lambda = 1215.67$ \AA), which spans photodissociation and excitation cross sections for a large number of molecules (including HCN and H$_2$O; see e.g., \citealt{vanDishoeck2006}) and contributes $\sim$88\% of the total UV flux on average \citep{France2014}. Many of the bright disks observed at high spatial resolution with ALMA do not have accompanying Ly$\alpha$ spectra (see e.g., \citealt{Bergner2021}), meaning that disk models cannot fully account for the Ly$\alpha$ properties of individual targets and may not reveal the impact of Ly$\alpha$ photodissociation on abundance ratios like CN/HCN, for example. The feature is challenging to observe, as the line centers where the fluxes peak are always masked by circumstellar and interstellar absorption. However, broad emission line wings are detected at velocities up to $\sim$1000 km s$^{-1}$ and carry signatures of the radiation fields incident on the disks \citep{Schindhelm2012}. Characterizing these observed features is critical for interpreting infrared and sub-mm observations of UV-sensitive molecular gas in protoplanetary disks.

The shapes of Ly$\alpha$ emission lines observed from accreting T Tauri stars fall into three general categories: a) P Cygni-like, where blueshifted absorption from outflowing material has removed photons from wavelengths at $\lambda < 1215.67$ \AA; b) inverse P Cygni-like, consistent with red-shifted absorption at $\lambda > 1215.67$ \AA \, by gas being accreted onto the central stars; and c) intermediate, symmetric profiles absorbed primarily near line center \citep{Arulanantham2021}. Absorption below the continuum is not observed at these velocities $\left(v > 300 \, \rm{km s}^{-1}\right)$. The ratios of flux from the red ($\lambda > 1215.67$ \AA) and blue ($\lambda < 1215.67$ \AA) Ly$\alpha$ emission line wings are strongly correlated with UV-fluorescent emission from electronic transitions of Ly$\alpha$-pumped H$_2$ and CO, demonstrating that the radiation is received by molecular gas in surface layers of the inner disks \citep{Schindhelm2012_UVCO, France2012, Hoadley2015}. The Ly$\alpha$ flux ratios are also significantly correlated with the infrared SED slopes of circumstellar disks with dust rings, gaps, and cavities \citep{Arulanantham2021}, measured between $\lambda = 13$ $\mu$m and $\lambda = 31$ $\mu$m \citep{Furlan2009}. Since the dust continuum emission at these wavelengths is determined by the temperature of the optically thick disk, which maps to the size of the emitting area, the infrared slopes trace the radii where dust grains have been removed from the disks and are not contributing to the total flux (see e.g., \citealt{Espaillat2010}). This implies that Ly$\alpha$ irradiation of the gas disks changes as the dust disks evolve, potentially altering the systems' chemical evolution over the timescales of planet formation.  

In order to further explore this connection between Ly$\alpha$ radiation and the circumstellar disks, we model \emph{HST} spectra from 42 targets included in the \textit{Hubble} UV Legacy Library of Young Stars as Essential Standards Director's Discretionary program (ULLYSES; \citealt{ULLYSES2020}). Ly$\alpha$ is the $n = 2-1$ resonant line of H I, which implies that \Lya photons typically scatter thousands to millions of times before reaching the observer because of the high optical depth in the ground state. Each scattering event leads to a slight frequency shift -- mostly simply due to the Doppler effect. In other words, \Lya escape through astrophysical media can be thought of as a double diffusion process through space and frequency. The frequency shift depends on the HI properties of the medium, in particular the kinematics and column density \citep{Adams1972,Neufeld1990,Bonilha1979} -- but also less intuitive properties such as the clumping factor, which describes the number of neutral gas clouds along the line of sight \citep{Neufeld1991,Hansen2006,Gronke2018}. We compare the observed spectra to models including resonant scattering for the first time, using a simplified shell geometry to represent the accretion shocks, disk winds, accretion flows, and disks. Optical and infrared data from the PENELLOPE Large Program on the ESO Very Large Telescope \citep{Manara2021}, which provides spectra acquired contemporaneously with the ULLYSES observations on \emph{HST}, are included in our analysis when available to assemble a comprehensive multi-wavelength view of interactions between the CTTSs and disks. This work was done as part of the Outflows and Disks around Young Stars: Synergies for the Exploration of Ullyses Spectra (ODYSSEUS; \citealt{Espaillat2022_ODYSSEUSI}) and PENELLOPE collaborations \citep{Manara2021}. 

\section{Targets \& Observations} 

Our sample consists of 42 low mass targets $\left(M_{\ast} = 0.14-1.71 \, M_{\odot} \right)$ with either archival \textit{HST} spectra or new observations from Data Release 3 of the ULLYSES program \citep{RomanDuval2020}. This group includes objects in Chamaeleon I $\left(N = 10 \right)$, Lupus $\left(N = 13 \right)$, Orion OB1 $\left(N = 7 \right)$, Taurus-Auriga $\left(N = 9 \right)$, and $\sigma$ Orionis $\left(N = 1 \right)$, TW Hya, and V4046 Sgr, covering a total protostellar age range of $\sim$1-12 Myr. All targets were observed with the G130M $\lambda$1291 mode on the Cosmic Origins Spectrograph (\textit{HST}-COS), which provides spectra from $\lambda = 1132-1436$ \AA \, at $R \sim 12,000-16,000$. Data reduction for the ULLYSES targets was done by the ULLYSES core implementation team at STScI \citep{RomanDuval2020}, which provides publicly available high level science products (HLSPs) for the entire sample\footnote{See \url{https://ullyses.stsci.edu/ullyses-data-description.html} for a full description of the available ULLYSES data products}. All of the ULLYSES data presented in this paper were obtained from the Mikulski Archive for Space Telescopes (MAST) at the Space Telescope Science Institute, via \dataset[10.17909/t9-jzeh-xy14]
{https://doi.org/10.17909/t9-jzeh-xy14}.

\begin{figure*}
    \centering
    \includegraphics[width=\linewidth]{ 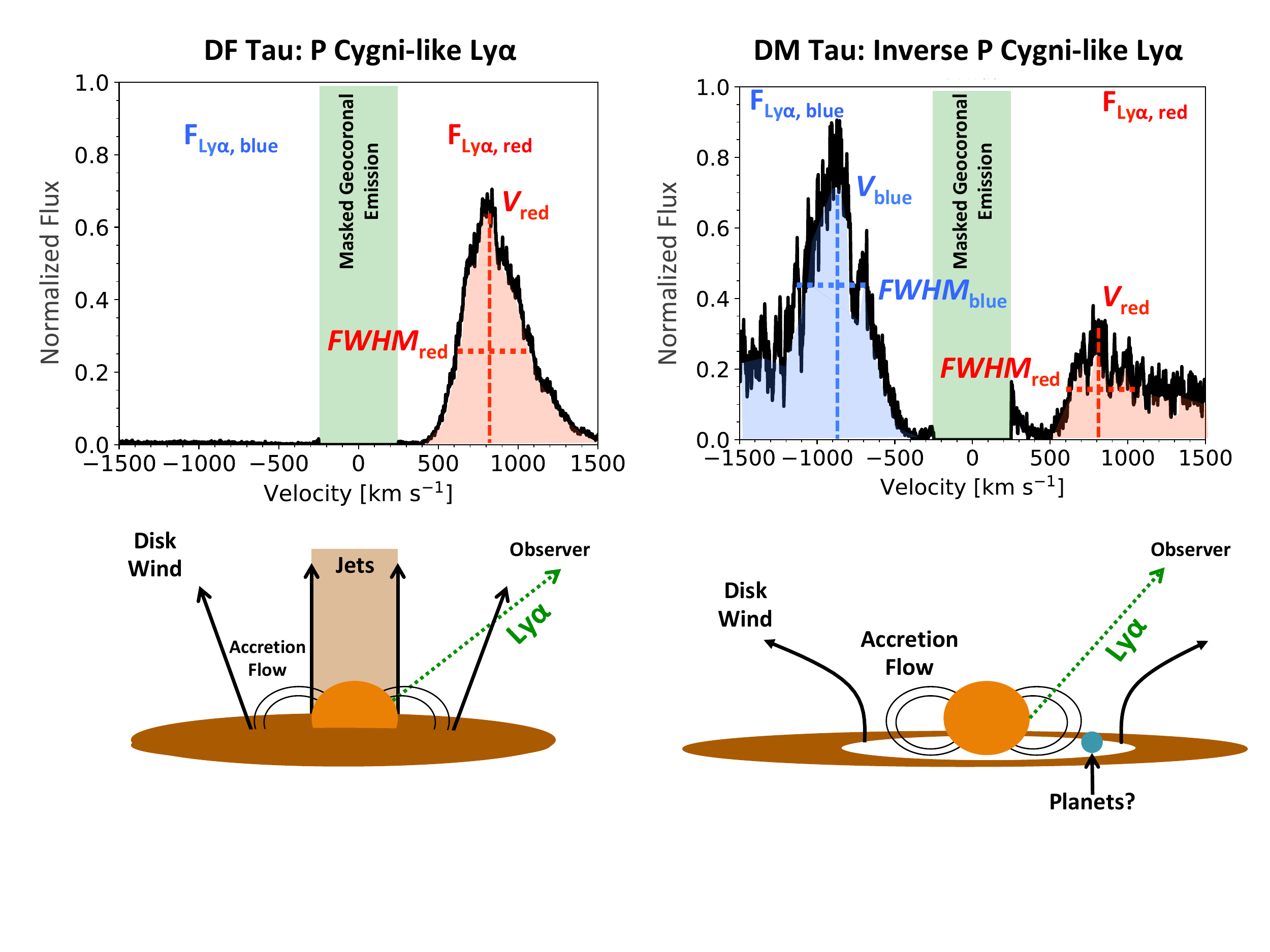}
    \caption{Characteristic Ly$\alpha$ emission lines observed in \emph{HST}-COS spectra from ULLYSES targets. We parameterize the observed Ly$\alpha$ profiles using the integrated fluxes at $v > 0$ ($F_{\rm{Ly}\alpha, \rm{red}}$) and $v < 0$ ($F_{\rm{Ly}\alpha, \rm{blue}}$) and the peak velocities and FWHMs of the red ($v_{\rm{red}}$, $FWHM_{\rm{red}}$) and blue ($v_{\rm{blue}}$, $FWHM_{\rm{blue}}$) emission line wings. DF Tau and DM Tau have the largest and smallest ratios of $F_{\rm{Ly}\alpha, \rm{red}} / F_{\rm{Ly}\alpha, \rm{blue}}$, respectively, within our sample. DF Tau shows P Cygni-like emission, where outflowing H I contained in jets and inner disk winds has suppressed the blue side of the profile. This scenario is illustrated in the schematic on the left, which also shows that the targets have smooth, full dust disks at this stage \citep{Arulanantham2021}. DM Tau shows opposite behavior, where infalling H I has absorbed photons on the red side of the line profile. The schematic on the right illustrates this stage, showing that large dust gaps consistent with models of planet-disk interactions have been cleared in the dust distributions (see e.g., DM Tau; \citealt{Hashimoto2021, Francis2022}). Low-velocity components of [O I] 6300 \AA \, and [Ne II] 12.81 $\mu$m emission lines are still observed, consistent with origins in MHD winds (see e.g., \citealt{Banzatti2019, Pascucci2020}) or photoevaporative flows \citep{Pascucci2009}. The green boxes in each panel mark velocities between $-250 < v < +250$ km s$^{-1}$, where the Ly$\alpha$ line profiles are dominated by geocoronal emission. We set these fluxes to zero throughout the analysis presented here and calculate velocities relative to line center at 1215.67 \AA.} 
    \label{fig:COS_LyA_demo}
\end{figure*}

The ULLYSES spectra include the Ly$\alpha$ line \citep{Schindhelm2012_UVCO, McJunkin2014} and fluorescent emission from H$_2$ and CO in the protoplanetary disks \citep{Herczeg2002, France2011_CO}. Since \emph{HST}-COS is effectively a slitless spectrograph \citep{Green2012}, Ly$\alpha$ photons from the sky background readily enter the aperture. This produces a narrow $\left(FWHM \sim 250 \, \text{km s}^{-1} \right)$ geocoronal emission line with a peak flux density of $\sim 10^{-12}$ erg s$^{-1}$ cm$^{-1}$ \AA$^{-1}$ at line center ($\lambda \sim 1216$ \AA)\footnote{See \url{https://www.stsci.edu/hst/instrumentation/cos/calibration/airglow} for \emph{HST}-COS calibration spectra of geocoronal Ly$\alpha$ emission.}. The Ly$\alpha$ emission from T Tauri stars is heavily absorbed by the ISM and much weaker than the geocoronal emission at line center. However, the line profiles from our targets are sufficiently broadened (FWHM $\sim$ 550-850 km s$^{-1}$; \citealt{Schindhelm2012}), such that Ly$\alpha$ photons from high velocities where geocoronal emission is not produced are readily detectable. These features appear double-peaked, where separations between line wings are much larger than both the instrumental resolution $\left( \Delta v \sim 18 \, \rm{km \, s}^{-1} \right)$ and the extent of the geocoronal emission $\left(FWHM \sim 250 \, \text{km s}^{-1} \right)$. For this reason, we do not fit for any radial velocity shifts from line center but rather measure the properties of the redder wing relative to the bluer wing (see Figure \ref{fig:COS_LyA_demo}). The continuum emission is effectively zero, so we do not fit it for the sake of model simplicity (see also \citealt{McJunkin2014}).  

Seventeen targets in our sample was also observed through the ESO PENELLOPE program \citep{Manara2021}, which uses the ESPRESSO, UVES, and X-Shooter instruments on the Very Large Telescope (VLT) to acquire contemporaneous UV, optical, and near-IR data for all ULLYSES targets. The spectra include H$\alpha$ emission, which also originates within the CTTS magnetospheres and the base of the disk winds (see e.g., \citealt{Lima2010}). Additionally, the near-IR X-Shooter \textit{K}-band observations can be used to trace hot inner disk material that is readily exposed to Ly$\alpha$ photons (see e.g., \citealt{McClure2013}). A detailed description of the data reduction procedures for the PENELLOPE program is available in \citep{Manara2021}.   

Table \ref{tab:targ_props} includes relevant published disk properties for all targets in our sample, including \textit{WISE} photometry \citep{Wright2010} from dataset \dataset[10.26131/IRSA142]{10.26131/IRSA142}, \textit{Herschel} 70 $\mu$m flux densities, and outer disk inclinations. Mass accretion rates based on Gaia DR3 distances for the systems in Orion OB1 are from \citet{Pittman2022}, a range for XX Cha based on its Gaia DR3 distance from \citet{Claes2022}, and the remaining values taken from the ULLYSES documentation (based on stellar positions from Gaia DR2). We note that upcoming work will extract updated measurements from all new spectra (Claes et al., Pittman et al., Wendeborn et al., in prep). All stellar coordinates, spectral types, masses, and extinction corrections adopted by the ULLYSES implementation team are provided at \url{https://ullyses.stsci.edu} and will be modified as the program is completed.  
\startlongtable
\begin{deluxetable*}{ccccccc}
\tablecaption{Sample of Targets \label{tab:targ_props}} 
\tablehead{
\colhead{Target Name} & \colhead{$\log{M_{acc}}$\tablenotemark{a}} & \colhead{\textit{W3}\tablenotemark{b}} & \colhead{\textit{W4}\tablenotemark{c}} & \colhead{$F_{70 \, \mu \text{m}}$\tablenotemark{d}} & \colhead{$i_{disk}$\tablenotemark{e}} & \colhead{ALMA/SMA\tablenotemark{e}} \\
\nocolhead{} & \colhead{($M_{\odot}$ yr$^{-1}$)} & \colhead{(mag)} & \colhead{(mag)} & \colhead{(mJy)} & \colhead{(degrees)} & \colhead{Dust Substructure} \\
}
\startdata
2MASS J11432669-7804454 & -8.71 & 8.676 & 7.052 & \nodata & \nodata & \nodata \\
2MASS J16083070-3828268 & -8.96 & 7.842 & 3.07 & $3400 \pm 27$ & 74 & Cavity at $r = 75$ au \\
AA Tau & -7.82 & 4.645 & 2.505 & $1300 \pm 300$ & $59.1 \pm 0.3$ & Rings at $r = 49, 95, 143$ au \\
CHX18N & -8.09 & \nodata & 3.584 & $300 \pm 100$ & \nodata & \nodata \\
CS Cha & -8.29 & 7.095 & 2.758 & $3200 \pm 600$ & 8 & Cavity at $r = 37$ au \\
CVSO 104 & -7.94\tablenotemark{$\ast$} & 7.56 & 5.571 & $77.41 \pm 2.35$ & $43^{+8}_{-4}$ & \nodata \\
CVSO 107 & -7.32\tablenotemark{$\ast$} & 7.419 & 5.128 & $105.34 \pm 3.32$ & \nodata & \nodata \\
CVSO 109A & -6.76\tablenotemark{$\ast$} & 6.523 & 4.527 & $69.38 \pm 2.69$ & \nodata & \nodata \\
CVSO 146 & -8.28\tablenotemark{$\ast$} & 7.072 & 4.597 & \nodata & \nodata & \nodata \\
CVSO 165A & -8.15\tablenotemark{$\ast$} & 6.571 & 4.626 & \nodata & \nodata & \nodata \\
CVSO 165B & -7.78\tablenotemark{$\ast$} & 6.571 & 4.626 & \nodata & \nodata & \nodata \\
CVSO 176 & -7.38\tablenotemark{$\ast$} & 7.896 & 6.221 & \nodata & \nodata & \nodata \\
CVSO 90 & -7.90\tablenotemark{$\ast$} & 7.622 & 5.345 & \nodata & \nodata & \nodata \\
DE Tau & -7.55 & 4.895 & 2.725 & $1300 \pm 300$ & $34.1 \pm 1.0$ & \nodata \\
DF Tau & -6.75 & 3.829 & 2.27 & $700 \pm 100$ & $24 \pm 5$ & \nodata \\
DK Tau & -7.47 & 3.599 & \nodata & $1170 \pm 120$ & $12.8^{+2.5}_{-2.8}$ & Smooth disk \\
DM Tau & -8.54 & 7.085 & 3.571 & $780 \pm 80$ & $35.2 \pm 0.7$ & Gap between $r = 4-20$ au \\
DN Tau & -8.0 & 5.166 & 3.039 & $700 \pm 100$ & $35.2^{+0.5}_{-0.6}$ & Gap (radius not reported) \\
HT Lup & -8.24 & 2.896 & 0.856 & $3500 \pm 49$ & $48.1 \pm 4.5$ & Spiral arms \\
IN Cha & -9.34 & \nodata & 5.316 & $<287.3$ & \nodata & \nodata \\
LkCa 15 & -8.51 & 5.696 & 3.565 & $1230 \pm 120$ & 55 & Cavity at $r = 76$ au \\
MY Lup & -9.67 & 5.164 & 2.833 & $1100 \pm 22$ & 73 & Rings at $r = 8, 20, 30, 40$ au \\
RECX-11 & -9.7 & 5.388 & 3.718 & $212 \pm 21$ & 70 & \nodata \\
RECX-15 & -9.0 & \nodata & 3.409 & $204 \pm 6$ & 60 & \nodata \\
RU Lup & -7.3 & 2.82 & 0.658 & \nodata & $18.5 \pm 2$ & Eight rings between $r = 14-50$ au \\
RW Aur & -7.50 & 3.375 & 1.448 & $2470 \pm 150$ & $55.51 \pm 0.13$ & Tidal streams \\
RY Lup & -8.19 & 3.647 & 1.404 & \nodata & 68 & Cavity at $r = 69$ au \\
Sz 10 & -8.7 & 7.816 & 5.847 & \nodata & \nodata & \nodata \\
Sz 111 & -9.12 & 8.922 & 5.731 & $1200 \pm 24$ & \nodata & \nodata \\
Sz 130 & -9.19 & 6.53 & 4.514 & \nodata & 53 & Smooth disk \\
Sz 45 & -8.09 & 6.948 & 4.156 & \nodata & \nodata & \nodata \\
Sz 69 & -9.51 & \nodata & \nodata & $3500 \pm 49$ & 69 & Smooth disk \\
Sz 71 & -9.06 & 5.716 & 3.671 & $240 \pm 34$ & 40 & Smooth disk \\
Sz 72 & -8.65 & 5.987 & 3.911 & \nodata & 53 & Smooth disk \\
Sz 75 & -7.67 & 4.342 & 2.379 & \nodata & \nodata & \nodata \\
Sz 76 & -9.26 & 7.366 & 5.274 & \nodata & \nodata & \nodata \\
Sz 77 & -8.79 & 5.362 & 3.437 & \nodata & \nodata & \nodata \\
T Cha & -8.40 & 4.663 & 2.567 & \nodata & 73 & Gap between $r = 18-28$ au \\
TW Hya & -8.7 & 4.54 & 1.516 & \nodata & 7 & Ten rings between $r = 1-52$ au \\
UX Tau & -8.0 & 5.71 & 2.176 & $3300 \pm 700$ & 40 & Cavity at $r = 31$ au \\
V4046 Sgr & -7.89 & 4.966 & 1.93 & \nodata & 34 & Cavity at $r = 31$ au \\
V510 Ori & -8.38 & 5.598 & 3.258 & \nodata & \nodata & \nodata \\
XX Cha\tablenotemark{$\ast \ast$} & -9.06 to -7.69 & \nodata & 3.569 & $190 \pm 40$ & \nodata & \nodata \\
\enddata
\tablecomments{ a) Mass accretion rates are taken from \citealt{Alcala2014, Alcala2017} (Lupus), \citealt{Manara2015, Manara2017} ($\rho$ Ophiuchus, Chamaeleon I), \citet{France2017}, and \citet{Xu2021}; b) \textit{WISE} photometry, 12 $\mu$m \citep{Wright2010}; c) \textit{WISE} photometry, 22 $\mu$m; d) \textit{Herschel} 70 $\mu$m flux densities from \citet{Mauco2016}; e) Sub-mm disk inclinations and dust substructure (unless otherwise indicated) from \citealt{vanderMarel2018} (2MASS J16083070-3828268), \citealt{Loomis2017} (AA Tau), \citealt{Francis2020} (CS Cha, LkCa 15, RY Lup, UX Tau, V4046 Sgr), \citealt{Frasca2021} (CVSO 104; H$\alpha$ modeling), \citealt{Simon2019} (DE Tau), \citealt{Shakhovskoj2006} (DF Tau), \citealt{Long2019} (DK Tau, DN Tau), \citealt{Kudo2018} (DM Tau), \citealt{Kurtovic2018} (HT Lup), \citealt{Huang2018} (MY Lup, RU Lup, TW Hya), \citealt{Lawson2004} (RECX-11, RECX-15; H$\alpha$ modeling), \citealt{Rodriguez2018} (RW Aur), \citealt{Ansdell2016} (Sz 130, Sz 69, Sz 71, Sz 72), \citealt{Hendler2018} (T Cha)}
\tablecomments{ *Mass accretion rates from \citet{Pittman2022}; **Range of mass accretion rates from \citet{Claes2022}}
\end{deluxetable*}

\section{Measured Properties of Ly$\alpha$ Emission Lines from CTTS}

\begin{figure*}
\centering
\includegraphics[width=\linewidth]{ 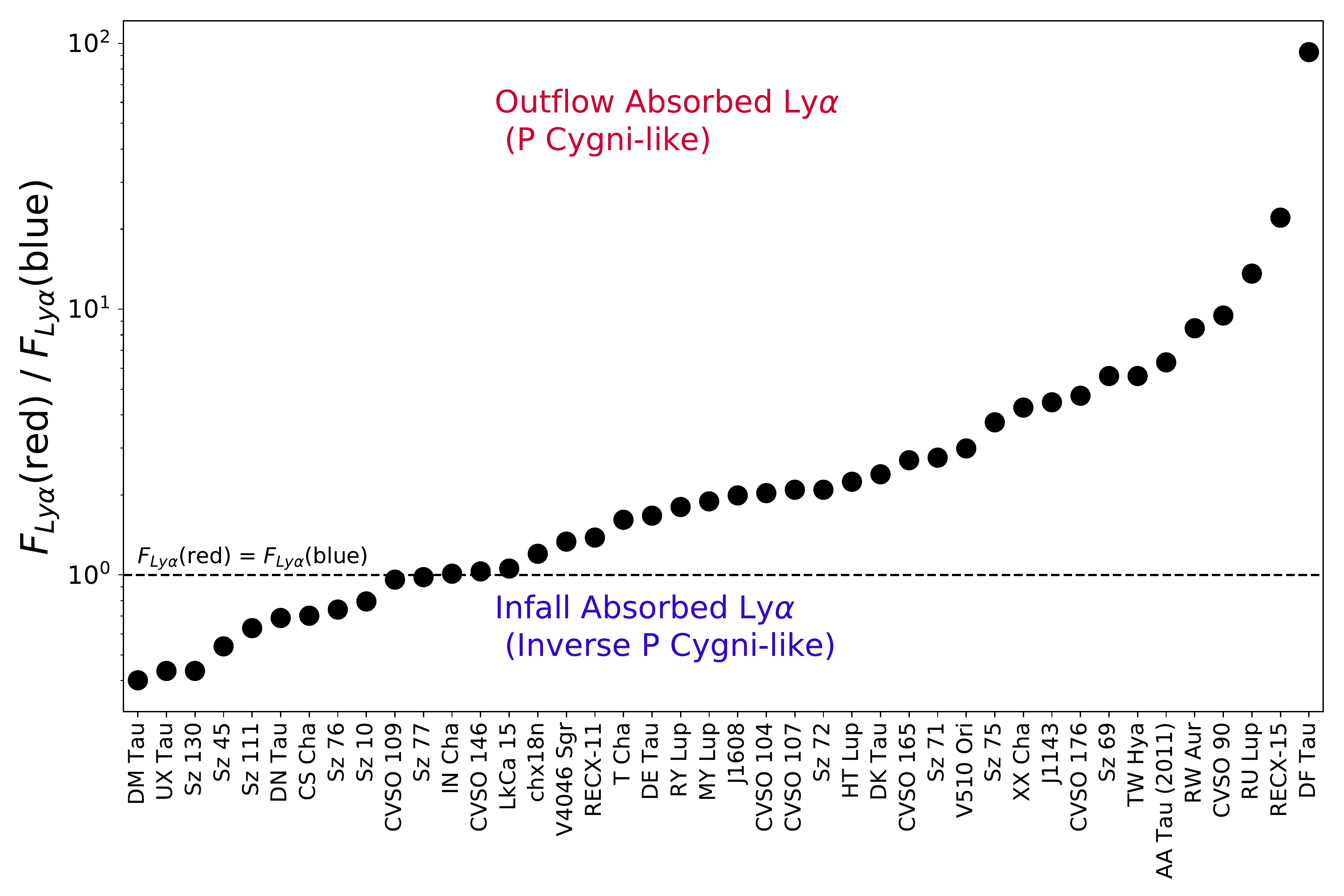}
\caption{Measured ratios of integrated Ly$\alpha$ flux from $\lambda \geq 1215.67$ $\left(F_{\text{Ly}\alpha} \text{(red)} \right)$ and $\lambda < 1215.67$ $\left(F_{\text{Ly}\alpha} \text{(blue)} \right)$ for all targets in our sample, which span three orders of magnitude across the parameter space. Roughly 70\% of targets have $F_{\text{Ly}\alpha} \text{(red)} / F_{\text{Ly}\alpha} \text{(blue)} > 1$, corresponding to outflow absorbed, P Cygni-like emission line profiles. The remaining 30\% have $F_{\text{Ly}\alpha} \text{(red)} / F_{\text{Ly}\alpha} \text{(blue)} < 1$, characteristic of infalling H I suppressing the red sides of the features (inverse P Cygni-like), although a small subset have symmetric profiles primarily absorbed near line center.}
\label{fig:measured_LyA_redtoblue}
\end{figure*}

The observed Ly$\alpha$ emission lines were parameterized by measuring six properties from the spectra in our sample: the integrated fluxes at $v > 0$ ($F_{\rm{Ly}\alpha, \rm{red}}$) and $v < 0$ ($F_{\rm{Ly}\alpha, \rm{blue}}$), and the peak velocities and FWHMs of the red ($v_{\rm{red}}$, $FWHM_{\rm{red}}$) and blue ($v_{\rm{blue}}$, $FWHM_{\rm{blue}}$) emission line wings (see Table \ref{tab:measured_LyA}). These parameters are marked in Figure \ref{fig:COS_LyA_demo}, which shows the observed Ly$\alpha$ emission lines from DF Tau and DM Tau and the rough dust disk morphologies detected for targets with similar Ly$\alpha$ profiles (see Section 5.1 for a more detailed discussion). Figure \ref{fig:measured_LyA_redtoblue} shows the measured ratios of red to blue fluxes for the full sample, which indicate whether blue-shifted outflows or red-shifted accretion flows are the dominant absorbers of Ly$\alpha$ photons. The measured flux ratios span three orders of magnitude, with $0.4 < F_{\text{Ly}\alpha, \text{red}} / F_{\text{Ly}\alpha, \text{blue}} < 93$. Approximately 70\% of targets have $F_{\text{Ly}\alpha, \text{red}} / F_{\text{Ly}\alpha, \text{blue}} > 1$, corresponding to P Cygni-like profiles with strong absorption from outflowing H I (e.g., DF Tau). The remaining $\sim$30\% have more flux on the blue sides of the line profiles, with red emission suppressed by infalling material (inverse P Cygni-like; e.g., DM Tau). A small subset of targets have $F_{\text{Ly}\alpha, \text{red}} / F_{\text{Ly}\alpha, \text{blue}} \sim 1$, corresponding to roughly symmetric Ly$\alpha$ emission lines (V4046 Sgr, CVSO 109A, Sz 77, IN Cha, CVSO 146, LkCa 15, CHX18N). These systems are among the weakest accretors in the sample (with the exception of the known binary CVSO 109; \citealt{Pittman2022}) and have lower integrated Ly$\alpha$ fluxes than the targets with larger accretion rates. The physical mechanisms responsible for producing intermediate Ly$\alpha$ flux ratios are discussed further in Section 5.1.    

\startlongtable
\centerwidetable
\begin{deluxetable*}{c|ccc|ccc|cc}
\tabletypesize{\footnotesize}
\tablecaption{Measured Properties of Ly$\alpha$ Emission Lines \label{tab:measured_LyA}} 
\tablehead{
\colhead{Target Name} & 
\colhead{$F_{\rm{Ly}\alpha, \, \rm{red}}$} &
\colhead{$F_{\rm{Ly}\alpha, \, \rm{blue}}$} & 
\colhead{$\frac{F_{\rm{Ly}\alpha, \, \rm{red}}}{F_{\rm{Ly}\alpha, \, \rm{blue}}}$} & \colhead{$v_{\rm{red}}$} & 
\colhead{$v_{\rm{blue}}$} & 
\colhead{$v_{\rm{red}} - v_{\rm{blue}}$} &
\colhead{$FWHM_{\rm{red}}$} &
\colhead{$FWHM_{\rm{blue}}$}
\\
\nocolhead{} & 
\colhead{[erg s$^{-1}$ cm$^{-2}$]} & 
\colhead{[erg s$^{-1}$ cm$^{-2}$]} & 
\nocolhead{} & 
\colhead{[km s$^{-1}$]} & 
\colhead{[km s$^{-1}$]} & 
\colhead{[km s$^{-1}$]} & 
\colhead{[km s$^{-1}$]} & 
\colhead{[km s$^{-1}$]}
}
\startdata
J11432669-7804454 & $9.3 \times 10^{-14}$ & $2.1 \times 10^{-14}$ & 4.4 & 389 & -330 & 719 & 198 & 232 \\
J16083070-3828268 & $3.9 \times 10^{-15}$ & $1.94 \times 10^{-15}$ & 2.0 & 542 & -359 & 901 & 423 & 260 \\
AA Tau & $2.6 \times 10^{-13}$ & $4.1 \times 10^{-14}$ & 6.3 & 683 & -971 & 1654 & 250 & 1230 \\
CHX18N & $2.8 \times 10^{-14}$ & $2.4 \times 10^{-14}$ & 1.2 & 566 & -439 & 1005 & 515 & 567 \\
CS Cha & $3.7 \times 10^{-12}$ & $5.3 \times 10^{-12}$ & 0.7 & 618 & -566 & 1184 & 308 & 371 \\
CVSO 104 & $4.3 \times 10^{-15}$ & $2.1 \times 10^{-15}$ & 2.0 & 602 & -575 & 1177 & 303 & 355 \\
CVSO 107 & $2.0 \times 10^{-14}$ & $9.6 \times 10^{-15}$ & 2.1 & 541 & -1160 & 1700 & 258 & 1408 \\
CVSO 109 & $6.7 \times 10^{-15}$ & $6.9 \times 10^{-15}$ & 0.96 & 599 & -719 & 1318 & 302 & 241 \\
CVSO 146 & $4.9 \times 10^{-15}$ & $4.8 \times 10^{-15}$ & 1.0 & 644 & \nodata & \nodata & 589 & \nodata \\
CVSO 165 & $5.9 \times 10^{-15}$ & $2.2 \times 10^{-15}$ & 2.7 & 686 & \nodata & \nodata & 323 & \nodata \\
CVSO 176 & $7.9 \times 10^{-15}$ & $1.7 \times 10^{-15}$ & 4.7 & 627 & \nodata & \nodata & 322 & \nodata \\
CVSO 90 & $1.9 \times 10^{-14}$ & $2.0 \times 10^{-15}$ & 9.5 & 663 & \nodata & \nodata & 353 & \nodata \\
DE Tau & $1.2 \times 10^{-13}$ & $7.0 \times 10^{-14}$ & 1.7 & 654 & -1192 & 1846 & 351 & 1450 \\
DF Tau & $5.1 \times 10^{-12}$ & $5.5 \times 10^{-14}$ & 93 & 876 & \nodata & \nodata & 400 & \nodata \\
DK Tau & $1.4 \times 10^{-13}$ & $5.7 \times 10^{-14}$ & 2.4 & 585 & \nodata & \nodata & 474 & \nodata \\
DM Tau & $3.4 \times 10^{-13}$ & $8.5 \times 10^{-13}$ & 0.4 & 1215 & -849 & 2064 & 1385 & 389 \\
DN Tau & $1.2 \times 10^{-13}$ & $1.7 \times 10^{-13}$ & 0.69 & 724 & -1146 & 1870 & \nodata & \nodata \\
HT Lup & $3.2 \times 10^{-15}$ & $1.4 \times 10^{-15}$ & 2.2 & 644 & \nodata & \nodata & 156 & \nodata \\
IN Cha & $2.5 \times 10^{-15}$ & $2.4 \times 10^{-15}$ & 1.0 & \nodata & \nodata & \nodata & \nodata & \nodata \\
LkCa 15 & $9.6 \times 10^{-14}$ & $9.1 \times 10^{-14}$ & 1.1 & \nodata & \nodata & \nodata & \nodata & \nodata \\
MY Lup & $6.8 \times 10^{-15}$ & $3.6 \times 10^{-15}$ & 1.9 & 501 & -365 & 865 & 318 & 409 \\
RECX-11 & $2.9 \times 10^{-12}$ & $2.1 \times 10^{-12}$ & 1.4 & 397 & -381 & 777 & 178 & 275 \\
RECX-15 & $1.9 \times 10^{-11}$ & $8.6 \times 10^{-13}$ & 22 & 410 & -345 & 755 & 243 & 182 \\
RU Lup & $2.8 \times 10^{-13}$ & $2.1 \times 10^{-14}$ & 14 & 428 & -481 & 909 & 246 & 264 \\
RW Aur & $2.7 \times 10^{-13}$ & $3.1 \times 10^{-14}$ & 8.5 & 620 & -386 & 1006 & 407 & 231 \\
RY Lup & $2.7 \times 10^{-13}$ & $1.5 \times 10^{-13}$ & 1.8 & 463 & -496 & 958 & 281 & 391 \\
Sz 10 & $4.4 \times 10^{-14}$ & $5.5 \times 10^{-14}$ & 0.79 & 555 & -472 & 1027 & 282 & 195 \\
Sz 111 & $8.4 \times 10^{-15}$ & $1.3 \times 10^{-14}$ & 0.63 & 670 & -734 & 1404 & 246 & 408 \\
Sz 130 & $4.4 \times 10^{-15}$ & $1.0 \times 10^{-14}$ & 0.44 & 531 & -593 & 1124 & 358 & 589 \\
Sz 45 & $4.7 \times 10^{-14}$ & $8.8 \times 10^{-14}$ & 0.54 & 602 & -677 & 1278 & 80 & 346 \\
Sz 69 & $6.3 \times 10^{-15}$ & $1.1 \times 10^{-15}$ & 5.6 & 720 & -400 & 1119 & 434 & 340 \\
Sz 71 & $1.5 \times 10^{-14}$ & $5.4 \times 10^{-15}$ & 2.8 & 529 & -594 & 1123 & 309 & 276 \\
Sz 72 & $2.4 \times 10^{-14}$ & $1.1 \times 10^{-14}$ & 2.1 & 625 & \nodata & \nodata & 213 & \nodata \\
Sz 75 & $5.0 \times 10^{-13}$ & $1.3 \times 10^{-13}$ & 3.8 & 649 & -1113 & 1762 & 321 & 1407 \\
Sz 76 & $4.2 \times 10^{-15}$ & $5.6 \times 10^{-15}$ & 0.74 & 430 & -764 & 1194 & 239 & 951 \\
Sz 77 & $2.1 \times 10^{-14}$ & $2.2 \times 10^{-14}$ & 0.98 & 590 & -714 & 1304 & 386 & 615 \\
T Cha & $2.9 \times 10^{-14}$ & $1.8 \times 10^{-14}$ & 1.6 & \nodata & -286 & \nodata & \nodata & 164 \\
TW Hya & $5.2 \times 10^{-11}$ & $9.2 \times 10^{-12}$ & 5.6 & 388 & -529 & 917 & 366 & 397 \\
UX Tau & $1.0 \times 10^{-13}$ & $2.3 \times 10^{-13}$ & 0.43 & 792 & -754 & 1546 & 344 & 464 \\
V4046 Sgr & $5.2 \times 10^{-12}$ & $3.9 \times 10^{-12}$ & 1.3 & 421 & -468 & 890 & 262 & 347 \\
V510 Ori & $1.3 \times 10^{-14}$ & $4.3 \times 10^{-15}$ & 3.0 & 710 & \nodata & \nodata & 306 & \nodata \\
XX Cha & $1.04 \times 10^{-13}$ & $2.4 \times 10^{-14}$ & 4.3 & 740 & -489 & 1229 & 356 & 323 \\
\enddata
\tablecomments{Uncertainties in $v_{\rm{red}}$, $v_{\rm{blue}}$, $FWHM_{\rm{red}}$, and $FWHM_{\rm{blue}}$ are dominated by the instrument spectral resolution ($\Delta v \sim 17$ km s$^{-1}$). Flux uncertainties are determined by the data calibration pipeline and are on the order of $\sim$5-10\%.}
\end{deluxetable*}

\begin{figure*}
    \centering
    \includegraphics[width=\linewidth]{ 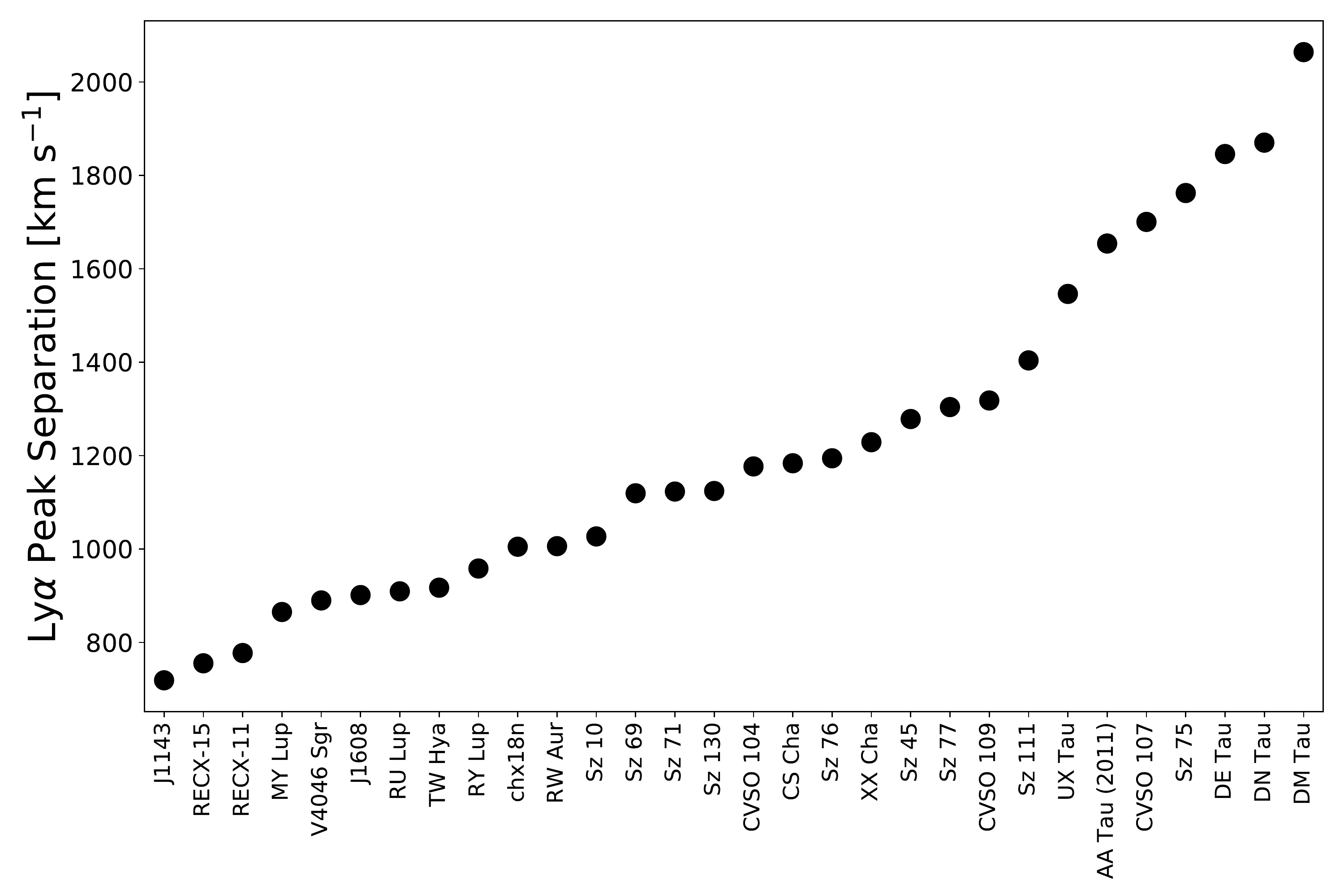}
    \caption{Measured separations between peaks of blue and red Ly$\alpha$ emission line wings for all targets in our sample with distinct double-peaked line profiles ($N = 30/42$, or $\sim$70\%). All targets sit well above the threshold of $v_{\rm{red}} - v_{\rm{blue}} \sim 300$ km s$^{-1}$, which roughly corresponds to $N_{\rm{H \, I}} = 10^{17}$ cm$^{-2}$ \citep{Verhamme2015, Verhamme2017}.}
    \label{fig:LyApeaksep}
\end{figure*}

Figure \ref{fig:LyApeaksep} shows the measured velocity separations between the peaks of the red and blue Ly$\alpha$ emission line wings $\left(v_{\rm{red}} - v_{\rm{blue}} \right)$, for all targets with distinct double-peaked line profiles ($N = 30/42$). Of the remaining 12 systems, nine have P Cygni-like profiles with very little Ly$\alpha$ flux above the continuum at $v < 0$, two have inverse P Cygni-like profiles that are almost entirely suppressed at $v > 0$, and one has a symmetric profile with peaks that are hidden by the geocoronal emission. Models of Ly$\alpha$ emission in the extragalactic context show that column densities of neutral, intervening H I are correlated with $v_{\rm{red}}$, $v_{\rm{blue}}$, and $v_{\rm{red}} - v_{\rm{blue}}$ measured from double-peaked profiles \citet{Verhamme2015, Verhamme2017}, which appear similar to the wings observed from T Tauri stars. The lower limit of the grid explored in the extragalactic work, below which the H I medium becomes optically thin, corresponds to $N_{\rm{H \, I}} = 10^{17}$ cm$^{-2}$ when $v_{\rm{red}} - v_{\rm{blue}} < 300$ km s$^{-1}$. Our sample ranges from $721 < v_{\rm{red}} - v_{\rm{blue}} < 2070$ km s$^{-1}$, indicative of much larger column densities ($N\left(\rm{H \, I} \right) > 10^{20}$ cm$^{-2}$). This lower limit is roughly consistent with the results from \citet{McJunkin2014}, who measured H I column densities between $\log_{10} N \left(\rm{H \, I} \right) \sim 19.6–21.1$ by fitting the red wings of the Ly$\alpha$ emission line profiles from a sample of T Tauri and Herbig Ae/Be stars.

\section{Description of Models}

In order to explore the physical mechanisms responsible for the range in Ly$\alpha$ emission line shapes observed in the ULLYSES sample, we fit the spectra with simple models of Ly$\alpha$ radiative transfer through H I in protoplanetary disk systems. The photons originate in the protostellar chromospheres and at the accretion shocks, then propagate through the accretion flows, inner disk winds and jets, circumstellar disks, outer disk winds, and ISM (see e.g., \citealt{McJunkin2014}). Although this medium includes multiple distinct populations of gas with a superposition of infalling, outflowing, and rotating velocity signatures, the goal of this work is to constrain the bulk properties of intervening gas along the line-of-sight to these sources. The two modeling frameworks used to reproduce the line profiles are described in the following sections.

\subsection{Shell Models With Ly$\alpha$ Emission and Absorption}

Previous work on the observed Ly$\alpha$ emission lines from T Tauri stars prioritized constraining column densities of H I along the line of sight \citep{McJunkin2014, Espaillat2022_ODYSSEUSI}, which could be accurately measured from the red wings of the line profiles. By fitting the data with models of superimposed broad and narrow Gaussian emission components and a single Voigt absorption feature, \citet{McJunkin2014} determined much lower ISM extinction values than measured from X-ray, optical, and IR observations. No correlations were detected between the best fit H I column densities and disk inclinations, implying that the bulk of the absorbing gas is contained in the ISM instead of within the circumstellar disks themselves. Since those model fits were optimized for the red sides of the observed Ly$\alpha$ profiles $\left(\lambda > 1216 \, \text{\AA} \right)$, the data at $\lambda < 1216$ \AA \, carrying signatures of stellar and disk winds were not explored (see also \citealt{Herczeg2004}). 

Other efforts to model the Ly$\alpha$ emission line wings used fluorescent emission from electronically excited CO as an additional constraint on the low velocity Ly$\alpha$ flux that reaches the protoplanetary disk surface but is hidden from \emph{HST} by geocoronal emission and interstellar absorption \citep{Schindhelm2012_UVCO}. Since Ly$\alpha$ photons from both the blue and red sides of the line profiles are required to excite the electronic transitions of CO, these Ly$\alpha$ models included a second Voigt absorption to represent outflowing H I in disk winds, in addition to the broad and narrow emission components and Voigt absorption from the ISM used in later work \citep{McJunkin2014}. The best-fit Ly$\alpha$ models constrain the total flux reaching the CO molecules, in order to reproduce the fluorescent CO emission lines. However, targets like DF Tau still required some additional absorption component to fully reproduce the observed Ly$\alpha$ profile, which is almost fully suppressed at $v < 0$ \citep{Schindhelm2012_UVCO}. 

While the superposition of emission and absorption components used in previous work to represent the stellar-disk-wind environment provides good fits to the Ly$\alpha$ profiles with high S/N, the large number of model parameters makes it difficult to fit similar models to the noisier targets from the ULLYSES sample. To explore the simplest possible framework for all targets in our sample, we first compare the data to models with a single Gaussian emission line and a Voigt absorption component that includes all intervening H I from the ISM, disk wind, and circumstellar disk. This simplest case model already includes five free parameters: the amplitude $\left(I_0 \right)$, FWHM, and central wavelength $\left(\lambda_0 \right)$ of the intrinsic Gaussian emission line, and the column density $\left(N_{\rm{out}} \right)$ and velocity $\left(v_{\rm{out}} \right)$ of absorbing H I. The full model profile is calculated as:
\begin{equation}
\begin{aligned}
I_{\rm{Ly}\alpha} \left(\lambda_i \right) = I_0 \times \exp{\left[\frac{- \left(\lambda_i - \lambda_0 \right)^2}{2 \times \left(\rm{FWHM} / 2 \sqrt{2 * \rm{ln}{2}}\right)^2} \right]} \\ \times \exp{\left[-\tau_{\rm{out}} \left(N_{\rm{out}}, v_{\rm{out}} \right) \right]},
\end{aligned}
\end{equation}
where $\tau_{\rm{out}} = \sigma_{\rm{Ly}\alpha} N_{\rm{out}}$ is the optical depth of the Voigt absorption and $\sigma_{\rm{Ly}\alpha}$ is the absorption cross section. Since the low velocity regions of the Ly$\alpha$ spectra $\left(|v| < 250 \, \text{km s}^{-1} \right)$ are hidden by geocoronal emission (and likely fully absorbed by the ISM), the central portions of the line profiles were masked by setting the fluxes to zero before comparing the data and models.

The best fit model parameters were determined by minimizing the mean square error (MSE):
\begin{equation}
\rm{MSE} = \frac{1}{N} \sum_{i=0}^N \left(y_i - I_{\rm{Ly}\alpha, i} \right)^2,
\end{equation}
where $y_i$ and $I_{\rm{Ly}\alpha, i}$ represent the observed and model fluxes, respectively, at each wavelength along the Ly$\alpha$ profiles. We chose the MSE test statistic because it does not include the flux measurement uncertainties, which are derived from the \emph{HST}-COS data reduction pipeline and are anomalously small for low S/N spectra (see e.g., \citealt{Arulanantham2021}). This biases test statistics that include the flux errors (e.g. $\chi^2$) toward models that fit the continuum well but not the high S/N emission line wings. An MCMC sampler \citep{emcee2013} with a Gaussian log-likelihood function was then used to determine the uncertainties on the model parameters that produced the smallest MSE, by exploring a set of uniform priors for each variable with 300 walkers over 1500 steps and a correction factor for the underpredicted flux uncertainties (see Figure \ref{fig:Sz111_noscattering_corner}). We assessed convergence of the chain by computing the acceptance fraction and re-initialized the sampler around the parameters that maximized the negative log-likelihood function if fewer than $\sim$30\% of samples were accepted.

The most robustly constrained prior is on the H I column densities, based on the results from \citet{McJunkin2014}. The smallest value measured in that sample was $\log N \left(H \, I \right) = 19.2$ for the weak accretor TWA 3A, and the largest was $\log N \left(H \, I \right) = 21.05$ for IP Tau. Here we explored values between $\log N \left(H \, I \right) = 18.5-22.0$ in order to capture the full posterior probability distributions. The Gaussian amplitudes were allowed to vary between $10^{-16} < I_0 < 10^{-11}$ erg s$^{-1}$ cm$^{-2}$ \AA$^{-1}$ and the FWHMs between $250 < \text{FWHM} < 2000$ km s$^{-1}$. Since the low velocity regions of the observed Ly$\alpha$ profiles are fully absorbed, we restricted the model velocities of absorbing gas to $|v_{out}| < 250$ km s$^{-1}$. These model parameters and priors for the MCMC sampler are summarized in Table \ref{tab:model_params_noscatt}.

\subsection{Resonant Scattering in a Spherical Shell}

In this work, we resorted to the highly simplified `shell-model' in order to include resonant scattering in fits to the observed \Lya spectra \citep{Ahn2001,Verhamme2015}. This framework consists of a \Lya (and adjacent continuum) emitting source surrounded by an isotropic shell of neutral hydrogen and dust. The source can be parametrized by the intrinsic \Lya equivalent width $EW_{\rm i}$ and the intrinsic width of the \Lya line $\sigma_{\rm i}$. The shell is characterized by its neutral hydrogen column density $N_{\rm HI}$, the (absorbing) dust optical depth $\tau_{\rm d}$, the (effective) temperature $T$ (which includes thermal as well as turbulent motion), and the outflowing (or infalling if $<0$) velocity $v_{\rm exp}$.

Despite its simplicity, the shell-model can fit a range of observed spectra \citep[in the extragalactic context, e.g.,][]{Verhamme2008,2017A&A...608A.139G} -- including the ones presented here. However, it is clear that a real scattering geometry is neither isotropic nor has constant velocity and density profiles as assumed in the shell model. There is, therefore, an ongoing debate in the literature regarding the usefulness of shell model fitting as well as the physical interpretation of the derived parameters \citep[e.g.,][]{Orlitova2018,Li2022}. What is clear from this debate is that if at all, shell model parameters cannot be interpreted literally but instead represent some weighted quantity probed by \Lya photons. In this case, we can interpret the model parameters as representative of the bulk intervening H I along the dominant photon trajectory, contained in accretion flows, jets and winds, protoplanetary disks, and the ISM.

In order to recover the physical parameters from the \Lya spectra observed with \emph{HST}-COS, one can compare the data to a suite of synthetic spectra which stem from radiative transfer calculations through simplified geometries. In this work, we used the improved fitting pipeline originally described in \citet{Gronke2015}, which employs an affine invariant Monte-Carlo algorithm to map out the likelihood landscape \citep{emcee2013}. First, the pipeline searches for the global maximum of the likelihood function. This is non-trivial due to the multi-modal and non-Gaussian nature of the likelihood function, but we use a `basinhopping' algorithm in which we randomly sample the discrete parameters ($N_{\rm HI}$, $v_{\rm exp}$, and $T$) and at each point use a COBYLA optimization algorithm to maximize efficiently (using gradients) over the smooth algorithm \citep{Powell1994}. This is repeated until the global maximum is not changed for 300 steps. Afterwards, the MC walkers are initialized around that maximum. If any of the walkers find a point with a greater likelihood, we restart the MC sampling and use that new global maximum as starting point.

Prior to feeding the ULLYSES data into the model fitting pipeline, we masked the observed spectra at velocities where geocoronal emission dominates the observed profiles $\left(|v| < 250 \, \rm{km \, s}^{-1} \right)$ by setting the fluxes to zero and increasing the associated uncertainties by $\sim$5 orders of magnitude. This effectively removes the geocoronal emission from the likelihood calculation without modifying the pipeline itself. However, this leaves the model fluxes at line center unconstrained, adding larger uncertainties to the model fits for the observed emission line wings. To correct for this, we added an overall normalization $A$ to the integrated Ly$\alpha$ flux as a free parameter. Figure \ref{fig:Sz111_scattering_corner} shows an example corner plot from the scattering model for Sz 111, illustrating the marginalized probability distributions of the parameters over the likelihood space sampled by the MCMC routines and showing intrinsic degeneracies between the model parameters. These degeneracies are further discussed in Section 5.3, and the model parameters and priors for the MCMC sampler are summarized in Table \ref{tab:model_params_scatt}. 

Double-peaked Ly$\alpha$ emission lines similar to those observed from T Tauri stars can also be reproduced by models where photons scatter through a multi-phase outflow \citep{Li2022}, with a velocity profile that decreases as a function of radial distance from the Ly$\alpha$ source \citep{Gronke2016}. Such models are highly dependent on the covering factor and column densities of intervening H I clumps, along with the power law index describing the wind velocity structure. These models consist typically of many more parameters than the shell model, and a full discussion is beyond the scope of this work. However, the models can be roughly divided into two subgroups, with a covering factor below or above a critical value\footnote{This critical number of clumps per line-of-sight depends on other parameters and can be analytically computed \citep{Gronke2017}. -- with the former and latter predicting high or low flux at line center, respectively.}. The resulting spectra also predict excess Ly$\alpha$ emission at line center, unfortunately, where geocoronal emission dominates any remaining signal from our sources. Since observational constraints on these parameters are difficult to extract with \emph{HST}-COS, spatially resolved spectra from e.g. \emph{HST}-STIS are required to trace the propagation of Ly$\alpha$ photons through the individual components of T Tauri environments using multi-phase models instead of the shell framework. Uncontaminated data from Ly$\alpha$ line center will also enable measurements of how much flux scatters into the molecular gas disk, providing a critical input for physical-chemical models (see e.g., \citealt{Bergin2003, Bethell2011}).

\section{Results}

\subsection{Ly$\alpha$ Properties Derived from Shell Models}

We fit two different simple shell models to the Ly$\alpha$ emission lines from \emph{HST}-COS spectra: one with a single Gaussian emission profile and Voigt absorption, and one with resonant scattering effects included. The observed double-peaked emission line wings are well reproduced with the resonant scattering models for 27/42 targets. Four of the remaining 15 systems had best-fit models without two distinct peaks, despite showing double-peaked profiles in the observed spectra (AA Tau, HT Lup, J16083070-3828268, Sz 77), and the rest had low S/N emission line wings for which the scattering models did not converge. 

\begin{figure*}
\gridline{\fig{ 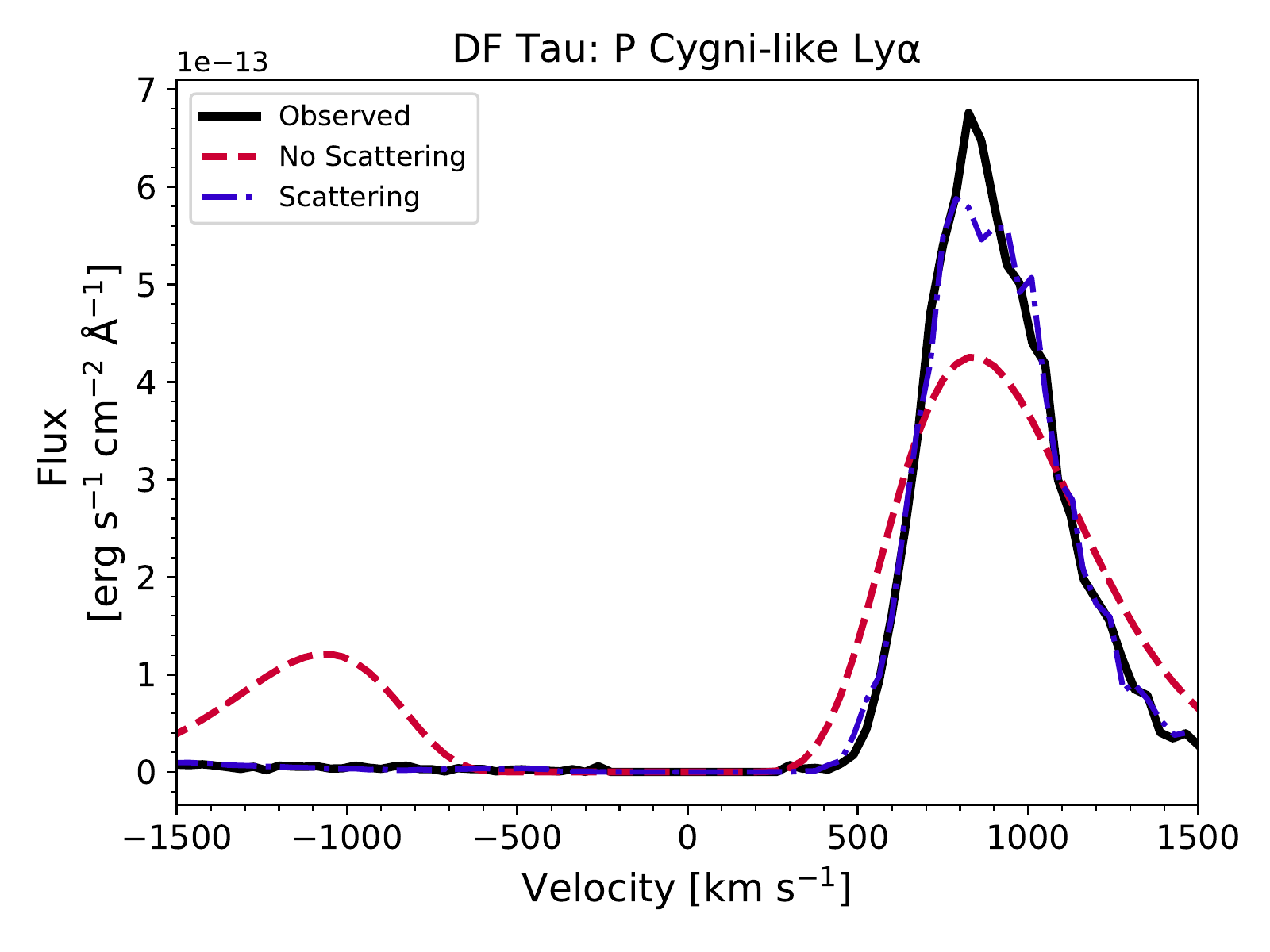}{0.5\textwidth}{ }
\fig{ 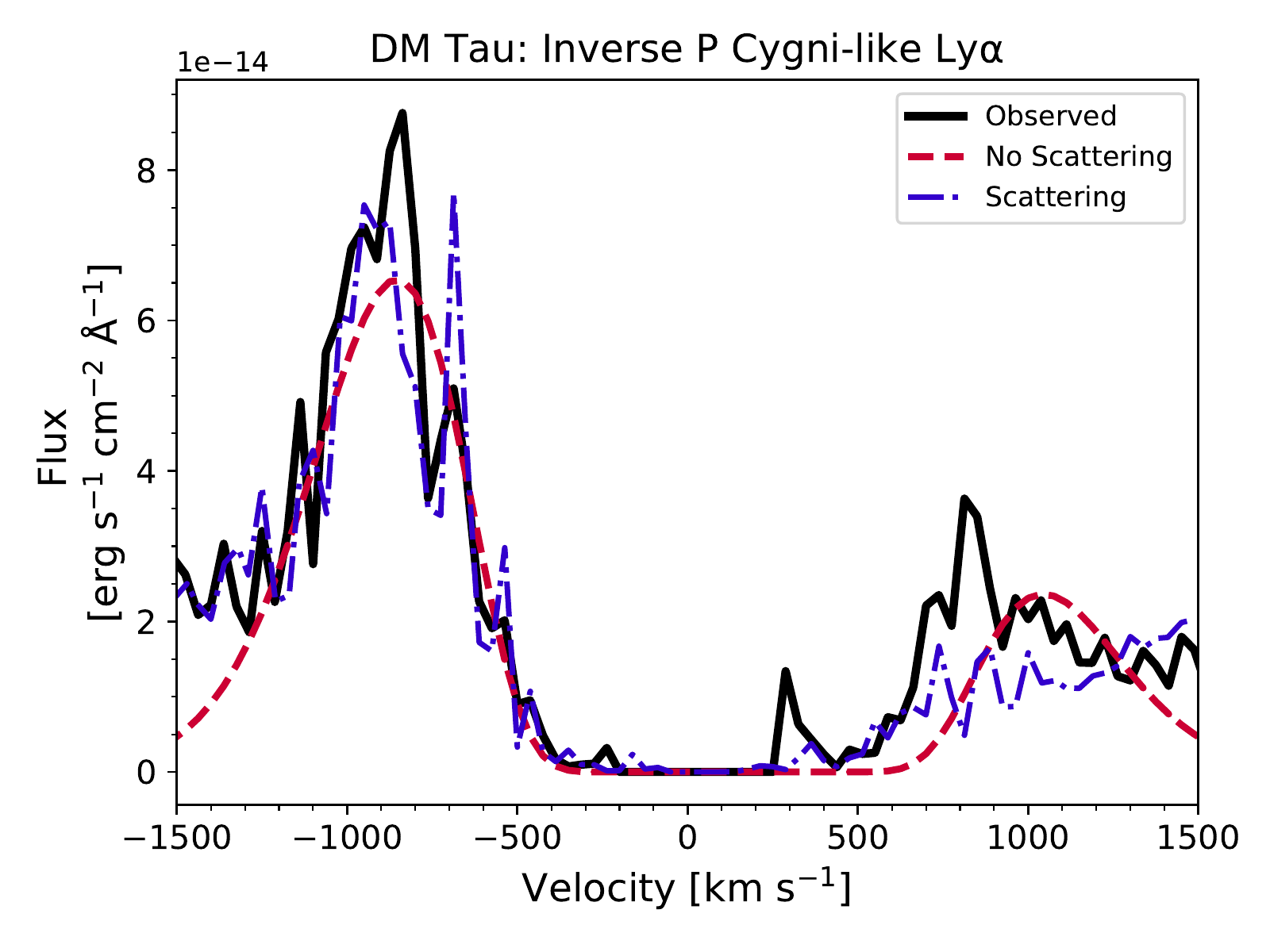}{0.5\textwidth}{ }}
\caption{Comparison of simple shell model fits from a target with P Cygni-like, outflow absorbed Ly$\alpha$ emission \emph{(DF Tau; left)} and a system with an inverse P Cygni-like, infall absorbed profile \emph{(DM Tau; right)}. Observed fluxes between $\pm$250 km s$^{-1}$ are masked by geocoronal emission, so these velocities were de-prioritized in the model fitting routines. The scattering models are better able to reproduce the observed spectra, since they allow the photons to be redistributed in frequency (velocity) space in addition to being removed from the physical system via absorption.}
\label{fig:LyAmodelcomp}
\end{figure*}

Figure \ref{fig:LyAmodelcomp} compares the best-fit models from the frameworks with and without resonant scattering for a target with P Cygni-like Ly$\alpha$ emission (DF Tau) and one with inverse P Cygni-like line wings (DM Tau). In the case of DF Tau, the best fit FWHM from the model without scattering ($FWHM = 1200^{+140}_{-90}$ km s$^{-1}$) is more than double the value from the resonant scattering model ($FWHM = 526 \pm 2$ km s$^{-1}$). The best-fit column density in the model without scattering is correspondingly smaller ($N\rm{(H I)} = 20.72^{+0.01}_{-0.2}$ versus $N\rm{(H I)} = 21.39 \pm 0.06$), but the absorber in this model still removes too much flux from the red wing while leaving too much behind in the blue wing. The scattering model, which allows the photons to be redistributed in frequency space in addition to being removed entirely via absorption, is much better able to dilute the flux in the blue wing while preserving the intensity and shape of the red peak that dominates other P Cygni-like line profiles. A similar effect is seen in the best-fit model parameters for DM Tau, although we note that the narrow FWHM predicted by the scattering model $\left(FWHM = 42^{+0.7}_{-0.6} \, \rm{km s}^{-1} \right)$ is smaller than expected from non-accreting systems. Again, the model including resonant scattering was better able to redistribute photons to both the red and blue high velocity emission line wings.  

\begin{figure*}
\gridline{\fig{ 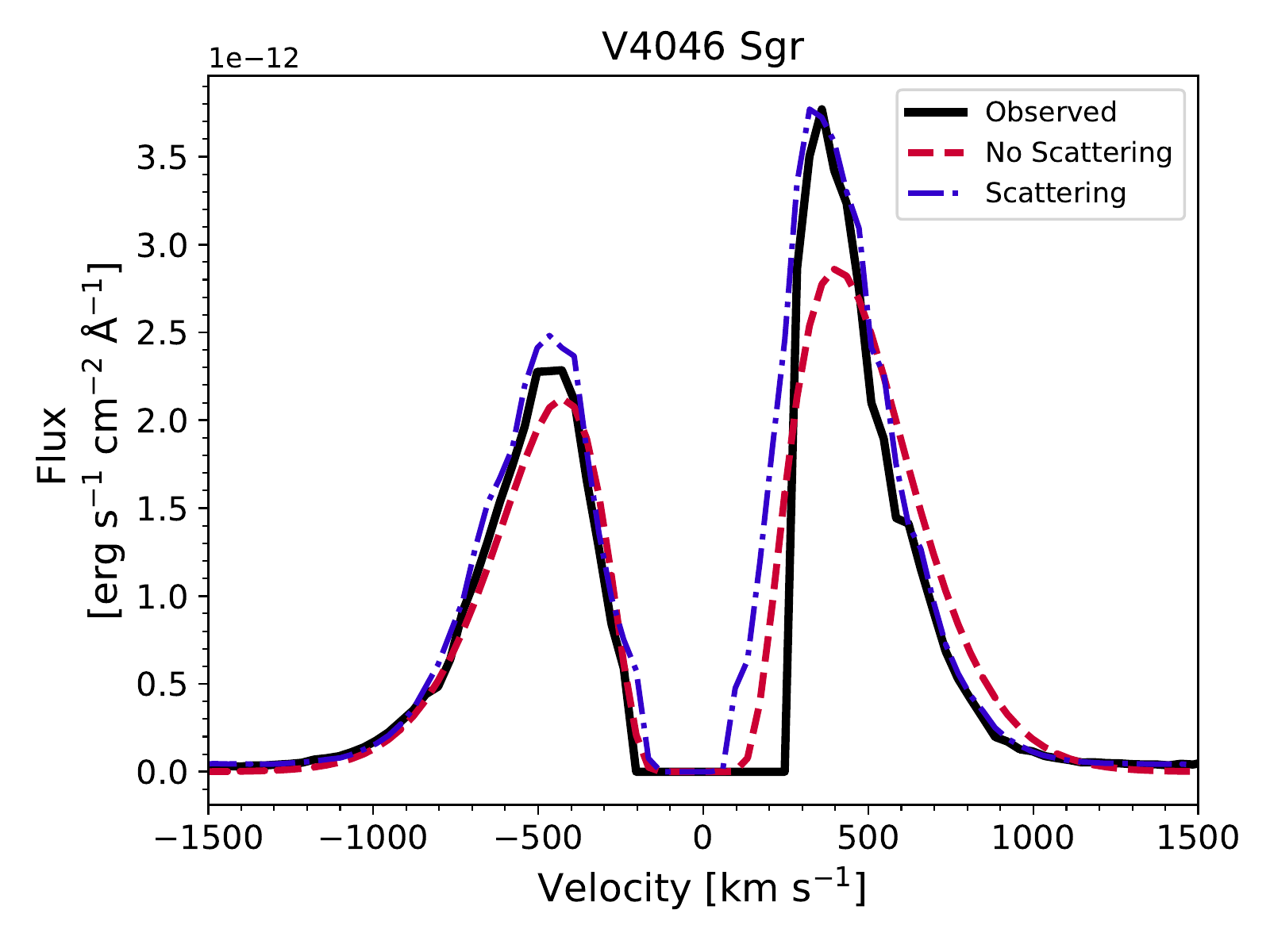}{0.5\textwidth}{ }
\fig{ 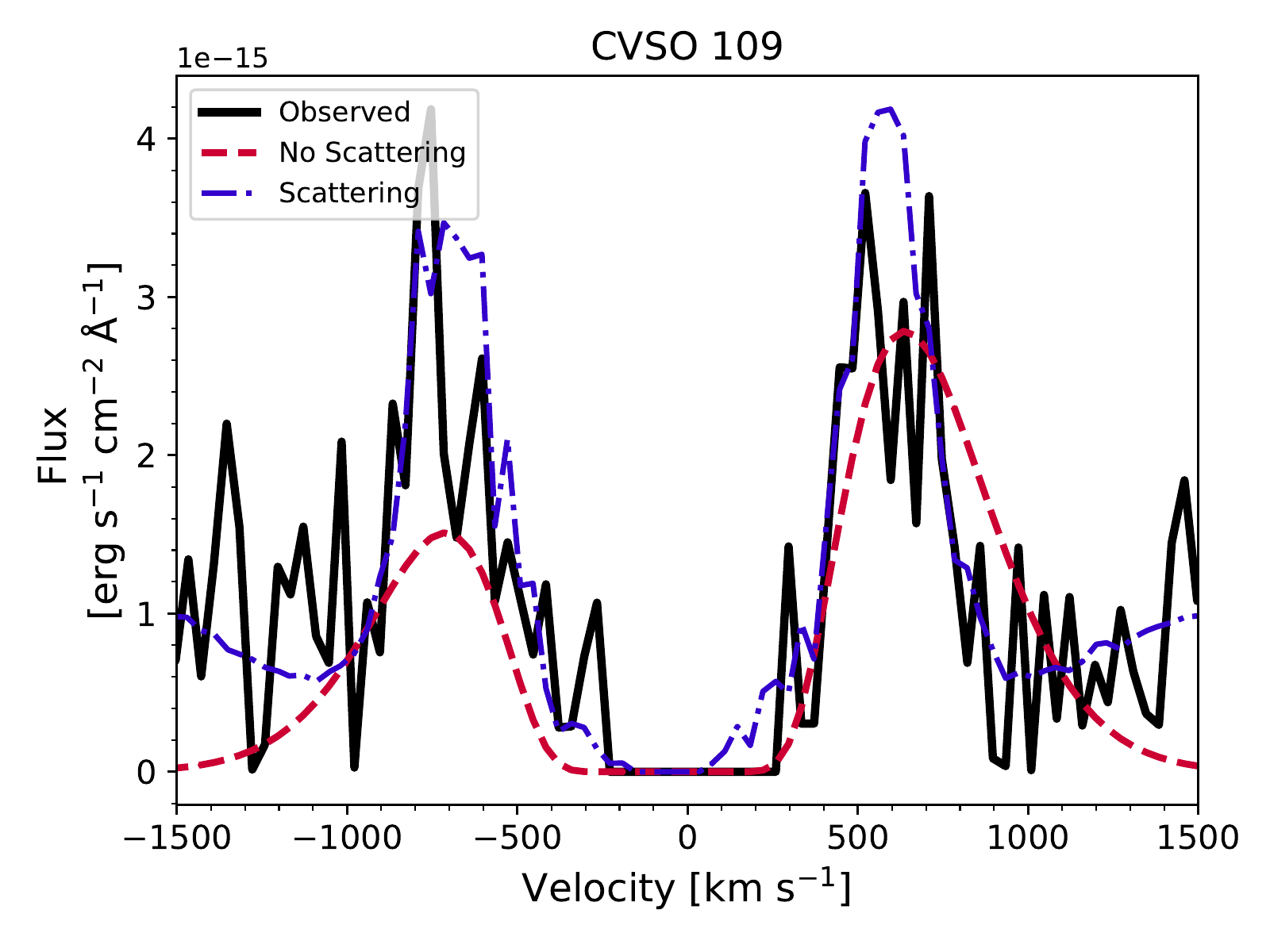}{0.5\textwidth}{ }}
\caption{Comparison of simple shell model fits from targets with intermediate values of $F_{\text{Ly}\alpha} \text{(red)} / F_{\text{Ly}\alpha} \text{(blue)} = 1.3$ (V4046 Sgr, left) and 0.96 (CVSO 109A, right). The other five sources with Ly$\alpha$ flux ratios between $\sim$0.8-1.4 do not have high enough S/N in the emission line wings to fit with the scattering models. Follow-up observations are required to further characterize these intermediate objects, three of which are known binaries (V4046 Sgr, CVSO 109, and Sz 77).}
\label{fig:LyAmodelcomp_intermediate}
\end{figure*}

We also consider two intermediate cases, V4046 Sgr and CVSO 109A, which have $F_{\text{Ly}\alpha} \text{(red)} / F_{\text{Ly}\alpha} \text{(blue)} = 1.3$ and 0.96, respectively (see Figure \ref{fig:LyAmodelcomp_intermediate}). Although the two targets have similar observed flux ratios, the integrated line wing fluxes from CVSO 109 are $\sim$3 orders of magnitude smaller than those measured from V4046 Sgr, and we note that CVSO 109A appears to generate less Ly$\alpha$ emission than the average CTTS \citep{Espaillat2022_ODYSSEUSI}. The remaining targets with Ly$\alpha$ flux ratios between $\sim$0.8-1.4 are all too noisy to fit with the Ly$\alpha$ scattering models (Sz 77, IN Cha, CVSO 146, LkCa 15, CHX18N) but could be reproduced with the models without scattering included. No correlations are detected between the measured Ly$\alpha$ properties (see Table \ref{tab:measured_LyA}) and mass accretion rates, or model H I velocities, column densities, and FWHMs for this subset of targets. Further observations are required to fully characterize the intermediate sources, three of which are known binary systems: V4046 Sgr \citep{deLaReza1986}, CVSO 109 \citep{Proffitt2021}, and Sz 77 \citep{Zagaria2022}.

\begin{figure*}
\centering
\includegraphics[width=\textwidth]{ 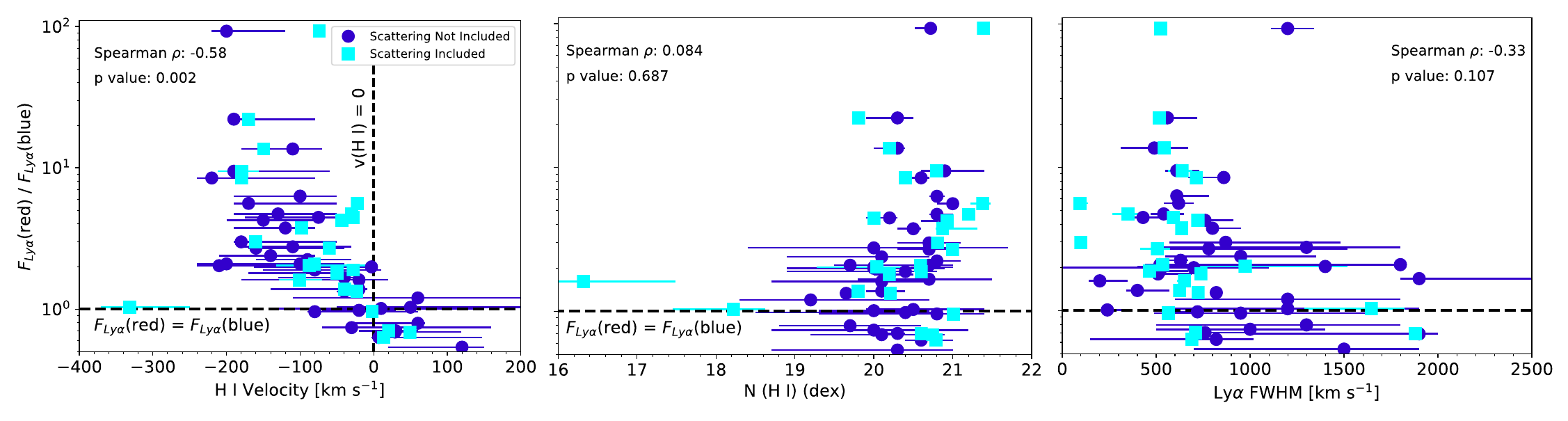}
\caption{Ratios of red to blue integrated Ly$\alpha$ fluxes versus model H I velocities \emph{(left)}, column densities \emph{(middle)}, and emission line FWHMs \emph{(right)}. The velocities of absorbing H I are statistically significantly correlated with the observed emission line profile shapes and generally consistent between models when scattering is and is not included. Outlier values in the upper right and middle panels (($N\left(\rm{H \, I} \right) = 16.3, 18.2$ for T Cha and CVSO 146 and $FWHM = 42, 94, 100$ km s$^{-1}$ for DM Tau, Sz 69, and V510 Ori) likely correspond to models that diverged from the more physical solutions with larger column densities and narrower FWHMs. Spearman $\rho$ and $p$ values are shown between the Ly$\alpha$ flux ratios and parameters from models with scattering included.}
\label{fig:model_param_comp}
\end{figure*}

Table \ref{tab:LyA_model_results} lists the H I velocities, column densities, and initial Ly$\alpha$ FWHMs from both sets of models for the full sample, which are compared to the ratios of red to blue Ly$\alpha$ flux $\left(F_{\text{Ly}\alpha} \text{(red)} / F_{\text{Ly}\alpha} \text{(blue)} \right)$ in Figure \ref{fig:model_param_comp}. The velocities of absorbing H I are the only parameters that are statistically significantly correlated with the observed emission line profile shapes, and values extracted from models when scattering is and is not included are almost always consistent within the 1$\sigma$ uncertainties. One has to bear in mind that degeneracies between the parameters exist, e.g. between $N$(H I), the intrinsic emission line width, and the effective temperature of the medium \citep{Li2022}. It is likely then that the best-fit scattering models with the smallest H I column densities ($N\left(\rm{H \, I} \right) = 16.3, 18.2$ for T Cha and CVSO 146, respectively) diverged from the solutions with larger column densities and narrower FWHMs. We report the solutions favored by the MCMC sampler in Table \ref{tab:LyA_model_results} for consistency across the sample, as the geocoronal emission hides any observational constraints on the intrinsic Ly$\alpha$ FWHMs for this sample. Similarly, scattering models returning the smallest FWHMs ($FWHM = 42, 94, 100$ km s$^{-1}$ for DM Tau, Sz 69, V510 Ori, respectively) could not be distinguished from those with larger FWHMs and smaller H I column densities. We explore alternate constraints on the intrinsic Ly$\alpha$ FWHMs in Section 6.3.

\startlongtable
\begin{deluxetable*}{c|cc|cc|cc}
\tabletypesize{\footnotesize}
\tablecaption{Shell Model Properties of Ly$\alpha$ Emission Lines \label{tab:LyA_model_results}} 
\tablehead{
\colhead{Target Name} & \multicolumn{2}{c}{N(H I)} & \multicolumn{2}{c}{FWHM} & \multicolumn{2}{c}{H I Velocity} \\
\hline
\nocolhead{} & \colhead{NS} & \colhead{S} & \colhead{NS} & \colhead{S} & \colhead{NS} & \colhead{S} 
}
\startdata
J11432669-7804454 & $20.2^{+0.1}_{-0.3}$ & $20^{+0.075}_{-0.066}$ & $430^{+110}_{-50}$ & $593^{+11}_{-9}$ & $-75^{+40}_{-100}$ & $-28^{+4}_{-2}$ \\
J16083070-3828268 & $20.0^{+0.9}_{-1.1}$ & \nodata & $700^{+400}_{-1100}$ & \nodata & $-3^{+4}_{-160}$ & \nodata \\
AA Tau & $20.8 \pm 0.1$ & \nodata & $610^{+170}_{-10}$ & \nodata & $-100^{+50}_{-90}$ & \nodata \\
CHX18N & $19.2^{+1.5}_{-0.9}$ & \nodata & $1200 \pm 600$ & \nodata & $60^{+150}_{-170}$ & \nodata \\
CS Cha & $20.3^{+0.1}_{-0.2}$ & $20.60 \pm 0.07$ & $760^{+230}_{-60}$ & $710 \pm 4$ & $30^{+30}_{-50}$ & $20 \pm 3$\\
CVSO 104 & $20.0^{+1.0}_{-0.2}$ & $20.04^{+0.43}_{-0.76}$ & $1400^{+0}_{-1000}$ & $976^{+546}_{-255}$ & $-210^{+190}_{-30}$ & $-88^{+90}_{-45}$ \\
CVSO 107 & $20.7 \pm 0.2$ & $20.59^{+0.08}_{-0.07}$ & $520^{+90}_{-80}$ & $536^{+15}_{-18}$ & $-100^{+50}_{-90}$ & $-81 \pm 4$ \\
CVSO 109 & $20.8^{+0.6}_{-1.5}$ & $21.01 \pm 0.08$ & $950^{+720}_{-460}$ & $566 \pm 29$ & $-80^{+140}_{-10}$ & $-2^{+1}_{-3}$ \\
CVSO 146 & $20.5^{+0.9}_{-0.6}$ & $18.2^{+0.4}_{-0.8}$ & $1200^{+700}_{-600}$ & $1648^{+178}_{-357}$ & $50^{+170}_{-100}$ & $-332^{+80}_{-40}$ \\
CVSO 165 & $20.7^{+0.3}_{-0.4}$ & $20.99 \pm 0.08$ & $780^{+740}_{-270}$ & $508^{+68}_{-92}$ & $-160^{+120}_{-10}$ & $-60^{+8}_{-6}$ \\
CVSO 176 & $20.8^{+0.2}_{-0.1}$ & $21.20 \pm 0.08$ & $540^{+110}_{-70}$ & $350^{+69}_{-83}$ & $-130^{+80}_{-60}$ & $-30 \pm 4$ \\
CVSO 90 & $20.9^{+0.5}_{-0.3}$ & $20.79 \pm 0.07$ & $610^{+120}_{-60}$ & $639^{+56}_{-76}$ & $-190^{+130}_{-5}$ & $-179^{+23}_{-32}$ \\
DE Tau & $20.7^{+0.8}_{-0.9}$ & \nodata & $1900^{+1100}_{-100}$ & \nodata & $-40^{+20}_{-90}$ & \nodata \\
DF Tau & $20.72^{+0.01}_{-0.2}$ & $21.39 \pm 0.06$ & $1200^{+140}_{-90}$ & $526 \pm 2$ & $-200^{+80}_{-20}$ & $-74^{+1}_{-3}$ \\
DK Tau & $20.1^{+0.6}_{-1.2}$ & \nodata & $950 \pm 400$ & \nodata & $-140^{+60}_{-70}$ & \nodata \\
DM Tau & $20.83^{+0.2}_{-0.04}$ & $21.58^{+0.08}_{-0.07}$ & $1100^{+80}_{-140}$ & $42^{+0.7}_{-0.6}$ & $80^{+100}_{-40}$ & $22^{+3}_{-2}$ \\
DN Tau & $20.1^{+0.8}_{-0.9}$ & $20.75^{+0.05}_{-0.04}$ & $1900^{+100}_{-900}$ & $1883^{+1}_{-3}$ & $29^{+90}_{-20}$ & $49 \pm 1$ \\
HT Lup & $20.80^{+0.3}_{-0.02}$ & \nodata & $630^{+40}_{-120}$ & \nodata & $-90^{+60}_{-70}$ & \nodata \\
IN Cha & $20.0 \pm 0.7$ & \nodata & $240^{+100}_{-1}$ & \nodata & $10 \pm 140$ & \nodata \\
LkCa 15 & \nodata & \nodata & \nodata & \nodata & \nodata & \nodata \\
MY Lup & $20.4^{+0.4}_{-1.5}$ & $20.60^{+0.07}_{-0.09}$ & $520^{+150}_{-10}$ & $465^{+16}_{-33}$ & $-80^{+90}_{-70}$ & $-28^{+5}_{-2}$ \\
RECX-11 & $20.1^{+0.3}_{-0.2}$ & $19.80 \pm 0.07$ & $400^{+170}_{-60}$ & $627^{+16}_{-19}$ & $-20^{+11}_{-120}$ & $-40 \pm 4$ \\
RECX-15 & $20.3^{+0.2}_{-0.4}$ & $19.81 \pm 0.08$ & $560^{+160}_{-10}$ & $517^{+5}_{-6}$ & $-190^{+110}_{-1}$ & $-170 \pm 4$ \\
RU Lup & $20.3^{+0.1}_{-0.3}$ & $20.20 \pm 0.07$ & $490 \pm 180$ & $544 \pm 3$ & $-110^{+40}_{-70}$ & $-150^{+4}_{-3}$ \\
RW Aur & $20.6^{+0.1}_{-0.2}$ & $20.40 \pm 0.07$ & $860^{+10}_{-190}$ & $716^{+17}_{-16}$ & $-220^{+140}_{-20}$ & $-180 \pm 4$ \\
RY Lup & $20.2^{+0.3}_{-0.2}$ & $20.20 \pm 0.07$ & $510^{+180}_{-50}$ & $739 \pm 11$ & $-50^{+30}_{-120}$ & $-50^{+4}_{-3}$ \\
Sz 10 & $19.7 \pm 0.9$ & \nodata & $1300^{+500}_{-800}$ & \nodata & $60^{+10}_{-90}$ & \nodata \\
Sz 111 & $20.6^{+0.4}_{-0.2}$ & $20.79^{+0.08}_{-0.07}$ & $820^{+200}_{-670}$ & $691^{+25}_{-24}$ & $7^{+140}_{-10}$ & $13 \pm 5$ \\
Sz 130 & $20.1^{+0.9}_{-0.6}$ & \nodata & $930^{+320}_{-450}$ & \nodata & $70^{+110}_{-50}$ & \nodata \\
Sz 45 & $20.3^{+0.7}_{-1.6}$ & \nodata & $1500^{+400}_{-800}$ & \nodata & $120^{+30}_{-100}$ & \nodata \\
Sz 69 & $21.0^{+0.05}_{-0.2}$ & $21.38^{+0.09}_{-0.2}$ & $620 \pm 80$ & $94^{+42}_{-26}$ & $-170^{+120}_{-20}$ & $-22^{+6}_{-4}$ \\
Sz 71 & $20.0^{+1.7}_{-1.6}$ & \nodata & $1300^{+500}_{-800}$ & \nodata & $-110^{+80}_{-70}$ & \nodata \\
Sz 72 & $19.7^{+1.3}_{-0.2}$ & \nodata & $1800^{+5}_{-1300}$ & \nodata & $-200^{+160}_{-40}$ & \nodata \\
Sz 75 & $20.5^{+0.1}_{-0.2}$ & $20.9^{+0.4}_{-0.1}$ & $800^{+150}_{-30}$ & $638 \pm 3$ & $-120^{+40}_{-70}$ & $-98 \pm 2$ \\
Sz 76 & $20.0^{+1.2}_{-1.3}$ & \nodata & $1000^{+400}_{-500}$ & \nodata & $-30^{+190}_{-40}$ & \nodata \\
Sz 77 & $20.4^{+0.4}_{-1.9}$ & \nodata & $720^{+30}_{-190}$ & \nodata & $-20 \pm 50$ & \nodata \\
T Cha & $20.1^{+0.6}_{-1.4}$ & $16.3^{+1}_{-0.4}$ & $200^{+150}_{-60}$ & $653^{+34}_{-39}$ & $-20^{+40}_{-160}$ & $-101^{+30}_{-28}$ \\
TW Hya & \nodata & \nodata & \nodata & \nodata & \nodata & \nodata \\
UX Tau & $20.7^{+0.3}_{-0.6}$ & $19.7^{+1.2}_{-0.5}$ & $830^{+140}_{-90}$ & $108^{+10}_{-19}$ & $50^{+20}_{-130}$ & $25^{+12}_{-6}$ \\
V4046 Sgr & $19.65^{+0.007}_{-0.04}$ & $20.21^{+0.07}_{-0.1}$ & $820^{+20}_{-10}$ & $724 \pm 2$ & $-40 \pm 10$ & $-23^{+2}_{-3}$ \\
V510 Ori & $20.7 \pm 0.4$ & $20.81 \pm 0.08$ & $870^{+610}_{-300}$ & $100^{+24}_{-17}$ & $-180^{+130}_{-10}$ & $-160^{+4}_{-5}$ \\
XX Cha & $20.9^{+0.02}_{-0.3}$ & $20.93^{+0.04}_{-0.1}$ & $760^{+150}_{-50}$ & $724^{+1}_{-2}$ & $-150^{+70}_{-50}$ & $-43^{+2}_{-1}$ \\
\enddata
\tablecomments{Best-fit Ly$\alpha$ shell parameters when considering a model without resonant scattering (NS) and with resonant scattering (S)}
\end{deluxetable*}

\subsection{Measured Disk \& Accretion Tracers from PENELLOPE Spectra}

The near-IR VLT/X-Shooter data from the PENELLOPE program cover the $K$ bandpass, which traces hot dust at small radii close to the YSOs ($T_{\rm{dust}} \sim 2000$ K, $r < 0.1$ au; see e.g., \citealt{McClure2013}). This component of the inner disk produces excess emission that fills in, or ``veils", absorption lines from the stellar photosphere, in addition to the excess emission generated by accretion (see e.g., \citealt{JohnsKrull2001, Fiorellino2021}). To explore this hot dust that is likely spatially adjacent to the Ly$\alpha$-generating accretion shocks, we computed the veiling in the $K$ band $\left(r_K \right)$ for all 17 targets with available PENELLOPE data and high S/N in the Ly$\alpha$ line wings. This process selects $K$ band data where telluric contamination was perfectly removed and emission lines were not present and compares the spectra to synthetic stellar templates. BT-Settl synthetic spectra from \citet{Allard2012} were used, with a compatible effective temperature to the targets (see Table \ref{tab:PENELLOPE_props}), solar metallicity, and assuming  a logg=4, a common value adopted for low-mass YSOs. Each synthetic template was degraded in resolution and rebinned according to the target spectra. Archival values of veiling $\left( r_K \right)$ are available for an additional eight targets in our sample, for a total of 25 measurements. Table \ref{tab:PENELLOPE_props} lists the means and standard deviations of the veiling values, along with integrated Br$\gamma$ fluxes and H$\alpha$ FWHMs for the PENELLOPE targets. The H$\alpha$ line widths were measured using the STAR-MELT Python package for fitting spectral features \citep{CampbellWhite2021}.

\startlongtable
\begin{deluxetable*}{ccccc}
\tablecaption{Measured Properties from PENELLOPE Spectra \label{tab:PENELLOPE_props}} 
\tablehead{
\colhead{Target Name} & \colhead{WTTS Template/SpT} & \colhead{$r_K$} & \colhead{Br$\gamma$ Flux} & \colhead{H$\alpha$ FWHM} \\
\nocolhead{} & \nocolhead{} & \nocolhead{} & \colhead{[$10^{-15}$ erg s$^{-1}$ cm$^{-2}$ nm$^{-1}$]} & \colhead{[km s$^{-1}$]} \\
}
\startdata
AA Tau & \nodata & $0.8 \pm 0.4$\tablenotemark{$\ast$} & \nodata & \nodata \\
CHX18N &  RX J1540.7-3756 / K6& $0.66 \pm 0.2$ & $< 11$ & 220 \\
CVSO 104 & \nodata & \nodata& $4 \pm 1$ & 235 \\
CVSO 107 & RX J1540.7-3756 / K6 & $0.48 \pm 0.25$ & $5.8 \pm 0.6$ & 200 \\
CVSO 109A & \nodata & $0.65 \pm 0.2$ & $13.6 \pm 0.8$ & 168 \\
CVSO 146 & RX J1540.7-3756 / K6 & $0.7 \pm 0.3$ & $10.9 \pm 0.6$ & 206 \\
CVSO 165 & RX J1540.7-3756 / K6  & $0.7 \pm 0.2$ & $8.9 \pm 0.6$ & 244 \\
CVSO 176 &  	TWA 7 / M2 & $1.3 \pm 0.3$ & $< 2.8$ & 211 \\
CVSO 90 & SO 879 / K7 & $2.2 \pm 0.4$ & $28.1 \pm 0.3$ & 244 \\
DF Tau & \nodata & $0.9 \pm 0.3$\tablenotemark{$\ast$} & \nodata & \nodata \\
DK Tau & \nodata & $1.8 \pm 0.6$\tablenotemark{$\ast$} & \nodata & \nodata \\
DM Tau & \nodata & $0$\tablenotemark{$\ast$} & \nodata & \nodata \\
DN Tau & \nodata & $0.1 \pm 0.1$\tablenotemark{$\ast$} & \nodata & \nodata \\
IN Cha & Par-Lup3-2 / M6 & $0.1 \pm 0.1$ & $< 2.8$ & 154 \\
RW Aur & \nodata & $>1.5$\tablenotemark{$\ast$} & \nodata & \nodata \\
Sz 10 & Par-Lup3-2 / M6 & $0.6 \pm 0.4$ & $5.1 \pm 0.4$ & 148 \\
Sz 45 &  TWA 25 / M0.5 & $0.25 \pm 0.2$ & $5 \pm 1$ & 150 \\
Sz 69 & SO 797 / M4.5 & $1.7 \pm 0.4$ & $10.8 \pm 0.7$ & 173 \\
Sz 71 & TWA 13B / M1 & $0.36 \pm 0.09$ & $10 \pm 2$ & 135 \\
Sz 72 & TWA 13B / M1 & $0.8 \pm 0.2$ & $81 \pm 1$ & 295 \\
Sz 75 & RX J1540.7-3756 / K6  & $2.4 \pm 0.5$ & $147 \pm 13$ & 213 \\
Sz 77 & RX J1540.7-3756 / K6 & $0.7 \pm 0.1$ & $< 7.3$ & 237 \\
TW Hya & \nodata & $0.1 \pm 0.1$\tablenotemark{$\ast$} & \nodata & \nodata \\
UX Tau & \nodata & $0.4$\tablenotemark{$\ast$} & \nodata & \nodata \\
XX Cha & TWA 9B / M3 & $0.3 \pm 0.3$ & $5 \pm 1$ & 136 \\
\enddata
\tablenotetext{*}{Values of $r_K$ taken from the literature: \citealt{Folha1999} (AA Tau, DF Tau, DK Tau, DN Tau, RW Aur, TW Hya); \citealt{Espaillat2010} (DM Tau, UX Tau)}
\end{deluxetable*}

\section{Discussion}

\subsection{Measured Ly$\alpha$ Properties Trace Dust Disk Evolution}

Previous work reported a strong correlation between the ratios of red to blue Ly$\alpha$ fluxes and infrared spectral indices \citep{Arulanantham2021}, which trace the presence of dust gaps in transition disks via the slopes of the SEDs between 13 and 31 $\mu$m \citep{Furlan2009, Espaillat2010}. This result indicates that the turnover from P Cygni-like to inverse P Cygni-like Ly$\alpha$ emission line profiles is a physical transition that occurs as the dust disks evolve, rather than an effect of the disk inclinations relative to the line of sight. However, the small sample size of $N = 12$ targets made it difficult to explore the causation behind the trend, prior to the release of data from the ULLYSES program. 

\begin{figure*}
\gridline{\fig{ 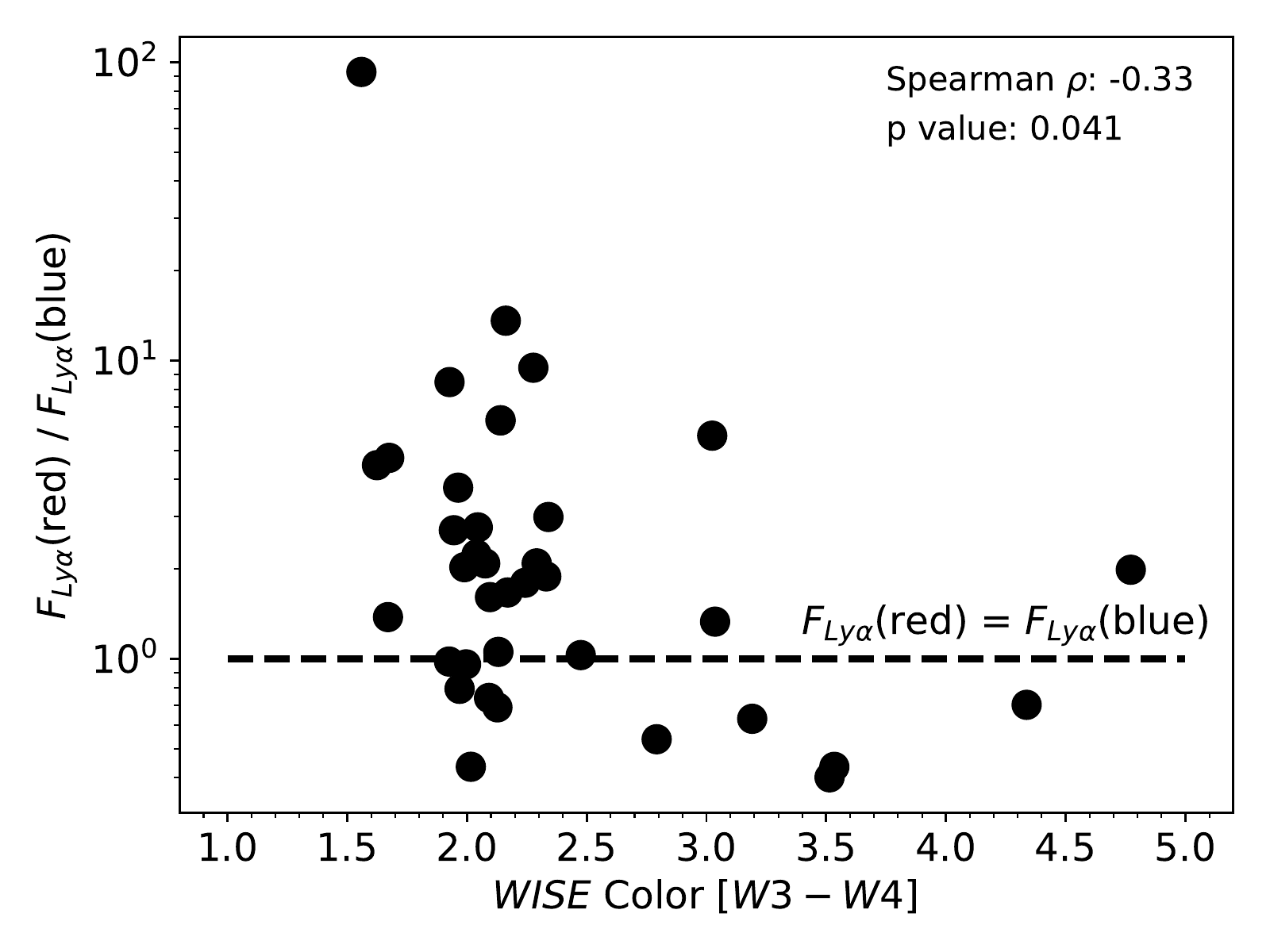}{0.5\textwidth}{ }
\fig{ 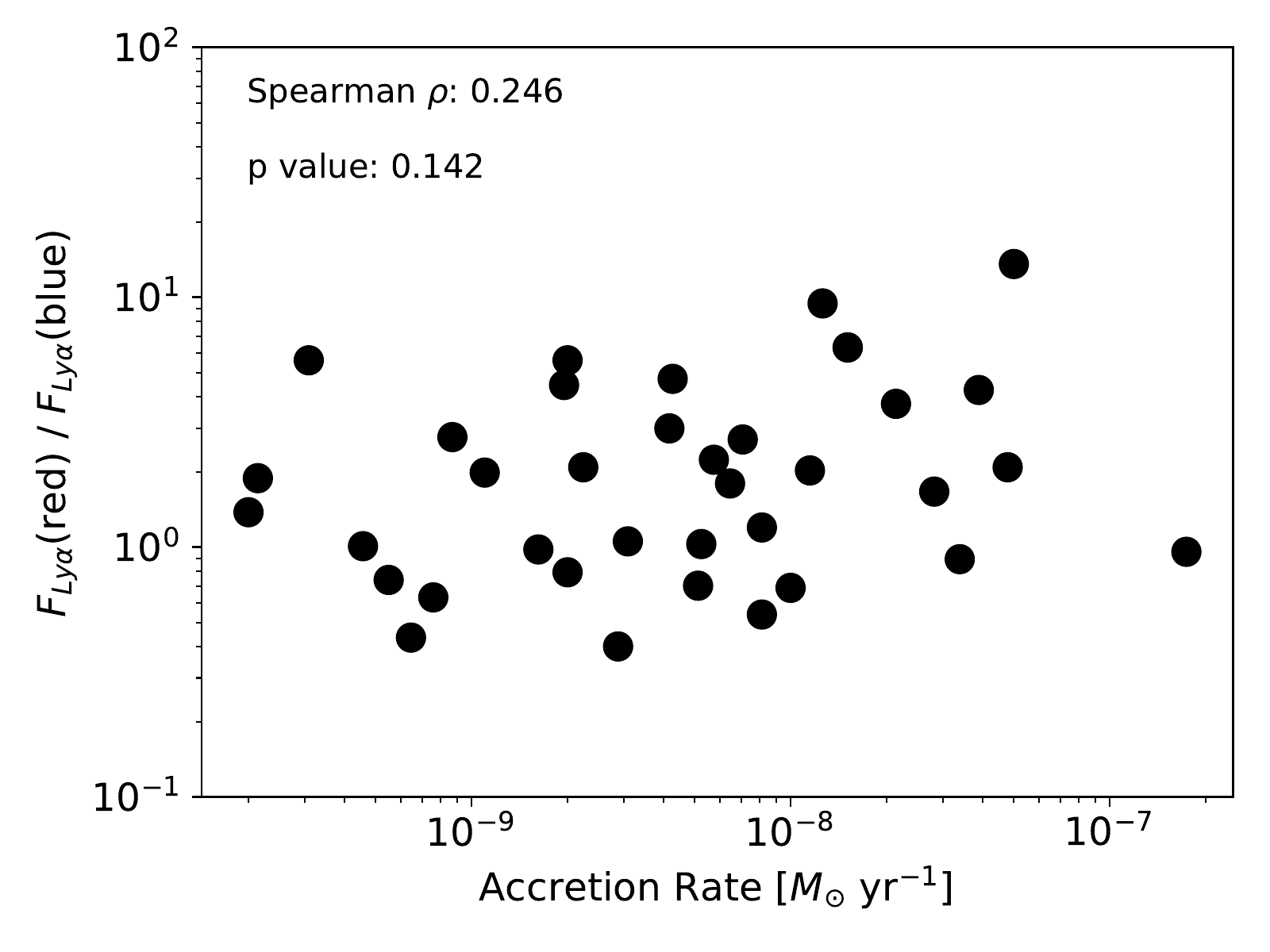}{0.5\textwidth}{ }}
\caption{Ratios of red to blue integrated Ly$\alpha$ fluxes versus \emph{WISE} colors between 12 (\emph{W3}) and 22 (\emph{W4}) $\mu$m \emph{(left)} and mass accretion rates \emph{(right)}. No correlations are detected between the flux ratios and accretion rates, implying that targets with P Cygni-like Ly$\alpha$ profiles are not necessarily undergoing more rapid accretion. Larger values of $\left[W3 - W4 \right]$ correspond to less flux from 10 $\mu$m silicate emission and optically thick dust continuum, both of which are captured in the $W3$ filter. These colors may trace the disk flaring in smooth disks and the locations of dust gaps in structured disks.}
\label{fig:accretion_W3W4}
\end{figure*}

Few additional ULLYSES targets have \emph{Spitzer} spectra from which the infrared spectral index between 13 and 31 $\mu$m can be calculated, but 36/42 of the circumstellar disks from our sample were detected at infrared wavelengths with \textit{WISE} (see e.g., \citealt{Koenig2014}). The remaining six were observed, but the photometry was flagged for contamination and is excluded from the following analysis. Figure \ref{fig:accretion_W3W4} compares the ratios of red to blue Ly$\alpha$ flux to the $\left[W3 - W4 \right]$ colors, which roughly capture the SED slopes between 12 and 22 $\mu$m. Large $\left[W3 - W4 \right]$ can also indicate contributions from circumstellar envelopes \citep{Koenig2014}, but the range of $\left[W1 - W2 \right]$ colors from our sample $\left(0.04 < \left[W1 - W2 \right] < 1.0 \right)$ place all targets in the Class II/transition disk regime defined in that work. 

We detect a statistically significant negative correlation between $\left[W3 - W4 \right]$ and the ratios of red to blue Ly$\alpha$ flux $\left(\rho = -0.3, p = 0.04 \right)$, such that larger positive values of $\left[W3 - W4 \right]$ correspond to inverse P Cygni-like Ly$\alpha$ emission lines. However, we note that the \emph{W3} filter is broad enough to capture both optically thin 10 $\mu$m silicate emission and the optically thick dust continuum. This trend then implies that targets with more blue than red Ly$\alpha$ flux have less flared disks and/or gaps in the dust distributions that place them in the transition disk category, as illustrated in Figure \ref{fig:COS_LyA_demo}.

We also compare the ratios of red to blue Ly$\alpha$ flux to 70 $\mu$m flux densities from the 23 targets that were also observed with \emph{Herschel} (Table \ref{tab:targ_props}; see e.g., \citealt{Mauco2016}). Disk models fit to the 70 $\mu$m emission return best-fit temperatures at $r = 10$ au ranging between $T = 24-236$ K, which are strongly correlated with the observed flux densities \citep{Ribas2017}. This relationship makes the \emph{Herschel} data important probes of large dust grains that continue to produce optically thick continuum emission even after near-IR excess from smaller grains disappears (see e.g., \citealt{Rebollido2015}). No significant correlation is detected between the \emph{Herschel} flux densities and Ly$\alpha$ flux ratios, perhaps indicating that the 70 $\mu$m emission remains relatively constant over the timescales when accretion broadened Ly$\alpha$ emission line wings are readily detectable. Although the observed Ly$\alpha$ flux ratios do not provide constraints on outer disk dust evolution, we conclude that the shapes of the emission line profiles are linked to the inner disk flaring angles and dust gaps within $r < 10$ au. 

Figure \ref{fig:measured_LyA_redtoblue} shows that $\sim$30\% of targets have accretion absorbed, inverse P Cygni-like Ly$\alpha$ emission line wings and therefore larger $\left[W3 - W4 \right]$ colors in Figure \ref{fig:accretion_W3W4}. This percentage is slightly higher than the fraction of transition disks with large sub-mm dust cavities (26\% with $r > 15$ au; see e.g., \citealt{Andrews2011}) and much larger than the fraction predicted from infrared SED analyses (10\%; see e.g. \citealt{Espaillat2014}), perhaps indicating that some targets in our sample have dust gaps opened between the inner and outer disks at radii smaller than 15 au. Although not all ULLYSES targets have been observed at sub-mm wavelengths with sufficient spatial resolution to detect dust substructure, four disks with inverse P Cygni-like Ly$\alpha$ profiles have dust cavities imaged at sub-mm wavelengths: DM Tau ($r = 19$ au; \citealt{Andrews2011}), UX Tau ($r = 30$ au; \citealt{Andrews2011}), Sz 111 ($r = 80$ au; \citealt{Ansdell2016}), and CS Cha ($r = 37$ au; \citealt{Francis2020}). Those same sources have a higher detection frequency for the 1600 \AA \, H$_2$O dissociation bump \citep{France2017}, implying that Ly$\alpha$ photons are more readily propagating through and photodissociating the molecular gas in these disks.

Signatures of fast inner disk winds and jets are no longer detected at the stage of disk evolution when inverse P Cygni-like Ly$\alpha$ emission is observed, although low velocity emission from [O I] and [Ne II] in slow photoevaporative winds is still detected (see e.g., \citealt{Banzatti2019, Pascucci2020}). Ly$\alpha$ photons must then only travel through the accretion flows, protoplanetary disks, and ISM before reaching the observer. The blue sides of the emission line profiles are no longer absorbed by fast outflowing H I, and red-shifted absorption within accretion flows becomes the dominant factor in shaping the observed Ly$\alpha$ features. However, we note that not all targets with dust sub-structure in our sample have inverse P Cygni-like Ly$\alpha$ profiles (see Table \ref{tab:targ_props}). Furthermore, models that produce dust sub-structure via photoevaporative winds or planet-disk interactions do not make predictions for the impact of these physical mechanisms on Ly$\alpha$ emission (see e.g., \citealt{Garate2021}), making it difficult to distinguish between scenarios in this dataset. Sub-mm observations at high spatial resolution for the remaining ULLYSES targets are required to further explore this relationship between Ly$\alpha$ flux ratios and dust disk evolution.      

\begin{figure*}
\gridline{\fig{ 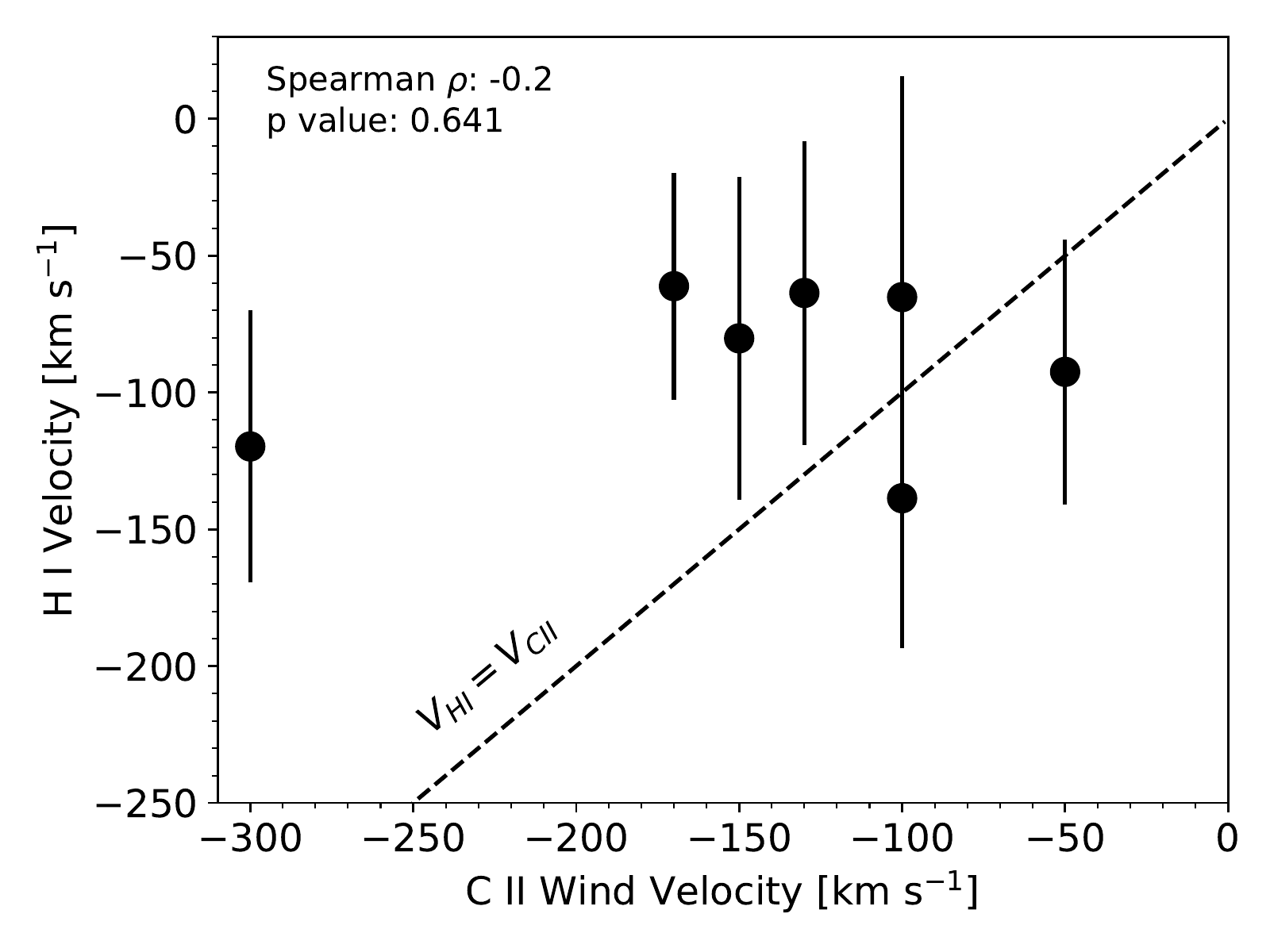}{0.5\textwidth}{ }
\fig{ 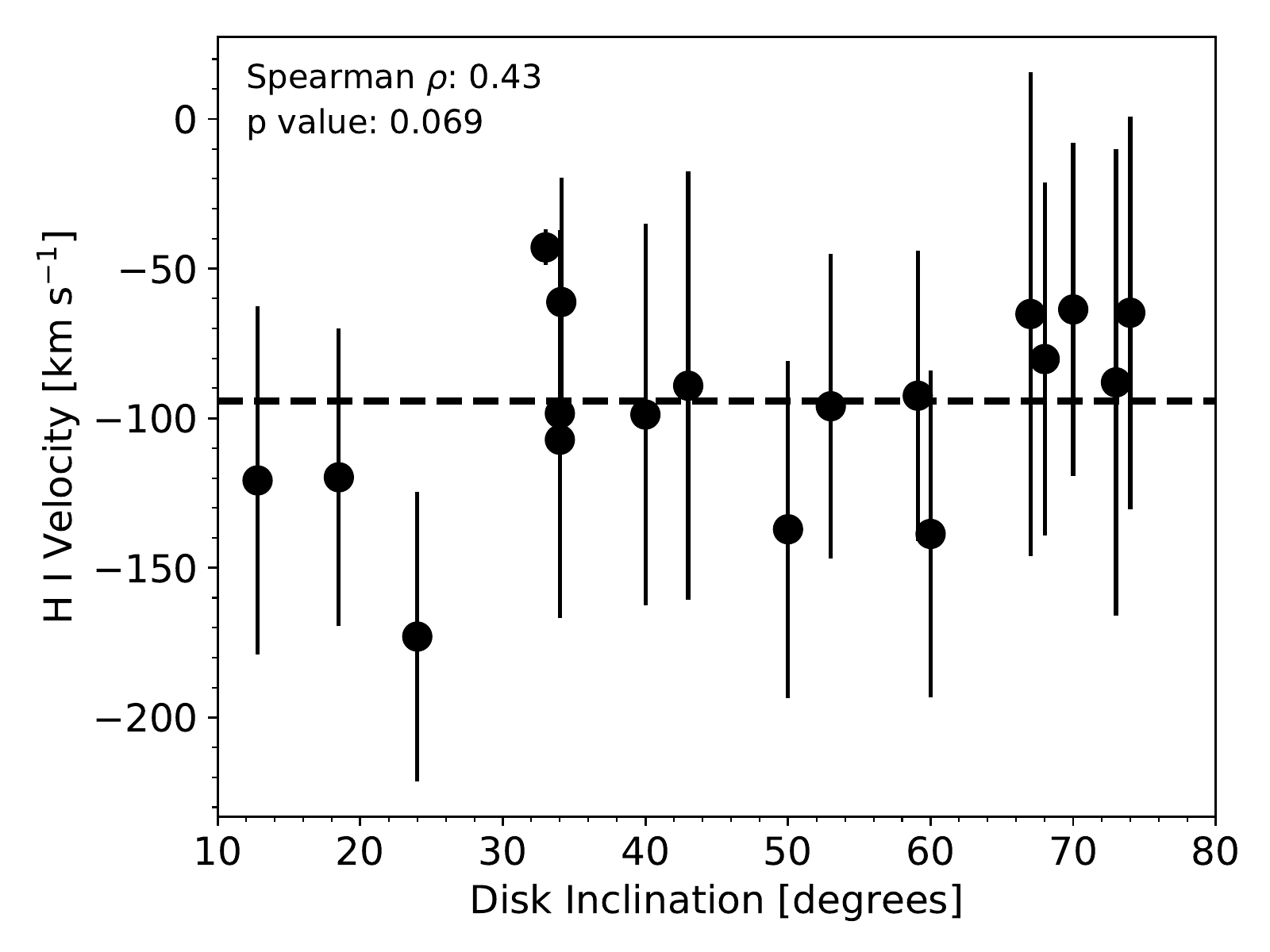}{0.5\textwidth}{ }}
\caption{Absorption velocities of outflowing H I from best-fit Ly$\alpha$ models versus C II wind velocities for the targets included in \citet{Xu2021} \emph{(left)} and circumstellar disk inclinations measured from sub-mm observations \emph{(right)}. The median H I velocity (-96 km s$^{-1}$) is marked as a black, dashed line in the right panel. No statistically significant correlation is detected between the velocities and disk inclinations, although we note that sub-mm observations are not available for most targets with H I velocities much larger or smaller than the median value. Uncertainties on the C II slow wind velocities are $<100$ km s$^{-1}$ \citep{Xu2021}.}
\label{fig:CIIvel_inc_vs_vHI}
\end{figure*}

Observations of the C II $\lambda$1335 doublet in \emph{HST}-COS spectra of T Tauri stars also show absorption signatures from fast, collimated jets and slow winds originating from the inner disks \citep{Xu2021}. Figure \ref{fig:CIIvel_inc_vs_vHI} compares the wind velocities derived from the C II features \citep{Xu2021} to the model H I velocities for targets with P Cygni-like, outflow-dominated Ly$\alpha$ profiles. Of the seven targets included in both samples, only three have C II and H I velocities that are consistent within the 1$\sigma$ uncertainties. All four of the remaining systems have smaller H I velocities than the C II measurements, corresponding to slower winds. The lack of correlation between H I and C II wind velocities $\left(\rho = -0.2, p = 0.6 \right)$ is also consistent with observations of Ly$\alpha$ emission lines from compact, star-forming green pea galaxies, which are always well reproduced by models with lower H I velocities than indicated by absorption against other low ionization features \citep{Orlitova2018}.  

The C II absorption velocities that \citet{Xu2021} associate with fast winds are inversely correlated with the disk inclinations, as expected for material originating in collimated jets with speeds that increase with distance from the source. Photons traveling from edge-on sources only pass through the slow bases of the jets but must propagate through faster regions to escape more face-on disks. Meanwhile, absorption from slow winds is only detected in disks with intermediate to high inclinations, but no significant correlation is detected between the wind velocities and disk inclinations. To explore whether the Ly$\alpha$-absorbing H I is more consistent with collimated jets or slow disk winds, Figure \ref{fig:CIIvel_inc_vs_vHI} compares the disk inclinations to the H I velocities for targets with P Cygni-like, outflow-dominated Ly$\alpha$ emission, which are not statistically significantly correlated $\left(\rho = 0.43, p = 0.07 \right)$. The targets in our sample that have been observed at sub-mm wavelengths have outer disk inclinations ranging from $i_d = 7^{\circ}$ to $i_d = 74^{\circ}$ (see Table \ref{tab:targ_props}). However, we note that sub-mm observations are not available for most targets with H I velocities much larger or smaller than the median value of the sample $v \sim -96$ km s$^{-1}$. Furthermore, misalignments between the inner and outer disk inclination angles are frequently detected (see e.g., \citealt{Davies2019, Bohn2022}). The high model H I velocities reported here are consistent with absorbing gas located very close to the stars (see e.g., \citealt{Pascucci2020}) and may be more correlated with the inner disk inclinations that are much harder to constrain. 

Of the five systems with sub-mm observations and inverse P Cygni-like, infall-dominated Ly$\alpha$ profiles, four have disk inclinations between $35^{\circ}-37^{\circ}$ that may be in alignment with the launching angle of MHD disk winds \citep{Banzatti2019}. This inclination angle places them at the edge of the population with slow winds absorbing the C II measured by \citet{Xu2021}. Targets with disk inclinations $\sim 35^{\circ}$ also show the largest blueshifts in both the broad and narrow low velocity components of [O I] 6300 \AA \, emission line profiles, as expected for coinciding disk inclinations and wind launching angles \citep{Banzatti2019}. It is possible then that Ly$\alpha$ photons from disks at intermediate inclinations do not scatter through the winds before escaping the system, leaving only the signatures of the accretion flows imprinted on the observed emission line profiles. Naturally, the simple, isotropic shell model does not capture these subtleties. In the future, we can model individual systems with more complex frameworks to study this point further. We note that three of the four targets with intermediate disk inclinations and P Cygni-like Ly$\alpha$ emission are known binaries (V4046 Sgr, HT Lup, and Sz 69), which likely complicates the scattering geometry.  

\subsection{Model H I Velocities Probe Kinematics of Accretion and Outflows}

Ly$\alpha$ photons are generated at the YSO accretion shocks (see e.g., \citealt{Hartmann2016, Schneider2020}), and previous work reports significant correlations between reconstructed Ly$\alpha$ emission line fluxes at the disk surfaces and mass accretion rates measured from the C IV $\mathbf{\lambda_0 = 1548, 1550}$ \AA \, resonance doublet \citep{France2014}. We compare the Ly$\alpha$ flux ratios to measured accretion rates for our sample, to explore whether the turnover in flux ratios occurs as accretion slows (see Figure \ref{fig:accretion_W3W4}). However, no clear relationships are detected, demonstrating that targets with P Cygni-like Ly$\alpha$ profiles are not necessarily undergoing more rapid accretion. We note that there is no statistically significant correlation between mass accretion rates and stellar masses for this sample $\left(\rho = 0.18, p = 0.3 \right)$. The measured accretion rates for the 13 disks with inverse P Cygni-like Ly$\alpha$ emission range from $5 \times 10^{-10} < \dot M < 3 \times 10^{-8} \, M_{\odot}$ yr$^{-1}$ (see Table \ref{tab:targ_props}), with one outlier and known binary at the high end ($\dot M = 1.72 \times 10^{-8} \, M_{\odot}$ yr$^{-1}$ for CVSO 109A; \citealt{Pittman2022}), confirming that accretion rates can remain high even as material is dissipated from the outer disks. This is consistent with sub-mm observations, which still show significant gas abundances in disks with high accretion rates and large dust cavities or gaps (see e.g., \citealt{Espaillat2014, Bruderer2013, Kudo2018, Francis2022}).

Previous modeling work found that the velocities of expanding (outflowing) or collapsing (accreting) shells are well constrained for emission lines with distinct, asymmetric double peaks \citep{Li2022}. Figure \ref{fig:model_param_comp} compares the observed Ly$\alpha$ flux ratios to the H I velocities derived from the shell models in this work that do and do not account for resonant scattering. The H I velocities from the two modeling frameworks are nearly always consistent with each other within the 1$\sigma$ uncertainties and are strongly correlated with the ratios of red to blue Ly$\alpha$ flux (Spearman $\rho = -0.95$, $p = 3 \times 10^{-13}$ without scattering; Spearman $\rho = -0.59$, $p = 2 \times 10^{-3}$ when scattering is included). Since almost all targets in our sample have double-peaked Ly$\alpha$ emission lines, we are able to interpret the model values as average outflow (or infall) velocities. 

\begin{figure*}
\gridline{\fig{ 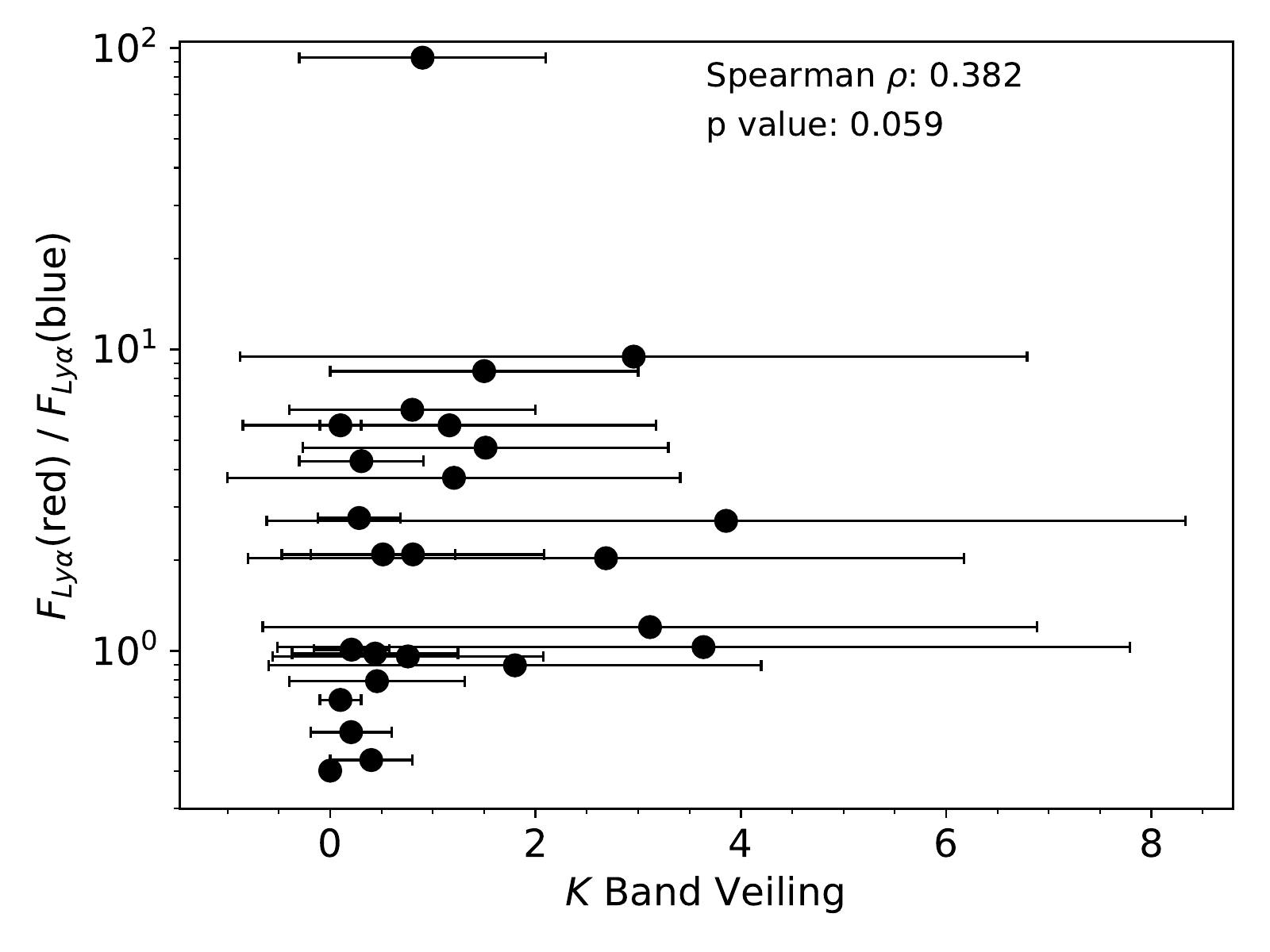}{0.33\textwidth}{(a)}
\fig{ 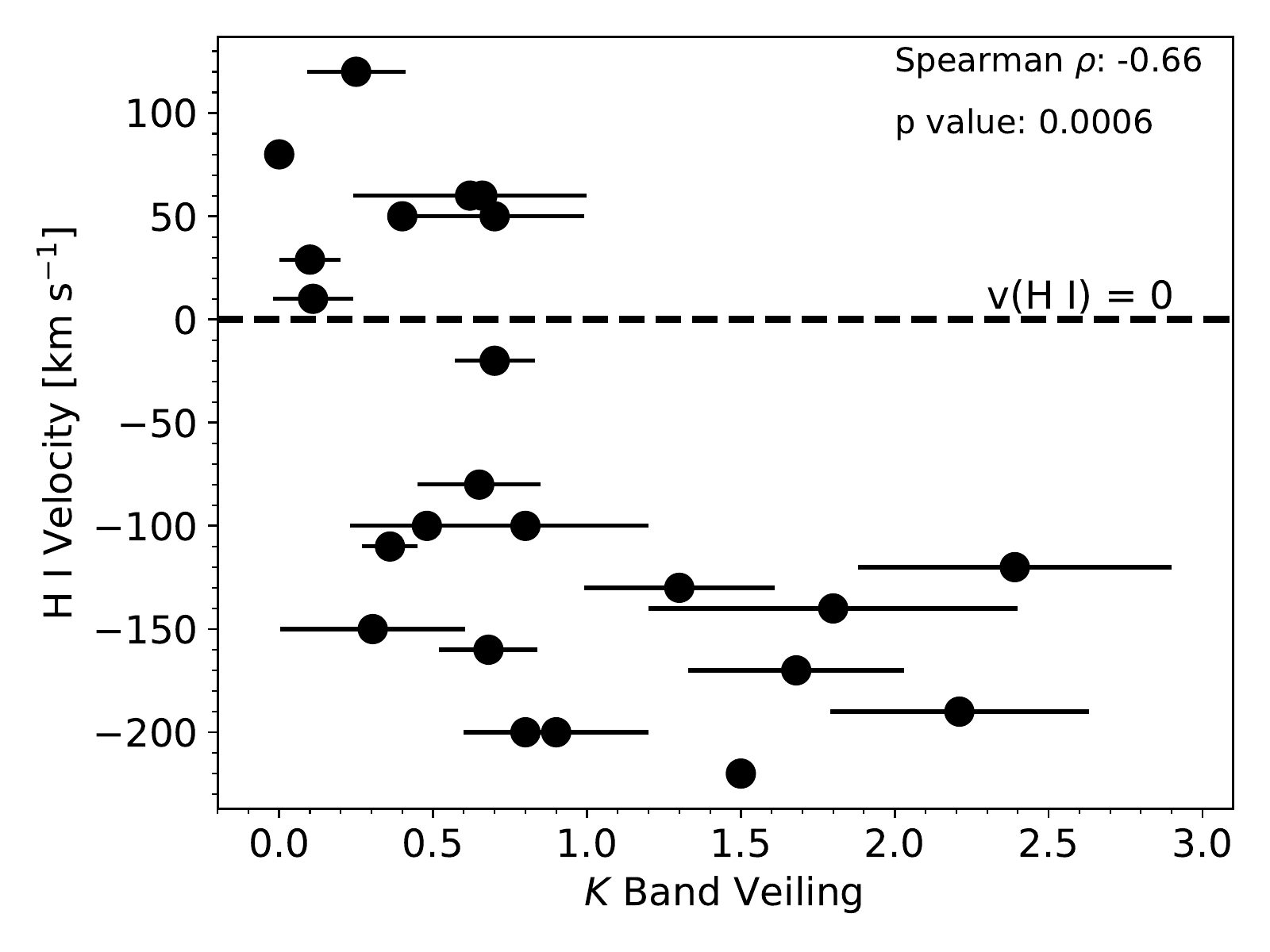}{0.33\textwidth}{(b)}
\fig{ 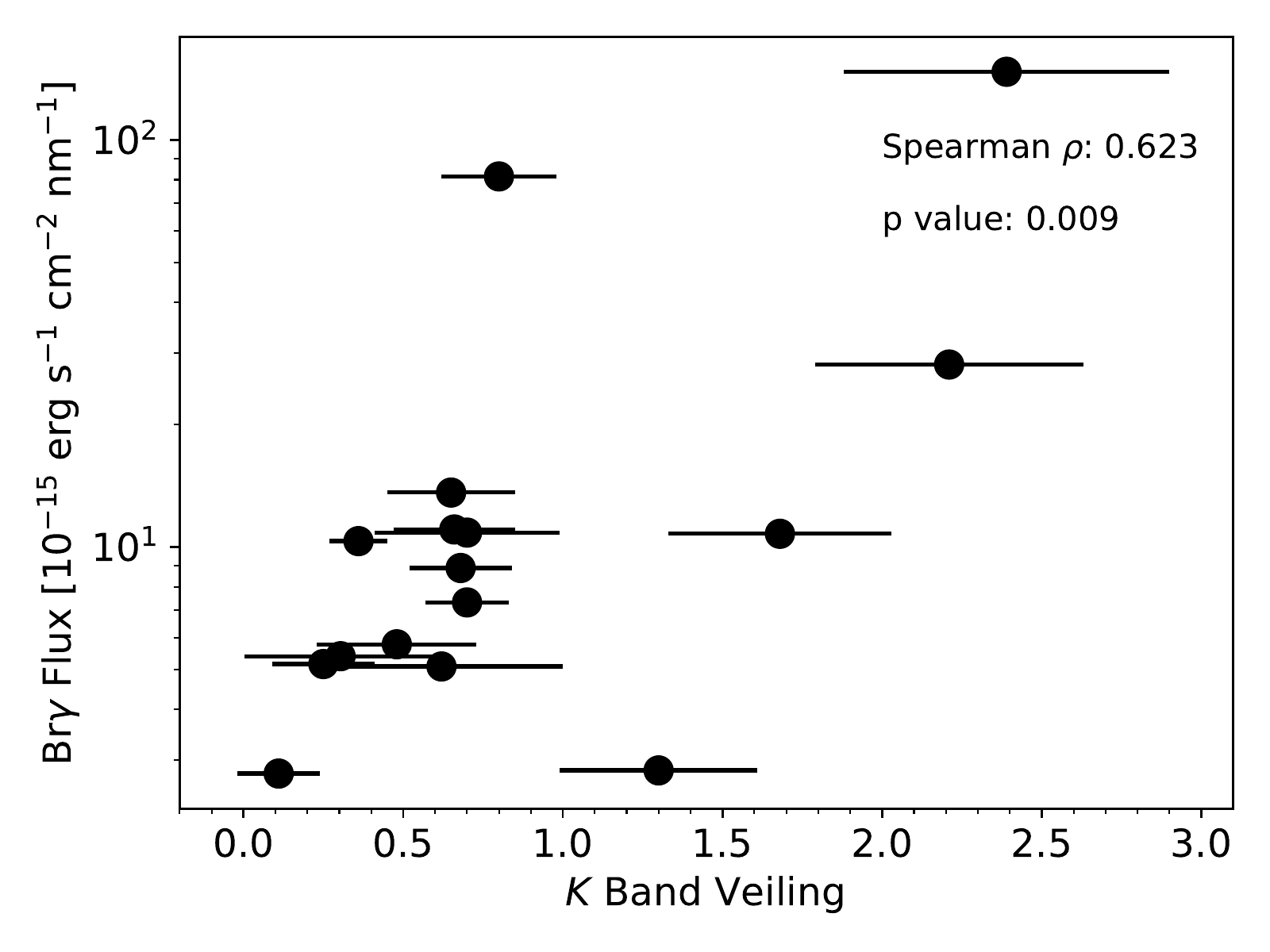}{0.33\textwidth}{(c)}}
\caption{\emph{K} band veiling $\left(r_K \right)$ measured from X-Shooter spectra acquired through the PENELLOPE program versus (a) measured ratios of red to blue Ly$\alpha$ flux, (b) model absorption velocities of outflowing and infalling H I, and (c) Br$\gamma$ fluxes measured from PENELLOPE spectra. Statistically significant correlations are detected between $r_K$ and $F_{\rm{Ly}\alpha} \rm{(red)} / F_{\rm{Ly}\alpha} \rm{(blue)}$ $\left( \rho = 0.60, p = 0.0006 \right)$, the H I velocities $\left(\rho = -0.66, p = 0.0006 \right)$, and Br$\gamma$ emission $\left(\rho = 0.62, p = 0.009 \right)$, indicating that the turnover in shape of Ly$\alpha$ profiles occurs for targets with spectra that are less veiled by hot dust $\left(T_{\rm{dust}} \sim 2000 \rm{K} \right)$.}
\label{fig:vout_vs_Kveil}
\end{figure*}

We compare the model H I velocities and Ly$\alpha$ flux ratios to veiling measured from \emph{K} band spectra acquired as part of the PENELLOPE program ($r_K$; see Figure \ref{fig:vout_vs_Kveil}), which quantifies how much the NIR spectral features are filled in, or veiled, by excess emission from hot gas that is accreting onto the central protostars and inner disk dust (see e.g., \citealt{McClure2013, Fiorellino2021}). The Ly$\alpha$ flux ratios and veiling measurements are statistically significantly positively correlated (Spearman $\rho = 0.60$, $p = 0.006$), showing that increased veiling generally corresponds to P Cygni-like Ly$\alpha$ emission lines. We also find a strong negative correlation between the H I velocities and the \emph{K} band veiling measurements (Spearman $\rho = -0.66$, $p = 0.0006$), consistent with the disappearance of $T \sim 2000$ K dust that fills in the intrinsic protostellar spectrum (see e.g., \citealt{Kidder2021}). This may indicate a depletion of inner disk material as outflowing winds slow by shifting to larger radii and infalling accretion flows become the dominant absorber of Ly$\alpha$ photons, although $r_K$ may also be influenced by the inner disk inclination or scale height that we cannot measure from these observations (see e.g., \citealt{JohnsKrull2001}). As an additional tracer of accretion from spectra acquired simultaneously, and therefore not impacted by variability, we also compare the \emph{K} band veiling to Br$\gamma$ emission measured from the PENELLOPE spectra in Figure \ref{fig:vout_vs_Kveil}. A statistically significant positive correlation is detected (Spearman $\rho = 0.623$, $p = 0.009$), although the Br$\gamma$ emission is not significantly correlated with the Ly$\alpha$ flux ratios (Spearman $\rho = 0.42$, $p = 0.16$). These trends will likely become more clear once simultaneous accretion rates have been measured for all ULLYSES targets (Pittman et al., Wendeborn et al., Claes et al., in prep). Despite the statistical significance of the correlations between Ly$\alpha$ flux ratios, model H I velocities, and $r_K$, we note that the small $p$ values reported here are dependent on the choice of stellar template used to measure the veiling and should not be interpreted as absolute.

\subsection{Constraints on turbulence within accretion shocks}

Figure \ref{fig:model_param_comp} demonstrates that no strong correlations are detected between the Ly$\alpha$ flux ratios and H I column densities (Spearman $\rho = -0.1$, $p = 0.6$ when scattering is not included; Spearman $\rho = 0.08$, $p = 0.7$ with scattering) or FWHMs of the intrinsic emission lines (Spearman $\rho = -0.2$, $p = 0.5$ without scattering; Spearman $\rho = -0.3$, $p = 0.1$ when scattering is included). \citet{Li2022} find that $N$(H I), emission line widths, and effective temperatures within the scattering medium can be degenerate parameters, which may be responsible for the lack of correlations with the observed flux ratios (see corner plot for Sz 111 in Appendix A). The intrinsic line widths are challenging to constrain, because of both the geocoronal emission masking velocities $<$250 km s$^{-1}$ and turbulence within the accretion shocks themselves. The Ly$\alpha$ photons must propagate through this highly turbulent region before reaching the protoplanetary disks, winds/jets, and ISM, which requires the intrinsic model emission line to be broader than the velocity dispersions of randomly moving clumps within the scattering medium \citep{Li2022}. When the clumps are accounted for in multi-phase models, the total H I column densities are expected to decrease, as the wider emission lines allow flux to be distributed to higher velocities with fewer scattering events.

\begin{figure*}
    \centering
    \includegraphics[width=\textwidth]{ 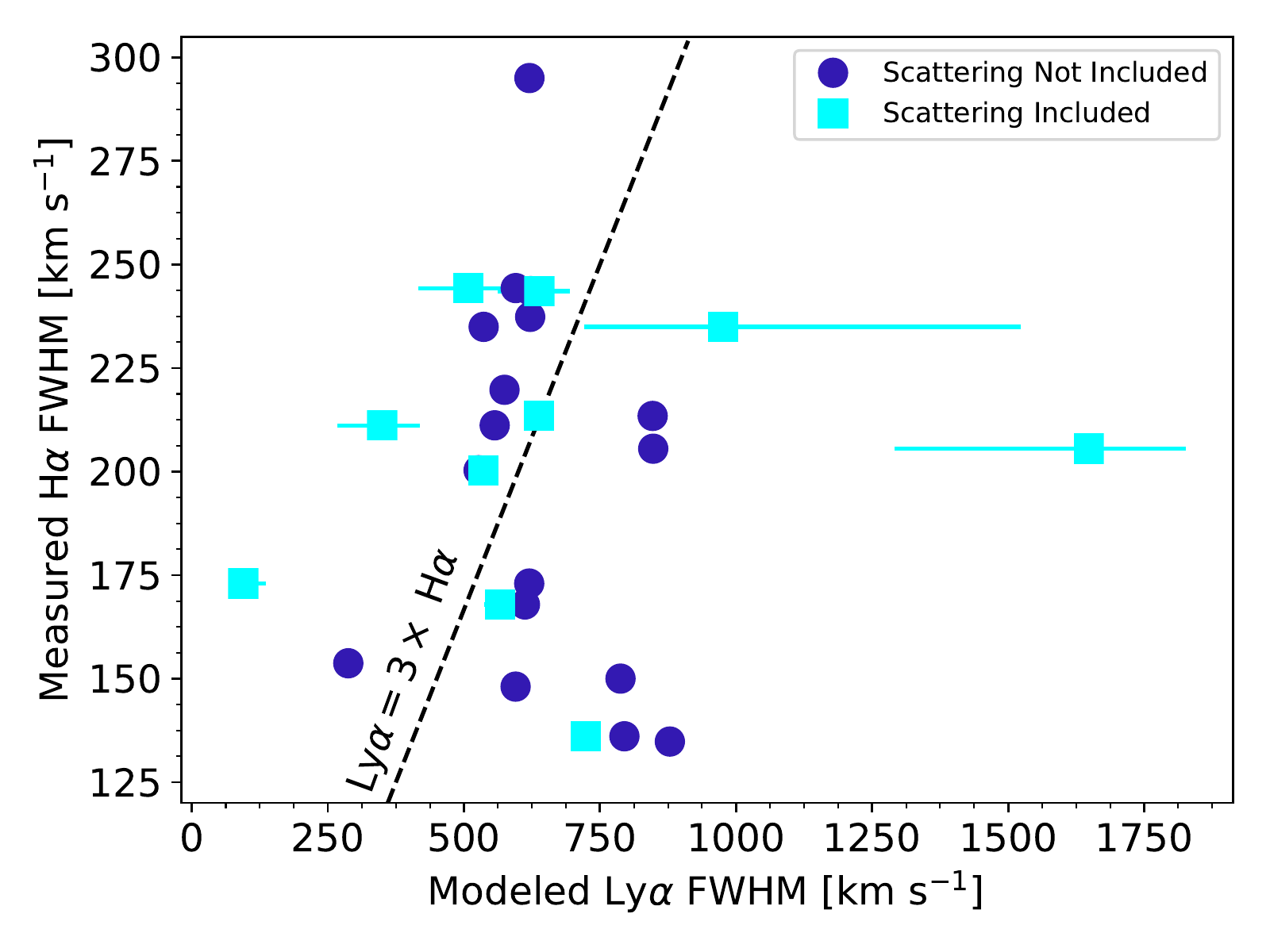}
    \caption{H$\alpha$ FWHMs measured from PENELLOPE data versus Ly$\alpha$ FWHMs extracted from models that do and do not include resonant scattering. The Ly$\alpha$ FWHMs are always larger than the measured H$\alpha$ values, with median ratios of $\text{FWHM}\left(\text{Ly} \alpha \right) / \text{FWHM} \left( \text{H} \alpha \right) \sim 3$ under both modeling frameworks. The discrepancy can be explained if the Ly$\alpha$ photons are propagating through a clumpy medium \citep{Li2022}, as expected within the accretion shocks surrounding TTSs.}
    \label{fig:HalphaLyAwidths}
\end{figure*}

Since H$\alpha$ scattering requires a large population of atomic hydrogen in the $n = 2$ level, the process occurs at a much lower rate than Ly$\alpha$ scattering and leads to emission lines with less turbulent broadening than the intrinsic Ly$\alpha$ profiles. To explore the impact of turbulence on the Ly$\alpha$ observations, Figure \ref{fig:HalphaLyAwidths} compares the Ly$\alpha$ FWHMs from the models presented in this work to H$\alpha$ emission line widths measured from the PENELLOPE spectra. The Ly$\alpha$ FWHMs are $\sim$3 times larger than the H$\alpha$ line widths, regardless of whether or not scattering is included in the models. This is consistent with the results of \citet{Orlitova2018}, whose models of Ly$\alpha$ emission lines from star-forming green pea galaxies had to have FWHMs that were $\sim$3 times broader than the Balmer lines from the same systems. This discrepancy requires the Ly$\alpha$ photons to propagate through a clumpy medium \citep{Gronke2017,Li2022}, as expected within the turbulent accretion shocks surrounding TTSs. 

\begin{figure*}
\gridline{\fig{ 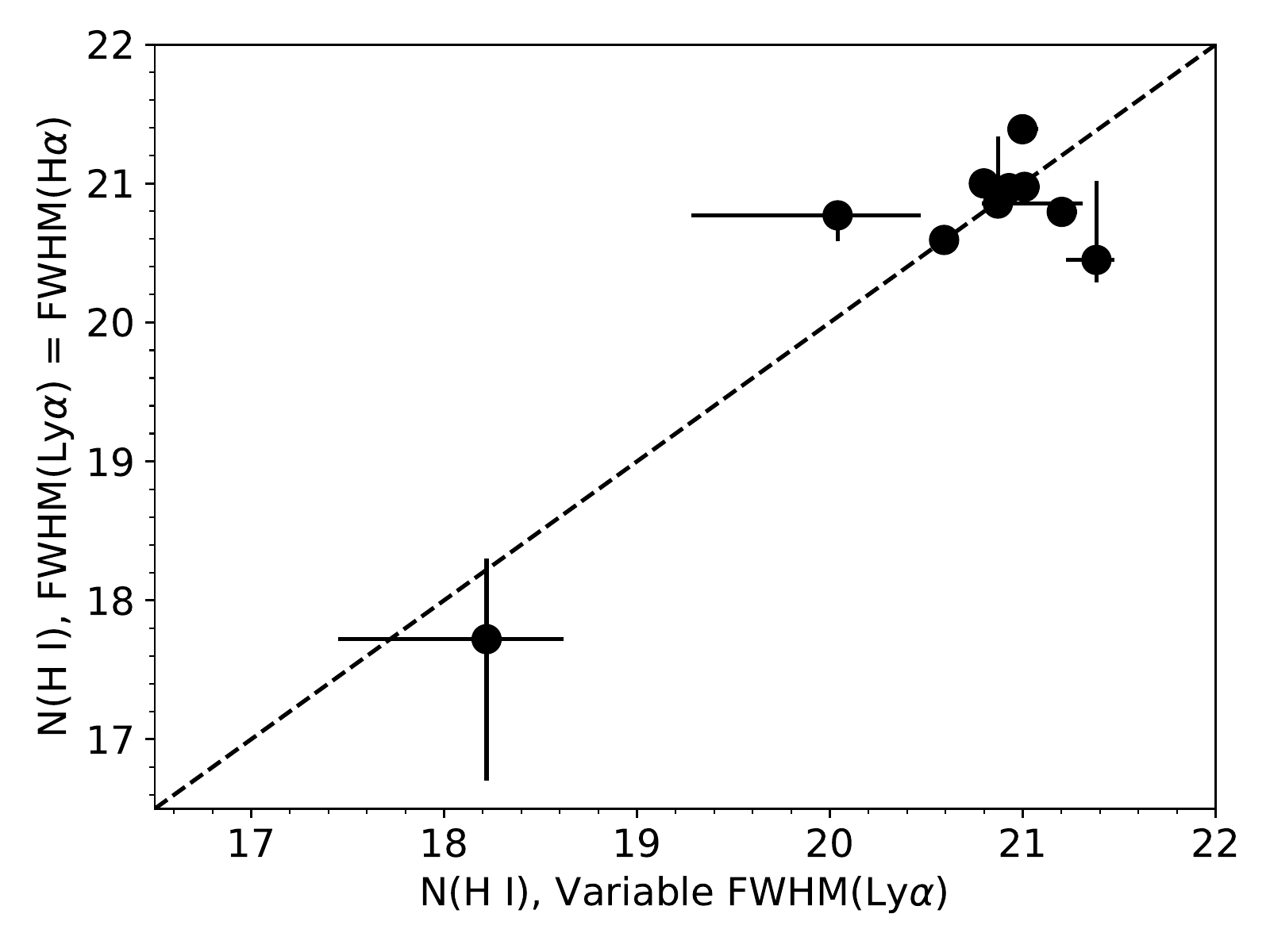}{0.5\textwidth}{ }
\fig{ 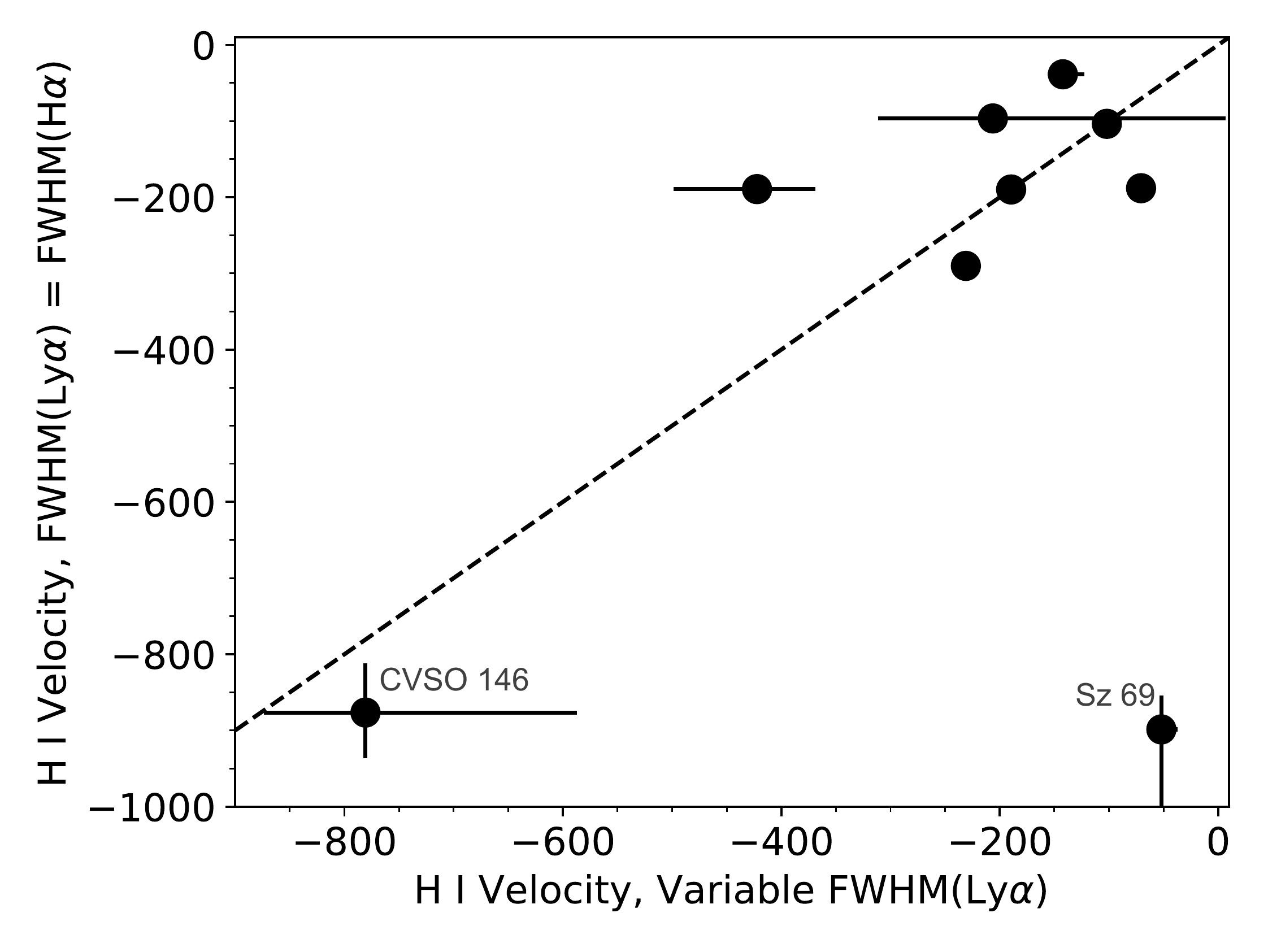}{0.5\textwidth}{ }}
\caption{Comparison of best-fit H I column densities \emph{(left)} and shell velocities \emph{(right)} in models when the Ly$\alpha$ line widths are held fixed at the measured H$\alpha$ FWHMs (y axes) or left as variable parameters (x axes). Despite changing the Ly$\alpha$ FWHMs by a factor of $\sim$3 on average between models, the column densities and velocities are roughly consistent, implying that the temperature of the scattering medium is the more strongly degenerate parameter. One target, Sz 69, has an H I velocity that is $\sim$17 times larger when the Ly$\alpha$ FWHM is held fixed. The value corresponds to the stronger of two peaks in the posterior distribution extracted from the MCMC sampler. However, the larger value is non-physical, as it exceeds the expected escape velocity by a factor of $\sim$3.}
\label{fig:Halpha_lims}
\end{figure*}

Alternatively, we explore whether the difference in emission line FWHMs is caused by degeneracies in model parameters by fitting a second set of shell models to the targets with PENELLOPE data and holding the Ly$\alpha$ line widths fixed at the measured H$\alpha$ FWHMs. In the case that H$\alpha$ scattering is minimal, this represents the ``true" intrinsic Ly$\alpha$ emission line width prior to passing through the turbulent accretion shock. Figure \ref{fig:Halpha_lims} compares the best-fit H I column densities and shell velocities from the two modeling frameworks. Despite decreasing the Ly$\alpha$ FWHMs by a factor of $\sim$3 on average to match the H$\alpha$ line widths, the best-fit H I column densities and velocities generally do not change significantly between models. This implies that the more significantly unconstrained pair of parameters is likely the effective temperature within the scattering medium and the H I column densities, rather than the H I column densities and intrinsic emission line widths (see also corner plot for Sz 111 in Appendix A). If a significant fraction of the intervening H I is located in the ISM, as reported by \citet{McJunkin2014}, the simple shell models used here will not accurately capture the temperature gradient of the scattering medium. However, the dominant velocity signatures setting the observed Ly$\alpha$ flux ratios are still consistent with outflowing and accreting H I within the protostellar/disk/wind environments.   

\subsection{Implications for Chemistry in Protoplanetary Disks}

Physical-chemical models that account for Ly$\alpha$ irradiation of protoplanetary gas disks identify several volatile-bearing species that are particularly sensitive to its effects. The large photodissociation cross sections near 1216 \AA \, for H$_2$O, HCN, and C$_2$H$_2$ can result in decreased abundances of all three species \citep{Bergin2003, Bethell2011}, depending on the fraction of the UV radiation fields comprised of Ly$\alpha$ emission \citep{Walsh2015}. Disks around M dwarfs are expected to receive less UV radiation than T Tauri or Herbig Ae/Be disks, resulting in relatively larger abundances of UV-sensitive molecules \citep{Walsh2015}.

We do not find statistically significant correlations between stellar masses and total observed Ly$\alpha$ fluxes $\left( \rho = 0.0486, p = 0.76 \right)$ or Ly$\alpha$ flux ratios $\left( \rho = -0.0797, p = 0.611 \right)$ from our sample of primarily K and M type stars, although the relationships between reconstructed Ly$\alpha$ fluxes at the disk surfaces and stellar properties will be explored in future work (France et al., in prep). The ratios of Ly$\alpha$/FUV continuum emission are strongly correlated with the mass accretion rates derived from C IV emission \citep{France2014}, and there is also no correlation between stellar mass and accretion rates for the sample presented here. However, the correlations between Ly$\alpha$ flux ratios, $\left[W3 - W4 \right]$ colors, and 1600 \AA \, H$_2$O dissociation ``bump" detection rates \citep{France2017} reported in this work are consistent with models predicting increased Ly$\alpha$ propagation with dust settling \citep{Bethell2011} and observations of CN/HCN abundance ratios peaking within UV-exposed dust gaps \citep{Bergner2021}.      

The ALMA Large Program ``Molecules with ALMA at Planet-forming Scales" (MAPS) unexpectedly found that the observed CN/HCN ratios from AS 209, GM Aur, IM Lup, HD 163296, and MWC 480 did not show any clear trends with stellar UV luminosity, despite the five targets spanning two orders of magnitude in integrated flux between 910-2000 \AA \citep{Bergner2021}. However, just two of these targets have Ly$\alpha$ spectra from HST: HD 163296 (P Cygni-like) and GM Aur (intermediate). AS 209 and MWC 480 were observed at low spectral resolution with IUE, and IM Lup does not have short wavelength UV coverage as of ULLYSES DR3 (see e.g., \citealt{Dionatos2019}). Furthermore, physical-chemical models have so far incorporated approximations of Ly$\alpha$ emission lines from different types of host stars \citep{Walsh2015}, as direct observations were not available. The true nature of Ly$\alpha$ as a driver of protoplanetary disk chemistry has not yet been studied from an observational or modeling perspective, but can now be fully explored using the ULLYSES data. 

\section{Summary \& Conclusions}

We have used two different modeling frameworks to reproduce observed Ly$\alpha$ emission lines in \emph{HST}-COS spectra of 42 CTTSs from the ULLYSES program, in an effort to explore the connection between the line profile shapes and outflow and accretion kinematics. Our results demonstrate that:
\begin{itemize}
    \item $\sim$70\% of targets in our sample have P Cygni-like Ly$\alpha$ emission lines, with blue velocity emission suppressed by outflowing H I. The remaining $\sim$30\% have inverse P Cygni-like profiles, a percentage that is roughly consistent with the observed fraction of sub-mm transition disks with dust cavities carved by planet-disk interactions or photoevaporative or MHD winds (see e.g., \citealt{Andrews2011}). The subset of our sample with inverse P Cygni-like profiles and spatially resolved sub-mm observations show dust gaps or cavities, and a statistically significant negative correlation is observed between the ratios of red to blue Ly$\alpha$ flux and $\left[W3 - W4 \right]$ $\left( \rho = -0.33, p = 0.04 \right)$. 
    \item The observed emission line wings for 27/42 targets are well reproduced by the models with resonant scattering in a simple shell. Although the geometry of T Tauri systems is more complex than the simple shell model used here, we interpret the best-fit parameters as average properties of the bulk intervening H I within inner disk winds, accretion flows, protoplanetary disks, and the ISM.
    \item The model H I velocities are strongly correlated with the \emph{K} band veiling when $r_K$ is measured against synthetic spectra ($\rho = -0.66$; $p = 0.0006$), tentatively indicating the Ly$\alpha$ line profiles change shape as the hot inner disk is depleted.   
    \item No statistically significant correlation is detected between the model H I velocities and outer disk inclinations, indicating that the trends reported here are not solely caused by the viewing angles. We note that a) sub-mm observations are not available for targets with H I velocities much larger or smaller than the median value $v \sim -94 \, \rm{km s}^{-1}$, and b) the Ly$\alpha$ emission lines are more strongly associated with the inner disks, which are frequently misaligned (see e.g., \citealt{Davies2019, Bohn2022}).
\end{itemize}
The resonant scattering models that were applied in this work use a simple shell geometry that may not be fully representative of the complex environment surrounding CTTSs. However, the trends reported here reveal that the ratios of red to blue Ly$\alpha$ fluxes observed with \emph{HST}-COS are closely linked to inner disk outflow and accretion kinematics, which go along with the evolution of the dust disks. The Ly$\alpha$ flux ratios across our sample have a range of three orders of magnitude, a spread in emission line shape that may cause significant variations in the photodissociation rates of Ly$\alpha$-sensitive molecules that will be detected with \emph{JWST} and ALMA (see e.g., \citealt{Bergin2003}). Spatially resolved UV spectra (e.g. from \emph{HST}-STIS) are required to fully explore multi-phase Ly$\alpha$ resonant scattering through the protoplanetary disks, accretion flows, outflowing winds, and ISM.   

\section{Acknowledgements}
Based on observations obtained with the NASA/ESA Hubble Space Telescope, retrieved from the Mikulski Archive for Space Telescopes (MAST) at the Space Telescope Science Institute (STScI). STScI is operated by the Association of Universities for Research in Astronomy, Inc. under NASA contract NAS 5-26555. Based on data obtained with ESO programmes 106.20Z8.002 and 106.20Z8.009. This work benefited from discussions with the ODYSSEUS team (HST AR-16129, \citealt{Espaillat2022_ODYSSEUSI}), https://sites.bu.edu/odysseus/. This research made use of Astropy,\footnote{http://www.astropy.org} a community-developed core Python package for Astronomy \citep{astropy2013, astropy2018}. This project has received funding from the European Research Council (ERC) under the European Union's Horizon 2020 research and innovation programme under grant agreement No 716155 (SACCRED). Funded by the European Union under the European Union’s Horizon Europe Research \& Innovation Programme 101039452 (WANDA). Views and opinions expressed are however those of the author(s) only and do not necessarily reflect those of the European Union or the European Research Council. Neither the European Union nor the granting authority can be held responsible for them.  J.F. Gameiro was supported by funda\c c\~ao para a Ci\^encia e Tecnologia (FCT) through the research grants UIDB/04434/2020 and UIDP/04434/2020.

\appendix

\section{Best-Fit Ly$\alpha$ Models for ULLYSES Targets}
\restartappendixnumbering

\begin{deluxetable*}{ccccc}
\tablecaption{Model Parameters and Prior Bounds for Ly$\alpha$ Without Scattering \label{tab:model_params_noscatt}} 
\tablehead{
\colhead{Gaussian Amplitude} & \colhead{Gaussian FWHM} & \colhead{Central Wavelength} &  \colhead{H I Column Density} & \colhead{H I Velocity} \\
\colhead{$I_0$} & \colhead{FWHM} & \colhead{$\lambda_0$} & \colhead{$N \left( \rm{H I} \right)_{\rm{out}}$} & \colhead{$v_{\rm{out}}$} \\
\hline
\colhead{(erg s$^{-1}$ cm$^{-2}$ \AA$^{-1}$)} & \colhead{(km s$^{-1}$)} & \colhead{(\AA)} & \colhead{(dex)} & \colhead{(km s$^{-1}$)} 
}
\startdata
$\left(10^{-16}, 10^{-11} \right)$ & (250, 2000) & (1214.5, 1216.5) & (18.5, 22.0) & (-250, 250) \\
\enddata
\tablecomments{All parameters were sampled over a continuous distribution.}
\end{deluxetable*}

\begin{deluxetable*}{cccc}
\tablecaption{Model Parameters and Prior Bounds for Ly$\alpha$ With Scattering \label{tab:model_params_scatt}} 
\tablehead{
\colhead{Shell Velocity} & \colhead{H I Column Density} & \colhead{Effective Temperature} & \colhead{Gaussian Width} \\
\colhead{$v_{\rm{exp}}$} & \colhead{$N \left( \rm{H I} \right)_{\rm{out}}$} & \colhead{$\log{T/K}$} & \colhead{$\sigma_i$} \\
\hline
\colhead{(km s$^{-1}$)} & \colhead{(dex)} & \nocolhead{} & \colhead{(km s$^{-1}$)}
}
\startdata
(-495, 495) & (15.9, 21.9) & (2.8, 6.0) & (1, 800)
\\
\hline
\hline
\colhead{Dust Optical Depth} & \colhead{Equivalent Width} & \colhead{Central Velocity} & \colhead{Integrated Flux Normalization} \\
\colhead{$\tau_d$} & \colhead{$EW_i$} & \colhead{$\Delta$} & \colhead{$A$} \\
\hline
\nocolhead{} & \colhead{(\AA)} & \colhead{(km s$^{-1}$)} & \nocolhead{} \\
\hline
(0, 5) & (1, 6000) & (-100, 100) & (0.7, 2.5) \\
\enddata
\tablecomments{$v_{\rm{exp}}$, $N \left( \rm{H I} \right)$, and $\log{T/K}$ were treated as discrete parameters. All other parameters were sampled over a continuous distribution.}
\end{deluxetable*}

\begin{figure*}
\gridline{\fig{ 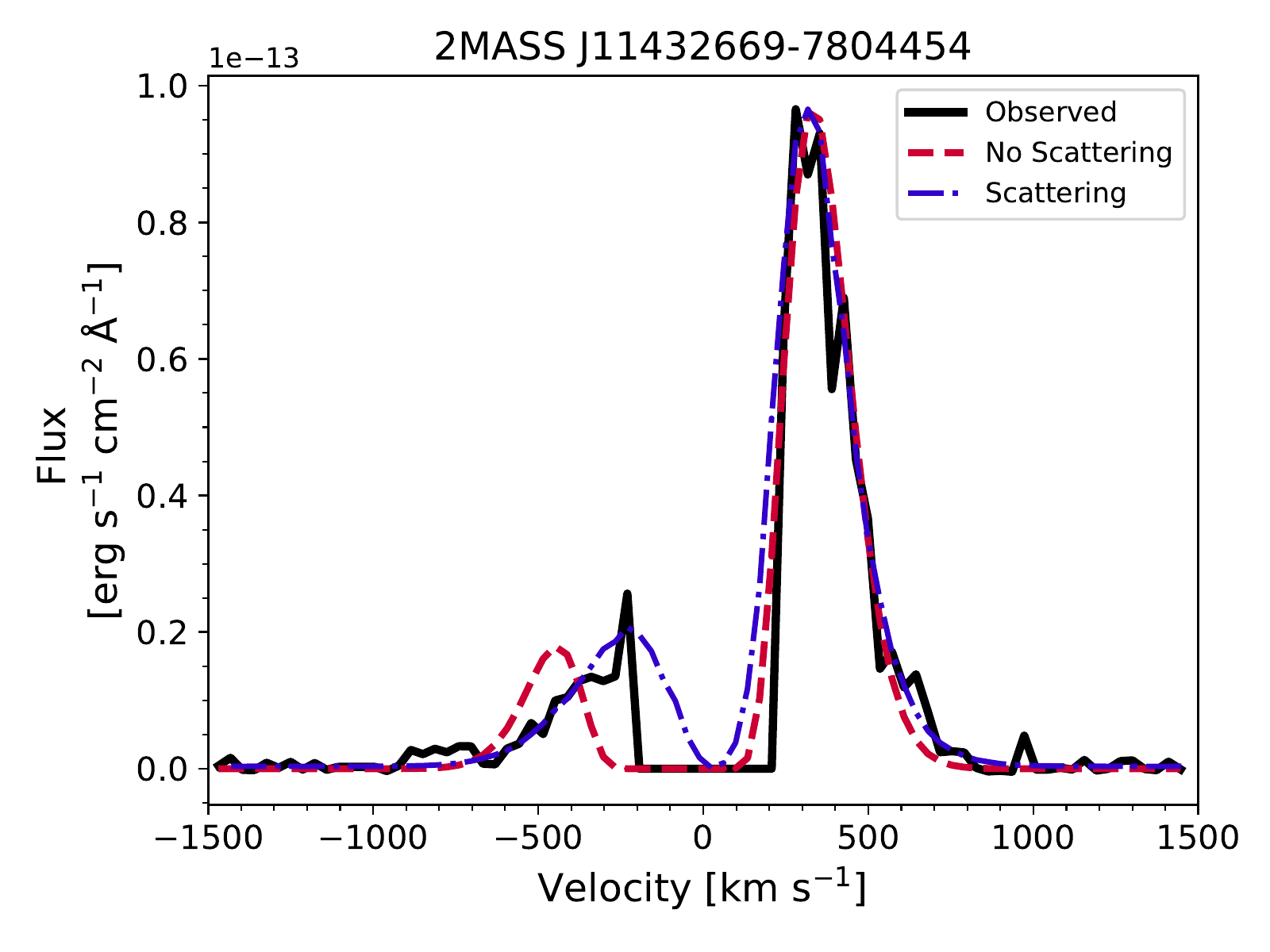}{0.5\textwidth}{(a) 2MASS J11432669-7804454}
          \fig{ 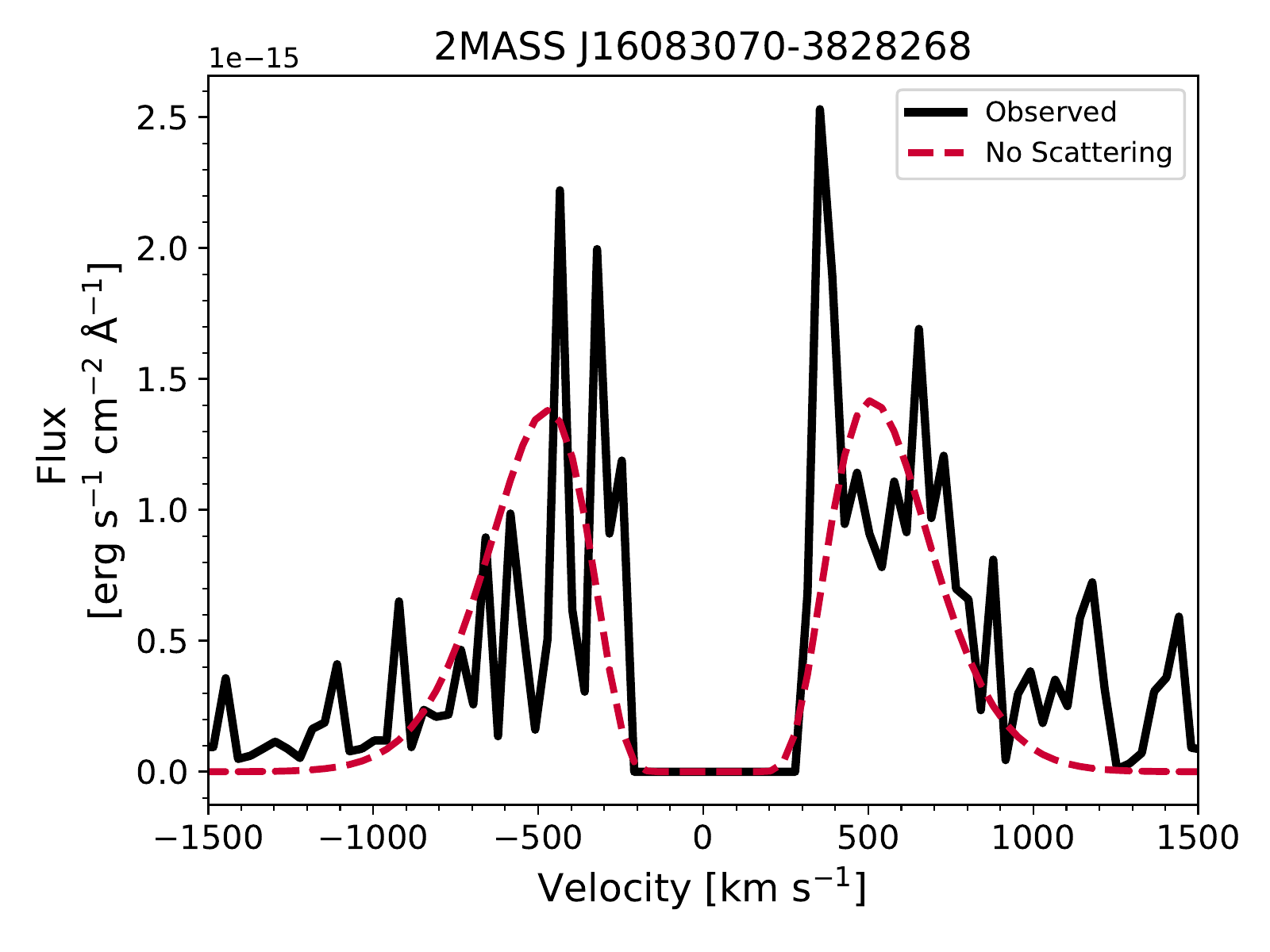}{0.5\textwidth}{(b) 2MASS J16083070-3828268}}
\gridline{\fig{ 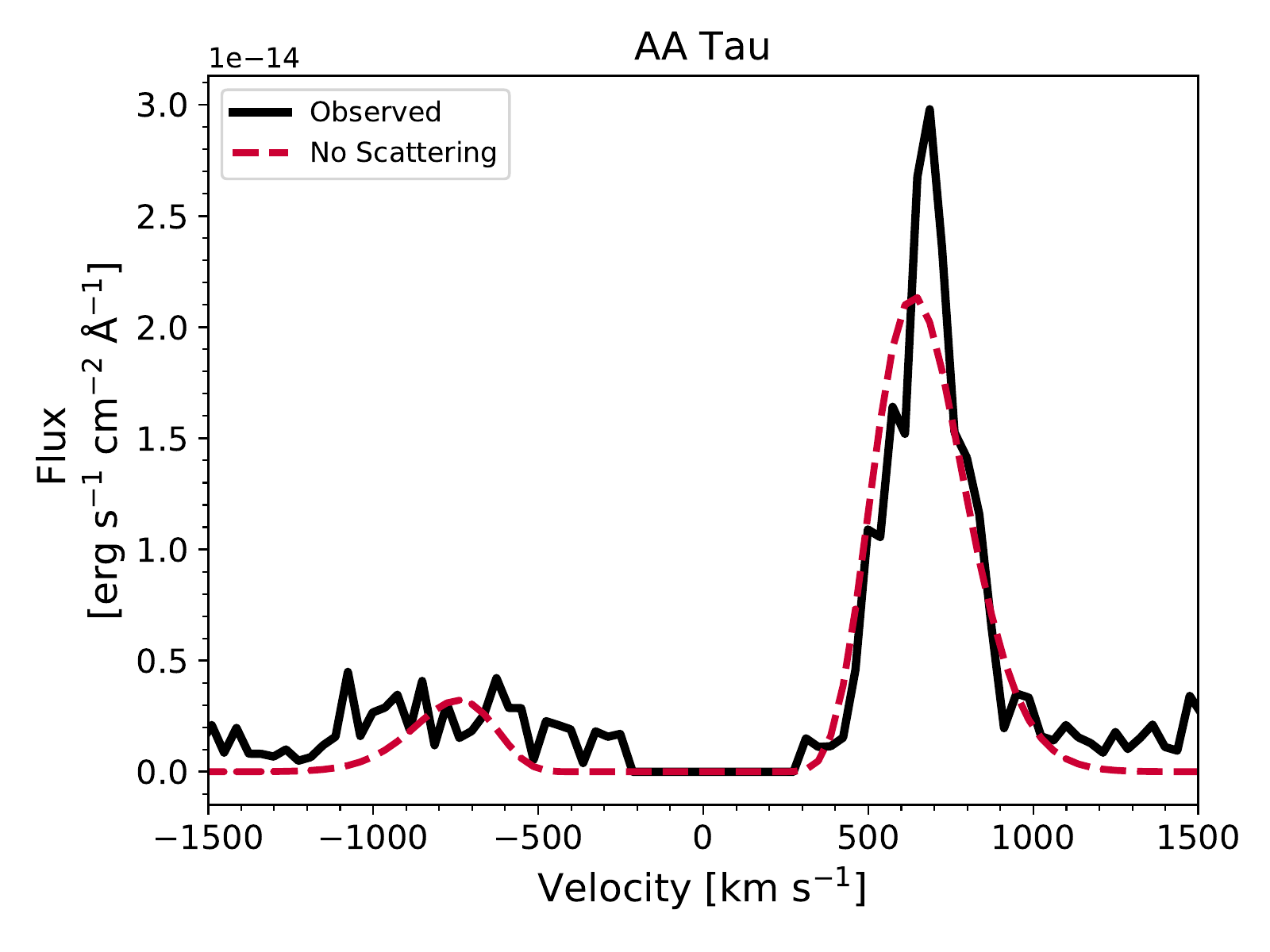}{0.5\textwidth}{(c) AA Tau}
\fig{ 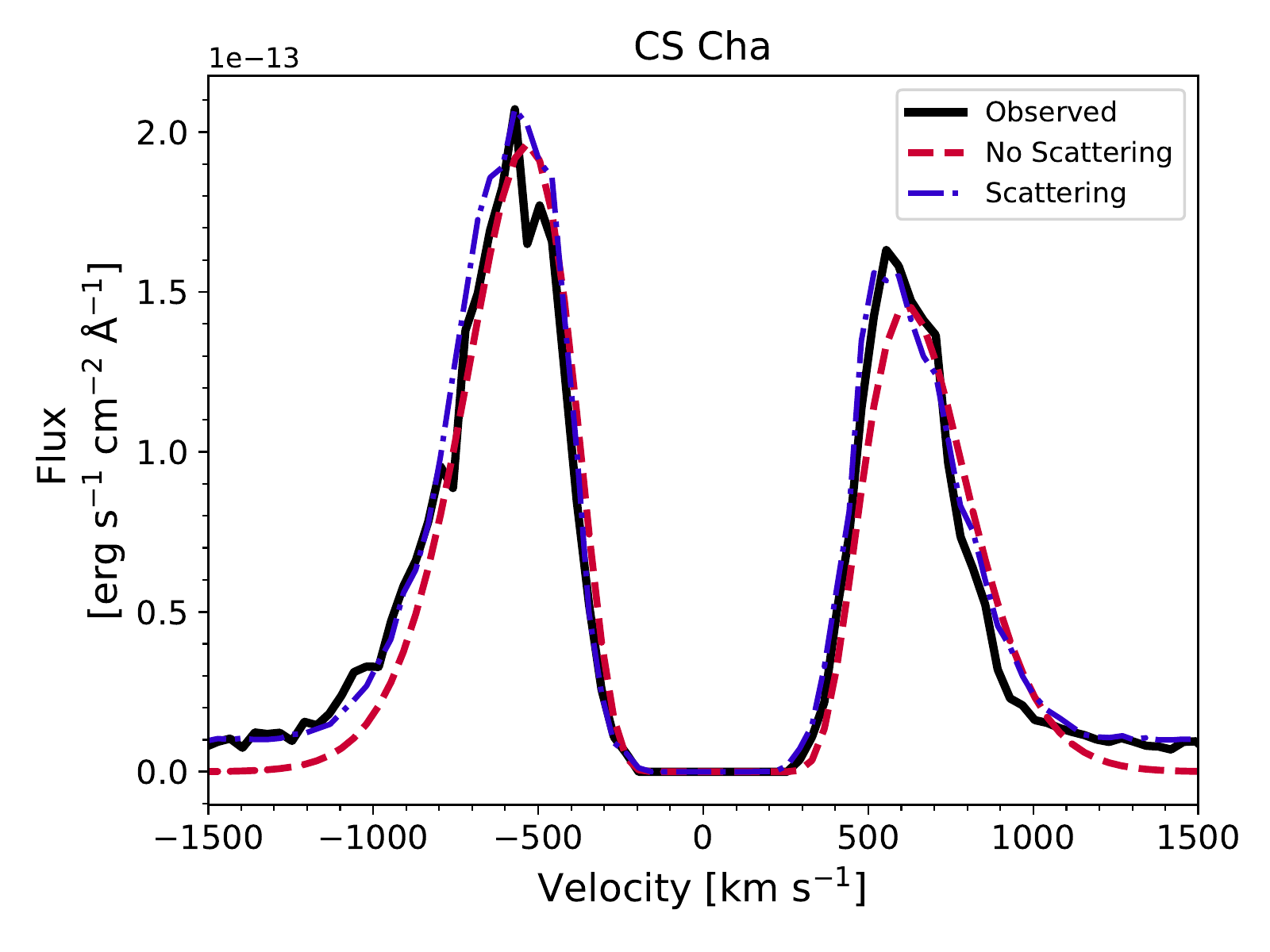}{0.5\textwidth}{(d) CS Cha}}
\gridline{\fig{ 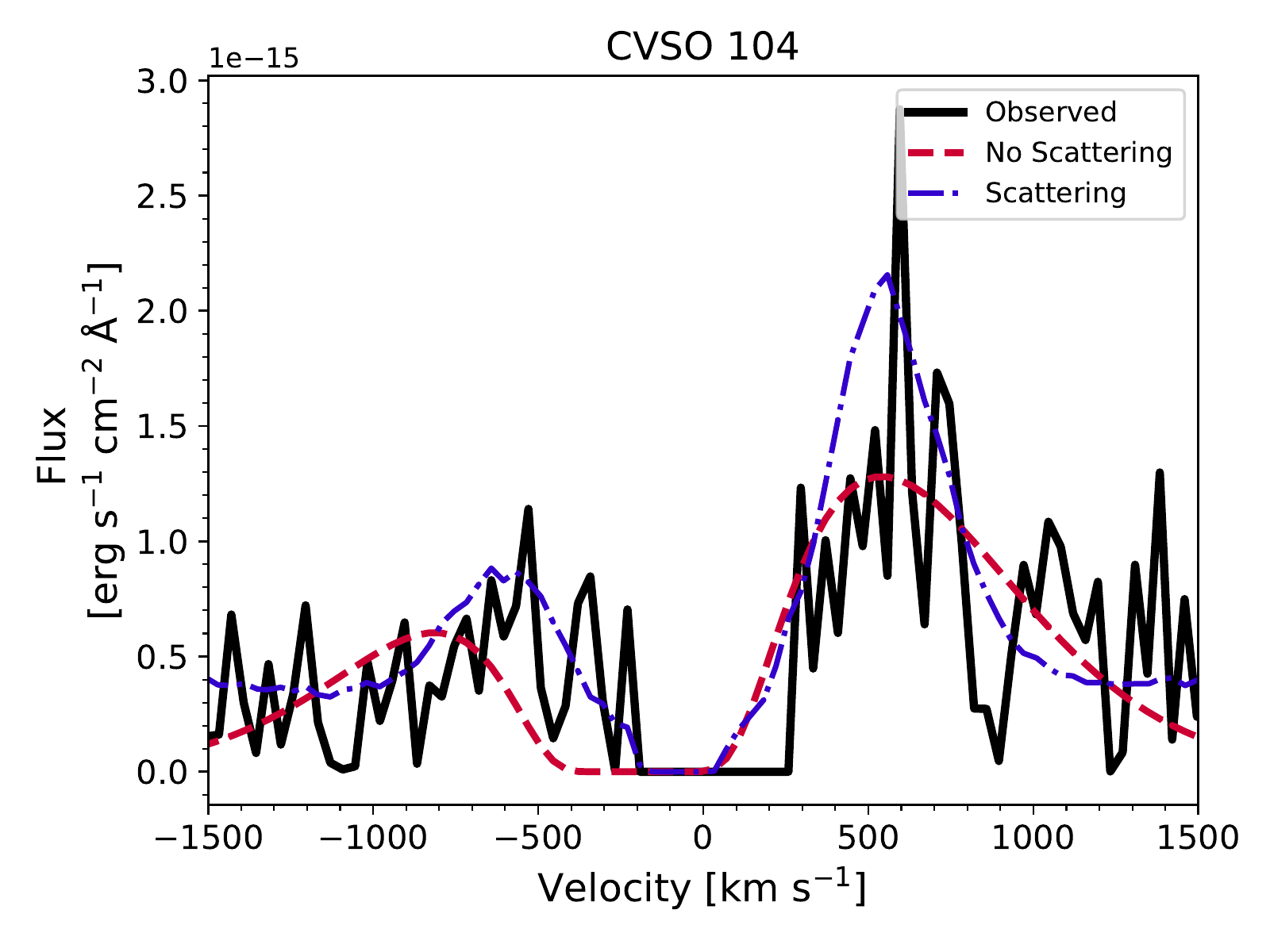}{0.5\textwidth}{(e) CVSO 104}
\fig{ 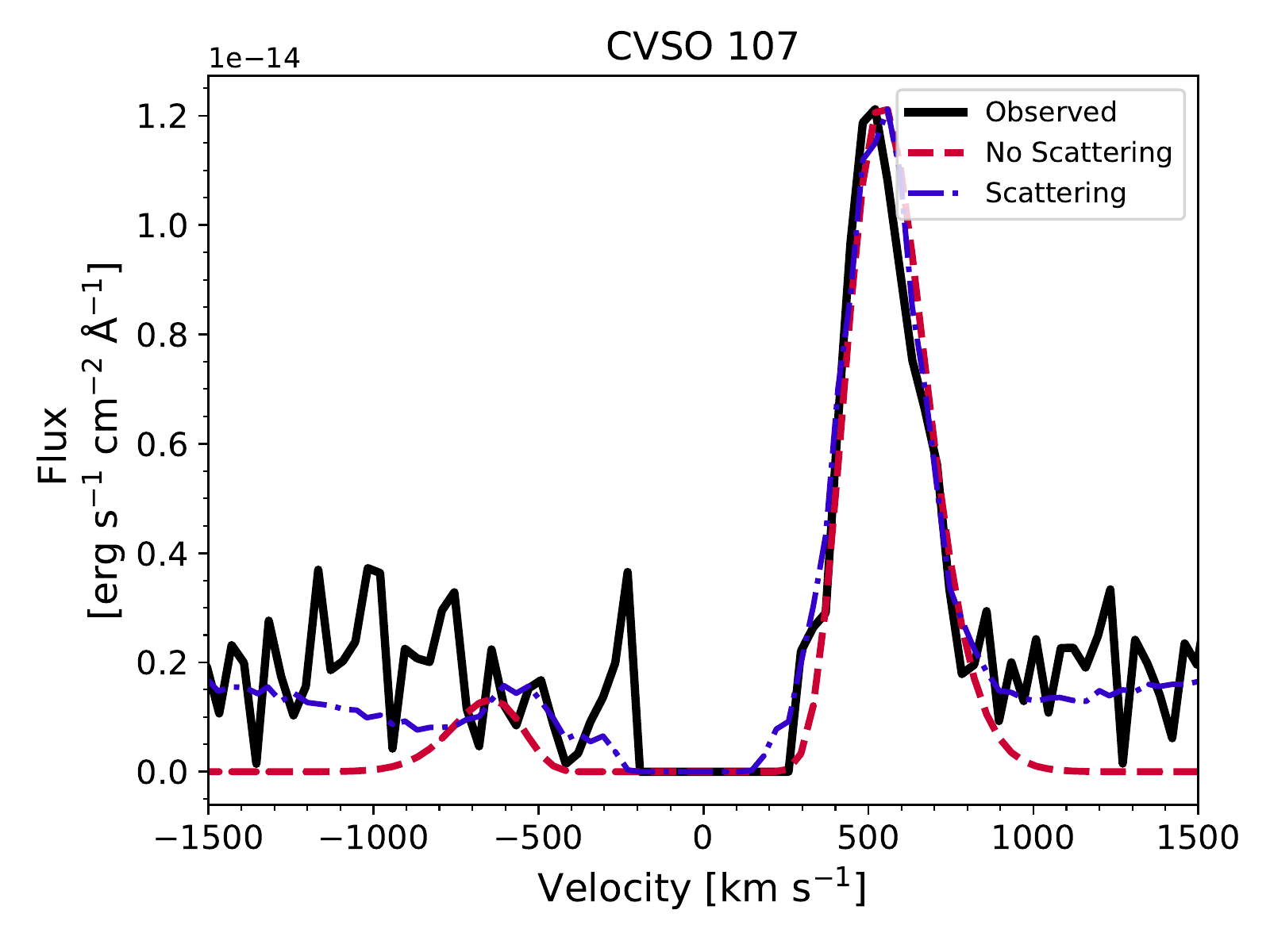}{0.5\textwidth}{(f) CVSO 107}}

\caption{Observed Ly$\alpha$ emission line (black, solid) and best-fit models with scattering (blue, dash-dotted) and without scattering (red, dashed) for all 44 targets in our sample. Scattering models are not shown for targets with low S/N Ly$\alpha$ spectra, for which the MCMC chains did not converge.}
\end{figure*}

\begin{figure*}
\gridline{\fig{ LyA_Scattering_arxiv/Figures/CVSO109_LyA_modelcomp.pdf}{0.5\textwidth}{(a) CVSO 109}
\fig{ 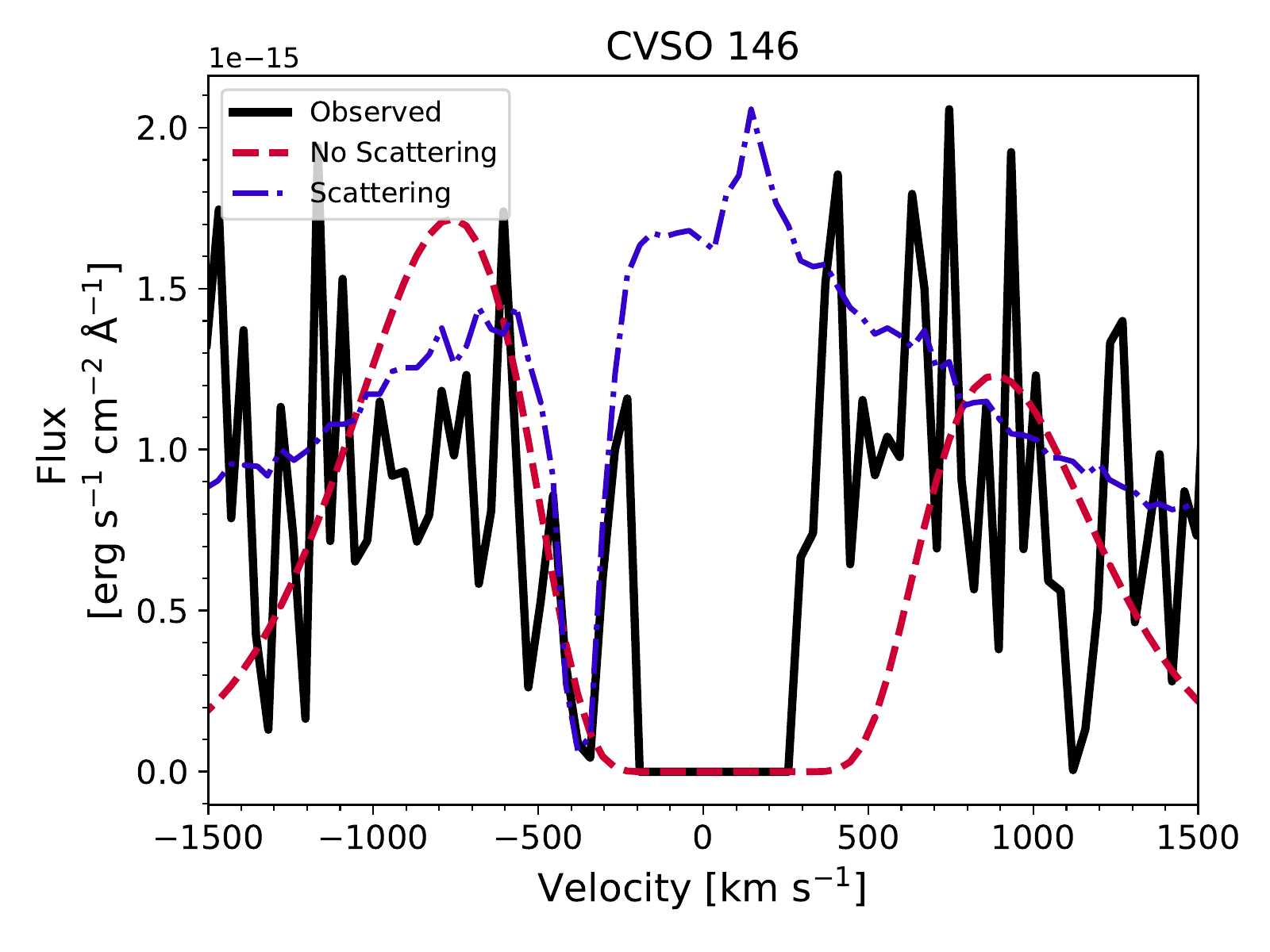}{0.5\textwidth}{(b) CVSO 146}}
\gridline{\fig{ 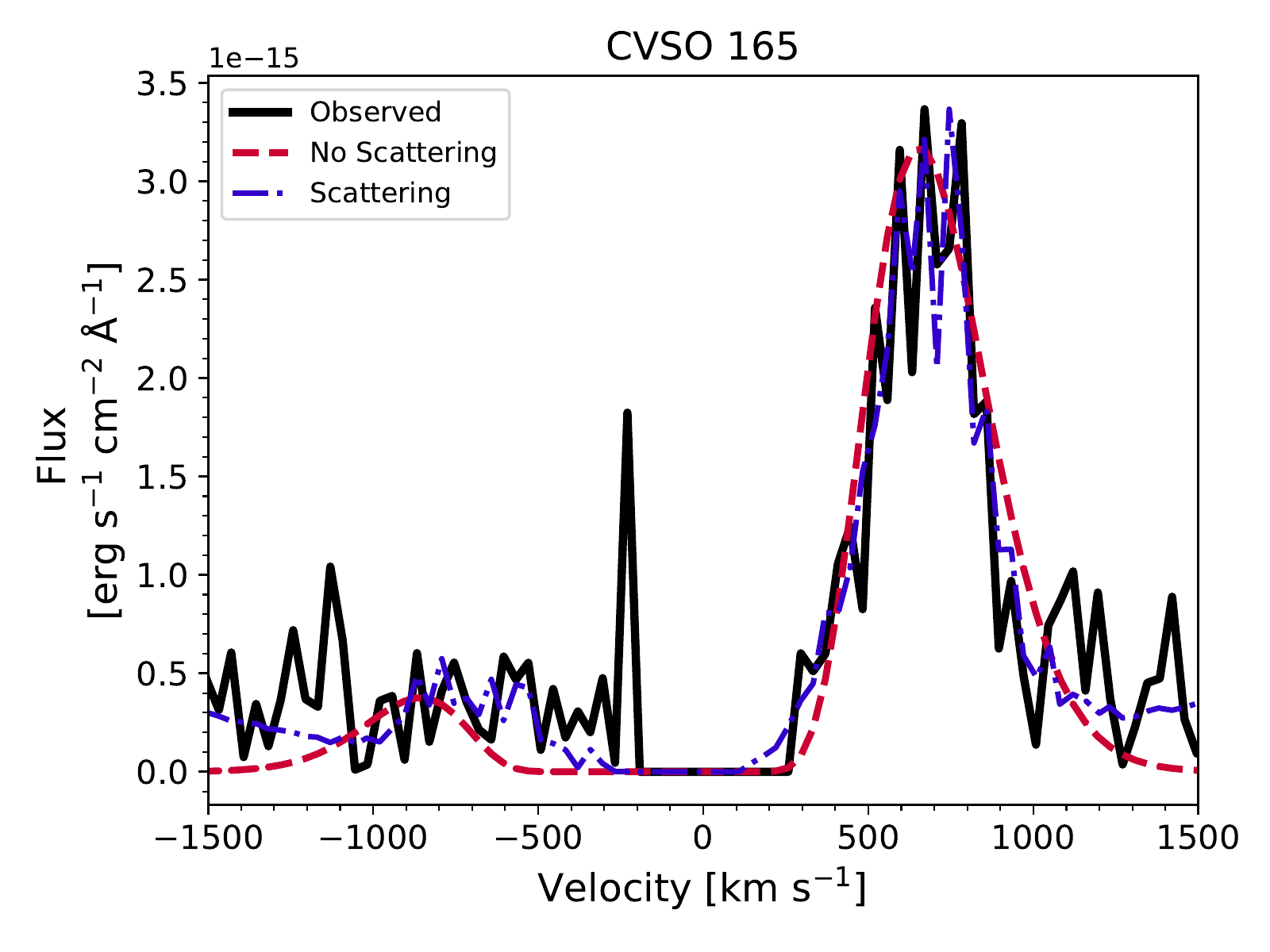}{0.5\textwidth}{(c) CVSO 165}
\fig{ 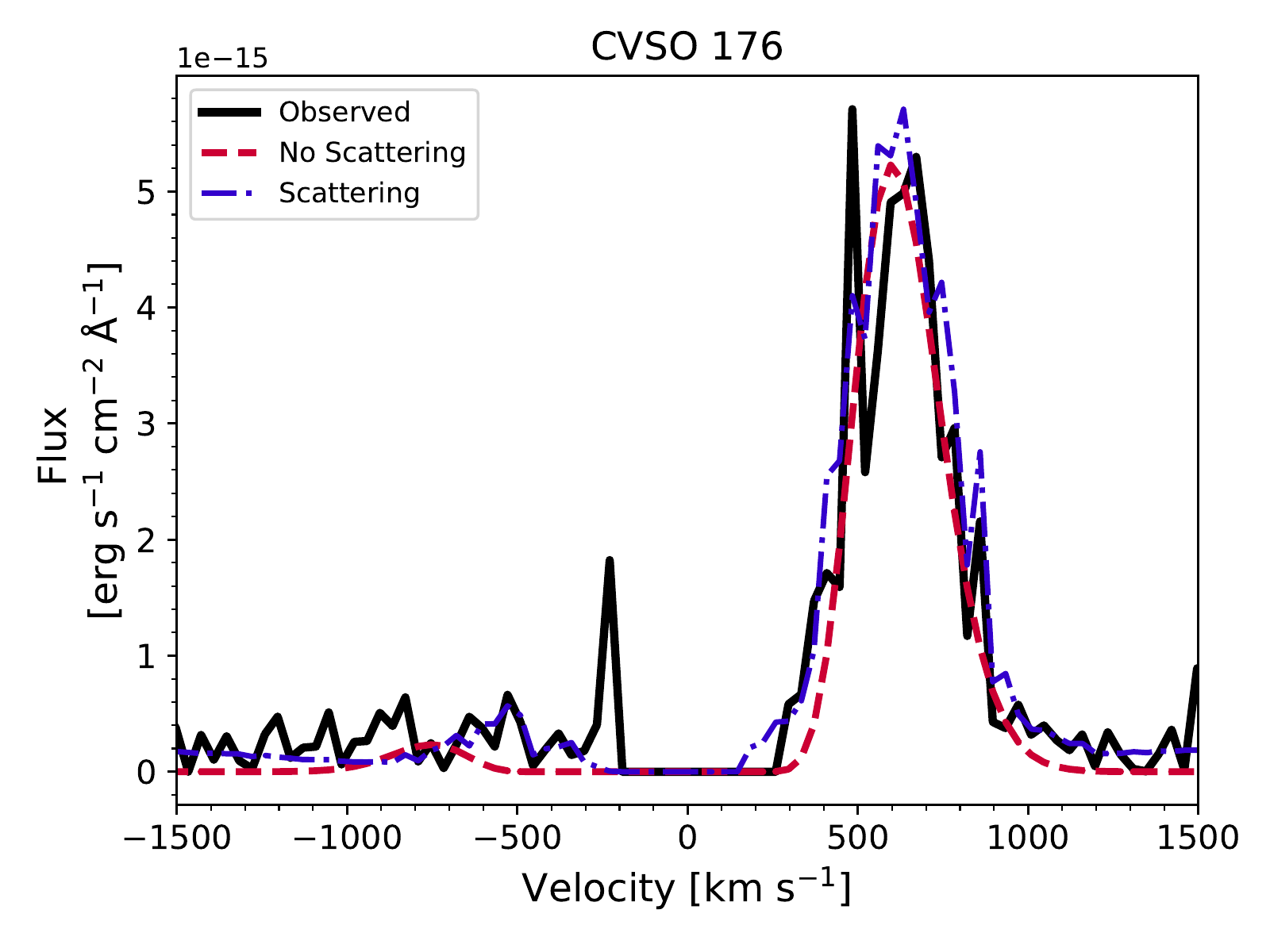}{0.5\textwidth}{(d) CVSO 176}}
\gridline{\fig{ 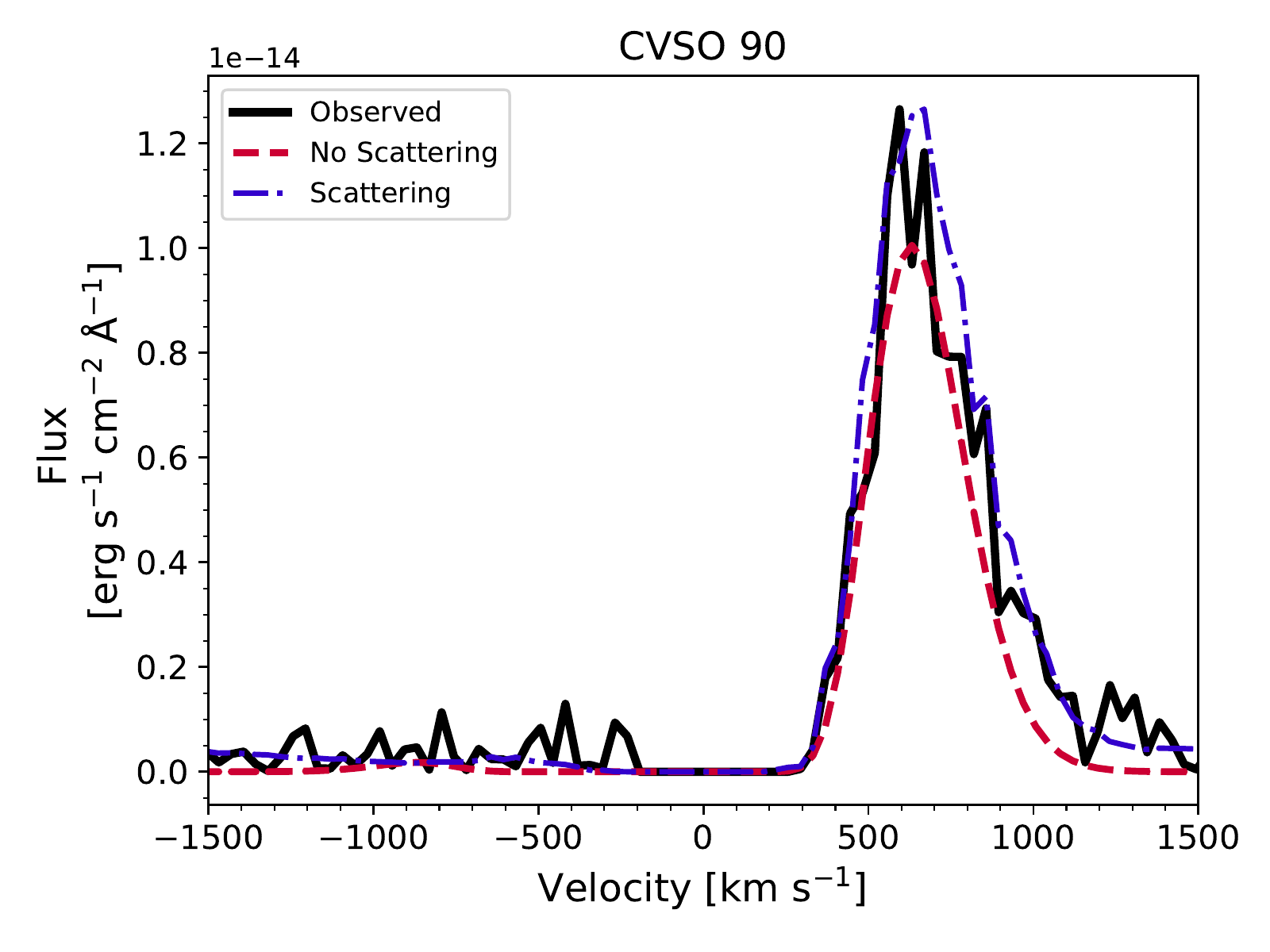}{0.5\textwidth}{(e) CVSO 90}
\fig{ LyA_Scattering_arxiv/Figures/DFTau_LyA_modelcomp.pdf}{0.5\textwidth}{(f) DF Tau}}
\caption{Observed Ly$\alpha$ emission line (black, solid) and best-fit models with scattering (blue, dash-dotted) and without scattering (red, dashed) for all 44 targets in our sample. Scattering models are not shown for targets with low S/N Ly$\alpha$ spectra, for which the MCMC chains did not converge.}
\end{figure*}

\begin{figure*}
\gridline{\fig{ LyA_Scattering_arxiv/Figures/DMTau_LyA_modelcomp.pdf}{0.5\textwidth}{(a) DM Tau}
\fig{ 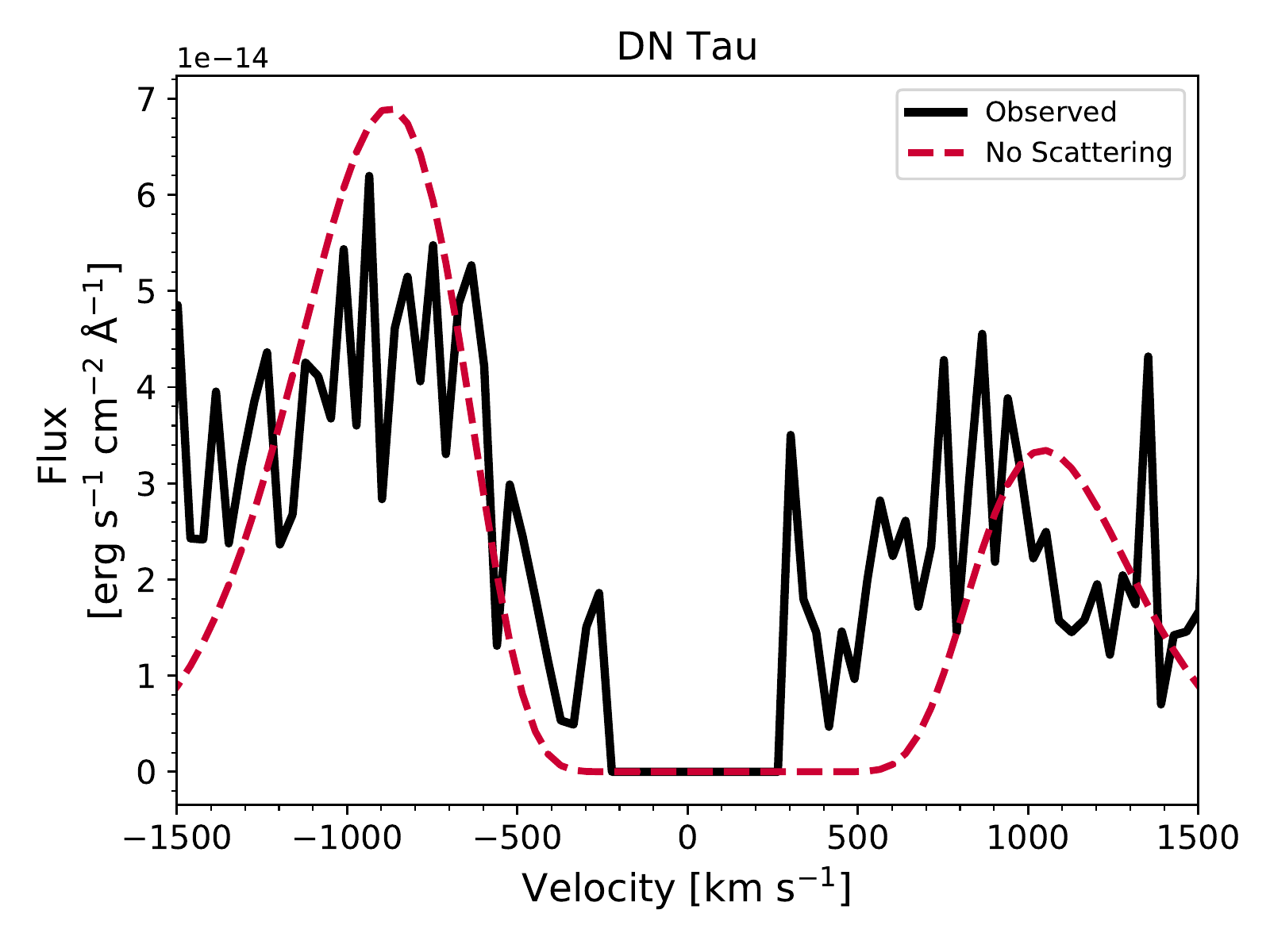}{0.5\textwidth}{(b) DN Tau}}
\gridline{\fig{ 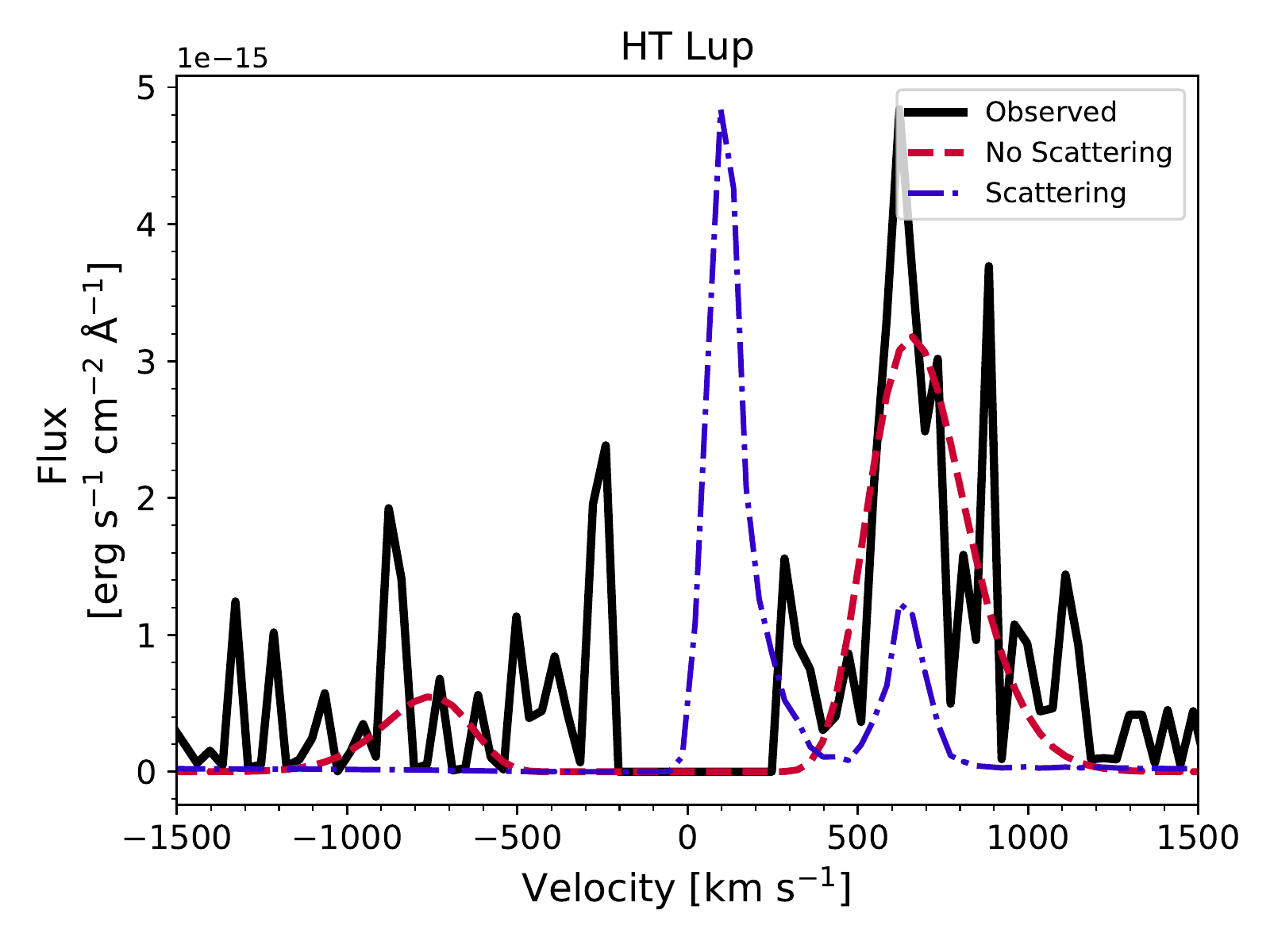}{0.5\textwidth}{(c) HT Lup}
\fig{ 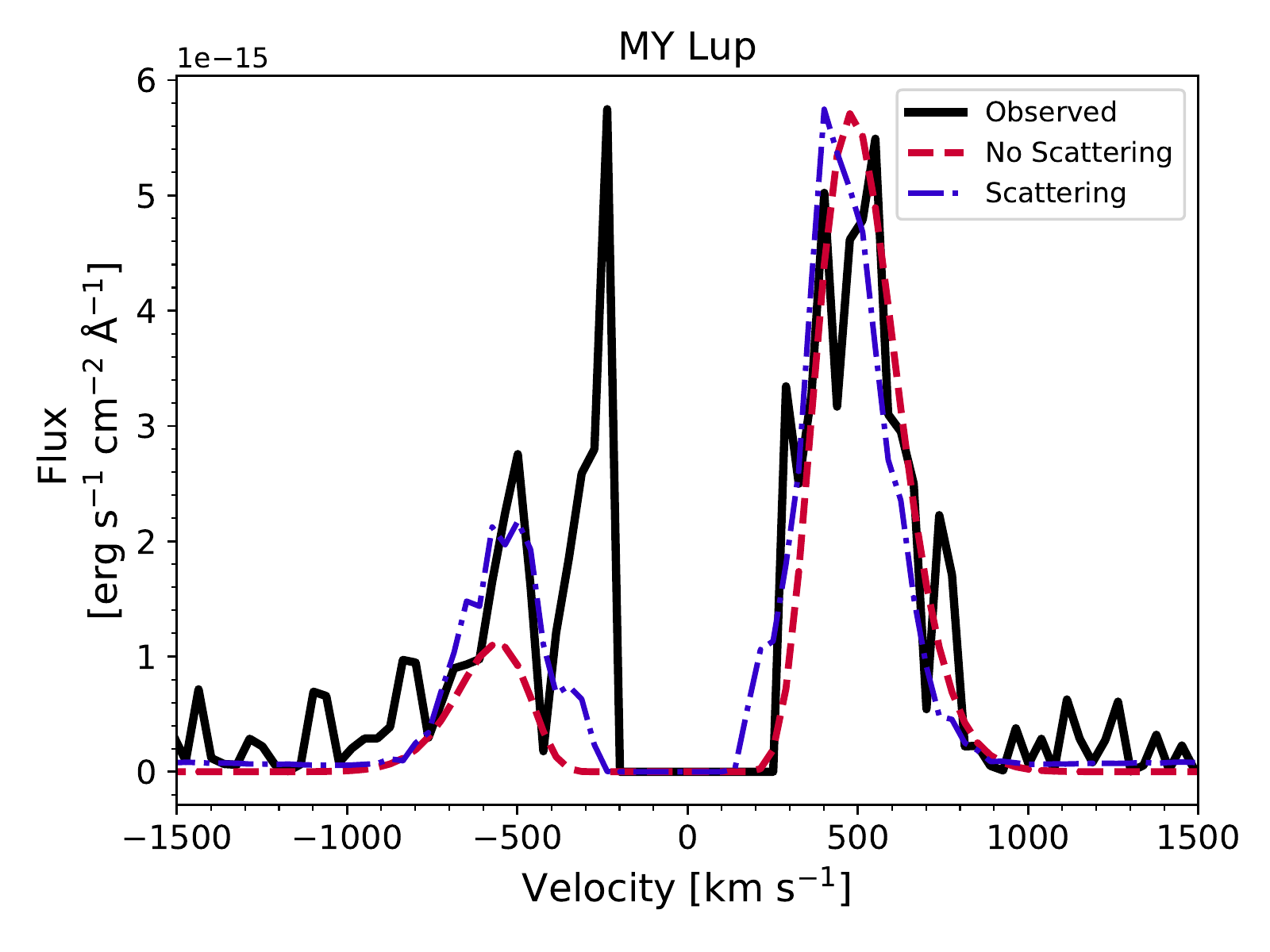}{0.5\textwidth}{(d) MY Lup}}
\gridline{\fig{ 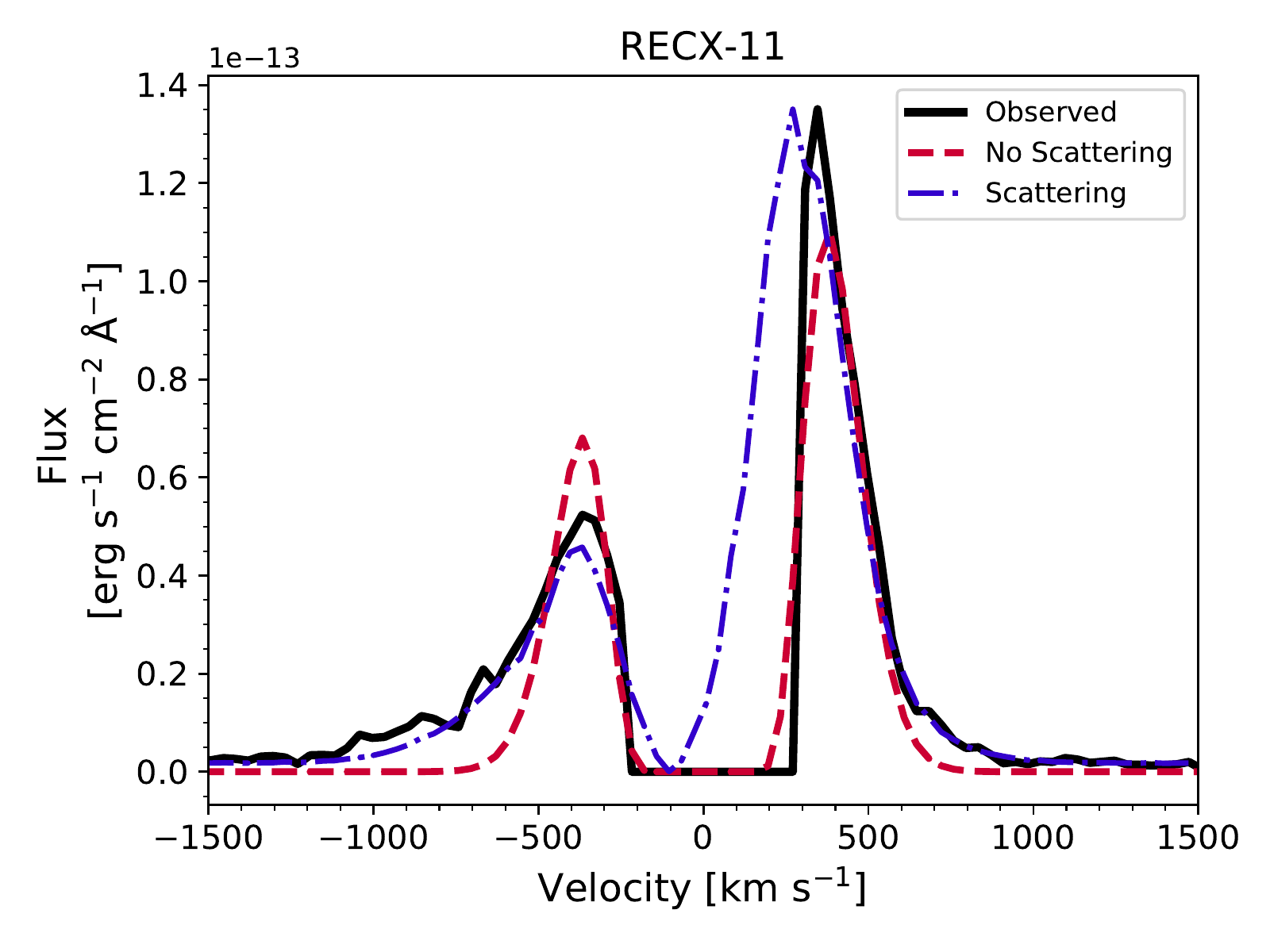}{0.5\textwidth}{(e) RECX-11}
\fig{ 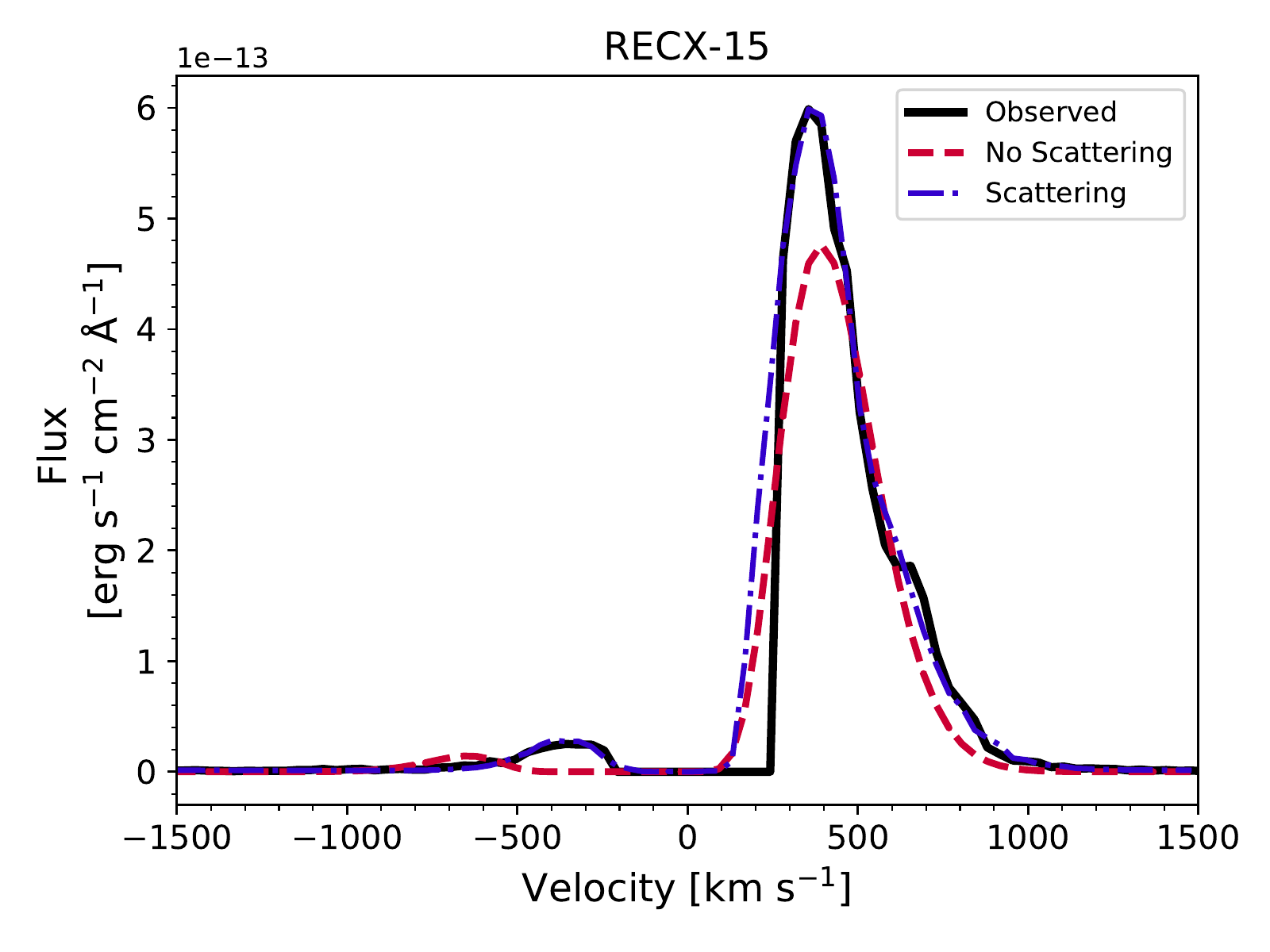}{0.5\textwidth}{(f) RECX-15}}
\caption{Observed Ly$\alpha$ emission line (black, solid) and best-fit models with scattering (blue, dash-dotted) and without scattering (red, dashed) for all 44 targets in our sample. Scattering models are not shown for targets with low S/N Ly$\alpha$ spectra, for which the MCMC chains did not converge.}
\end{figure*}

\begin{figure*}
\gridline{\fig{ 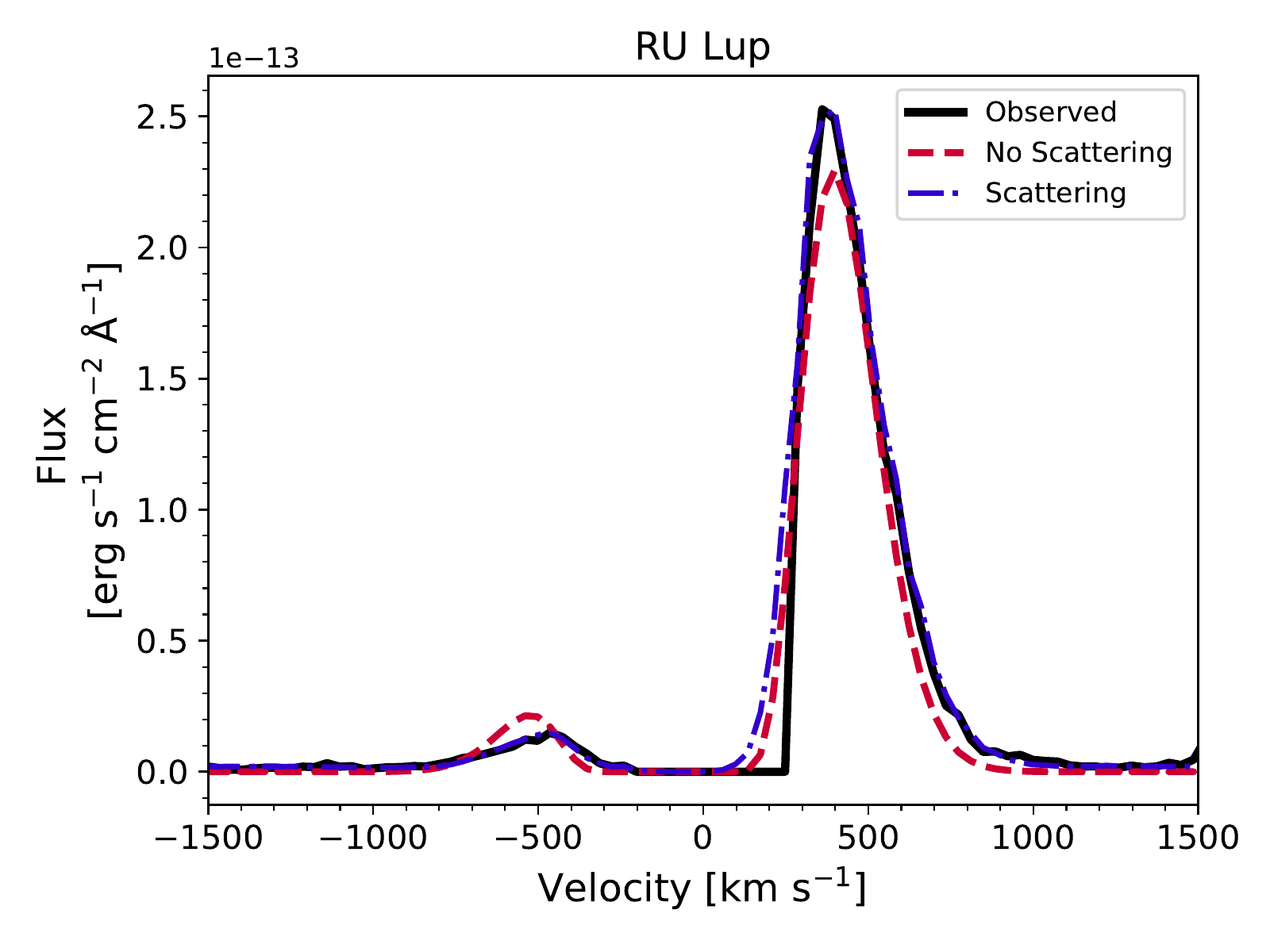}{0.5\textwidth}{(a) RU Lup}
\fig{ 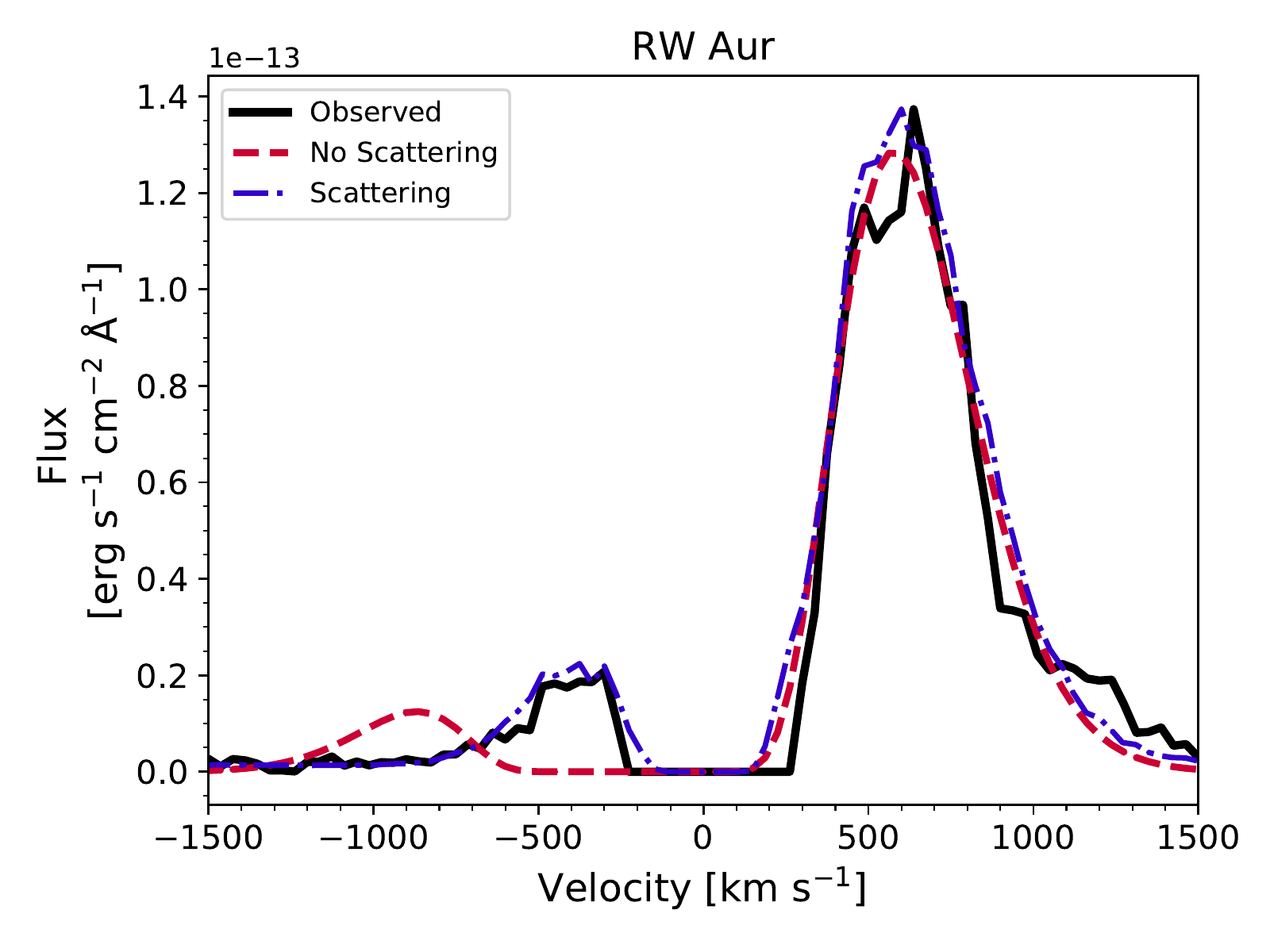}{0.5\textwidth}{(b) RW Aur}}
\gridline{\fig{ 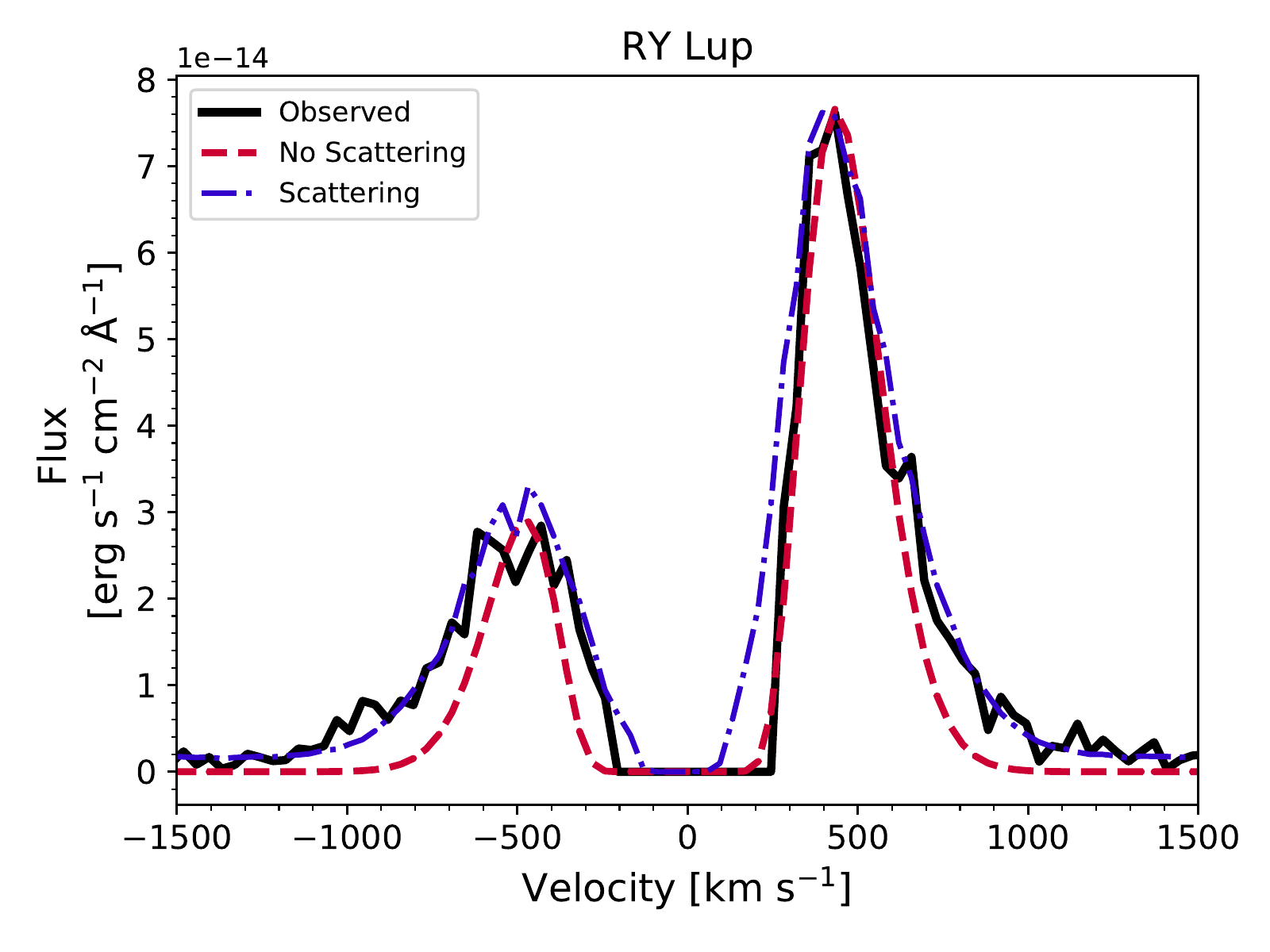}{0.5\textwidth}{(c) RY Lup}
\fig{ 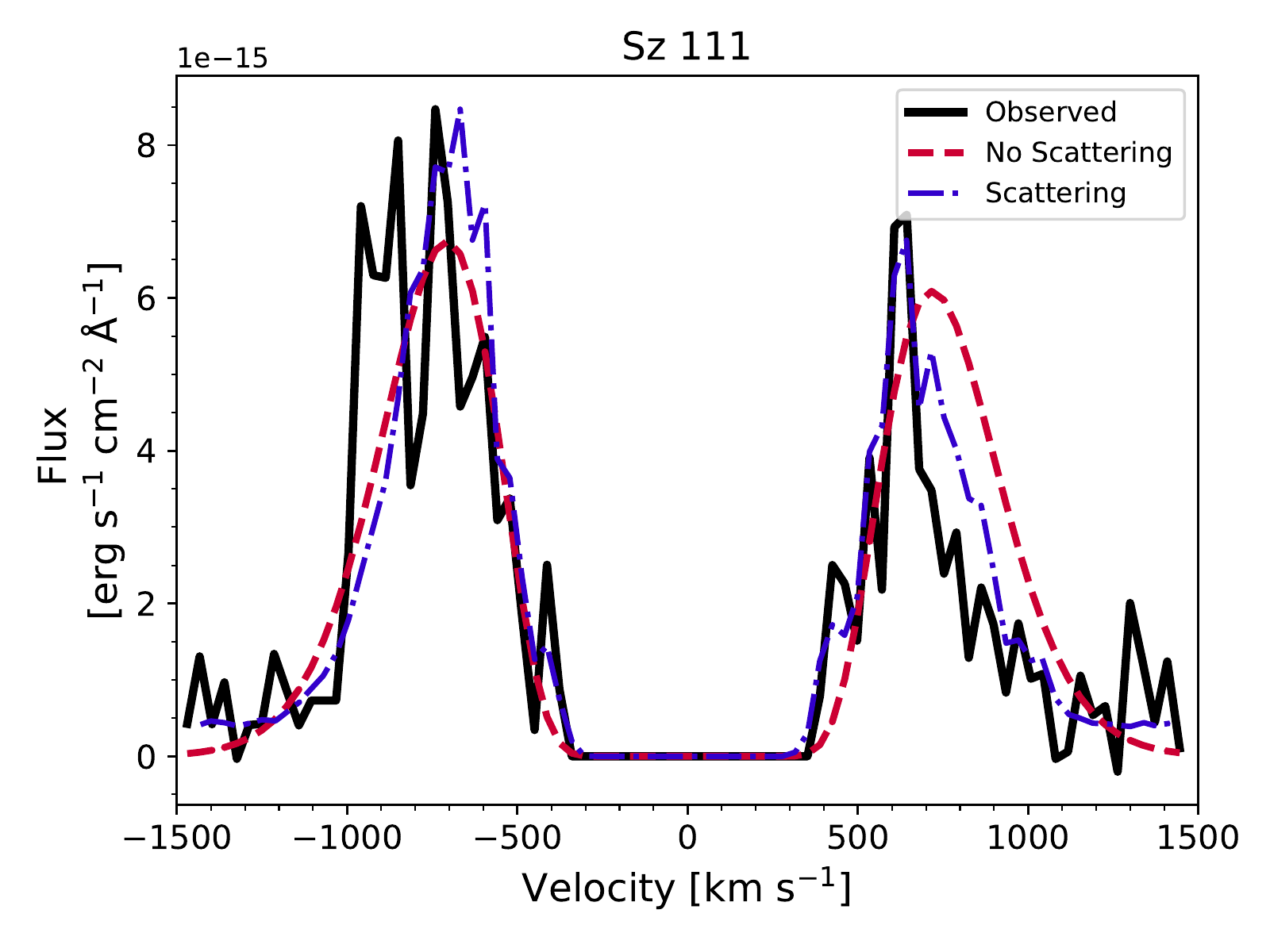}{0.5\textwidth}{(d) Sz 111}}
\gridline{\fig{ 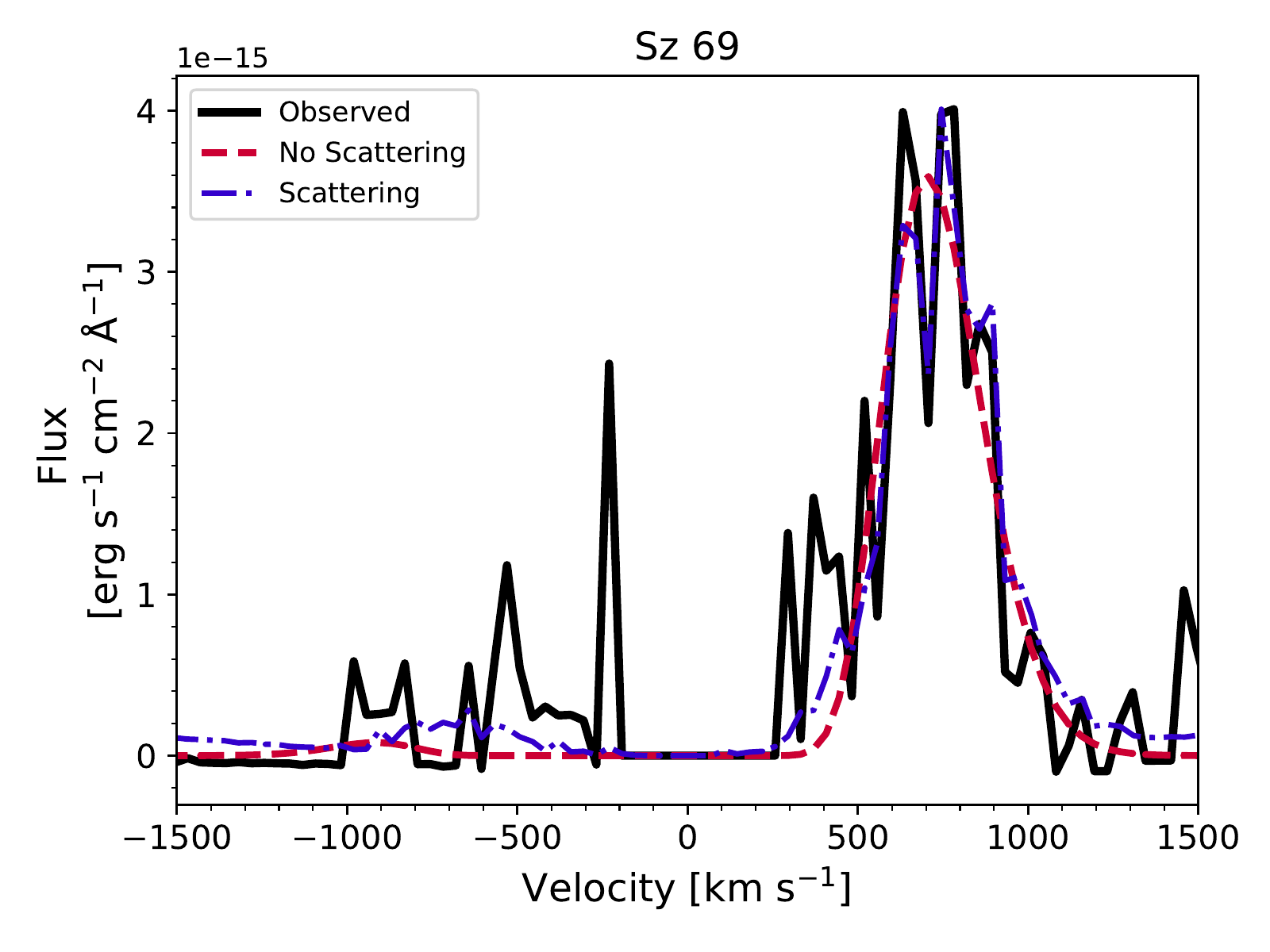}{0.5\textwidth}{(e) Sz 69}
\fig{ 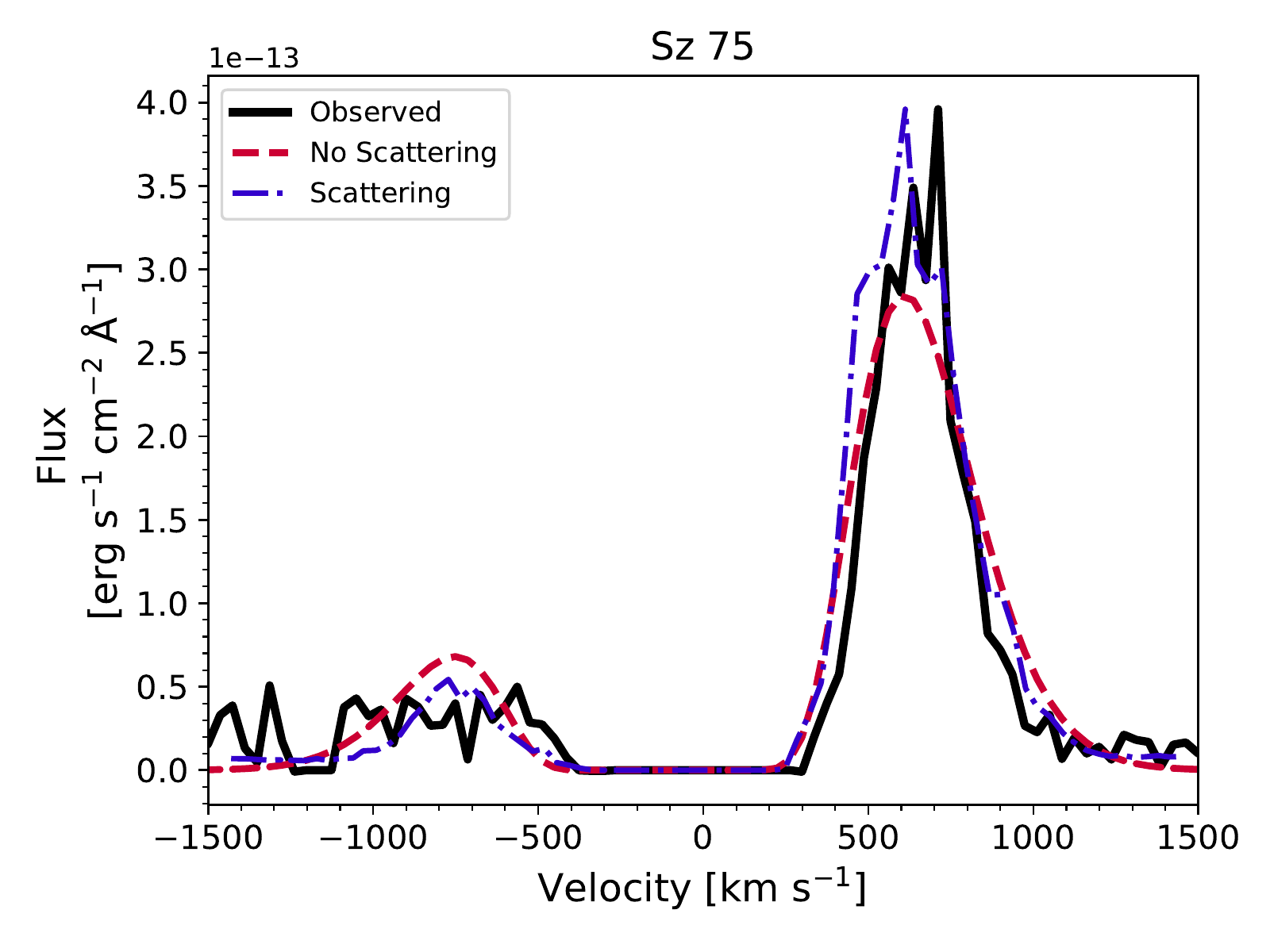}{0.5\textwidth}{(f) Sz 75}}
\caption{Observed Ly$\alpha$ emission line (black, solid) and best-fit models with scattering (blue, dash-dotted) and without scattering (red, dashed) for all 44 targets in our sample. Scattering models are not shown for targets with low S/N Ly$\alpha$ spectra, for which the MCMC chains did not converge.}
\end{figure*}

\begin{figure*}
\gridline{\fig{ 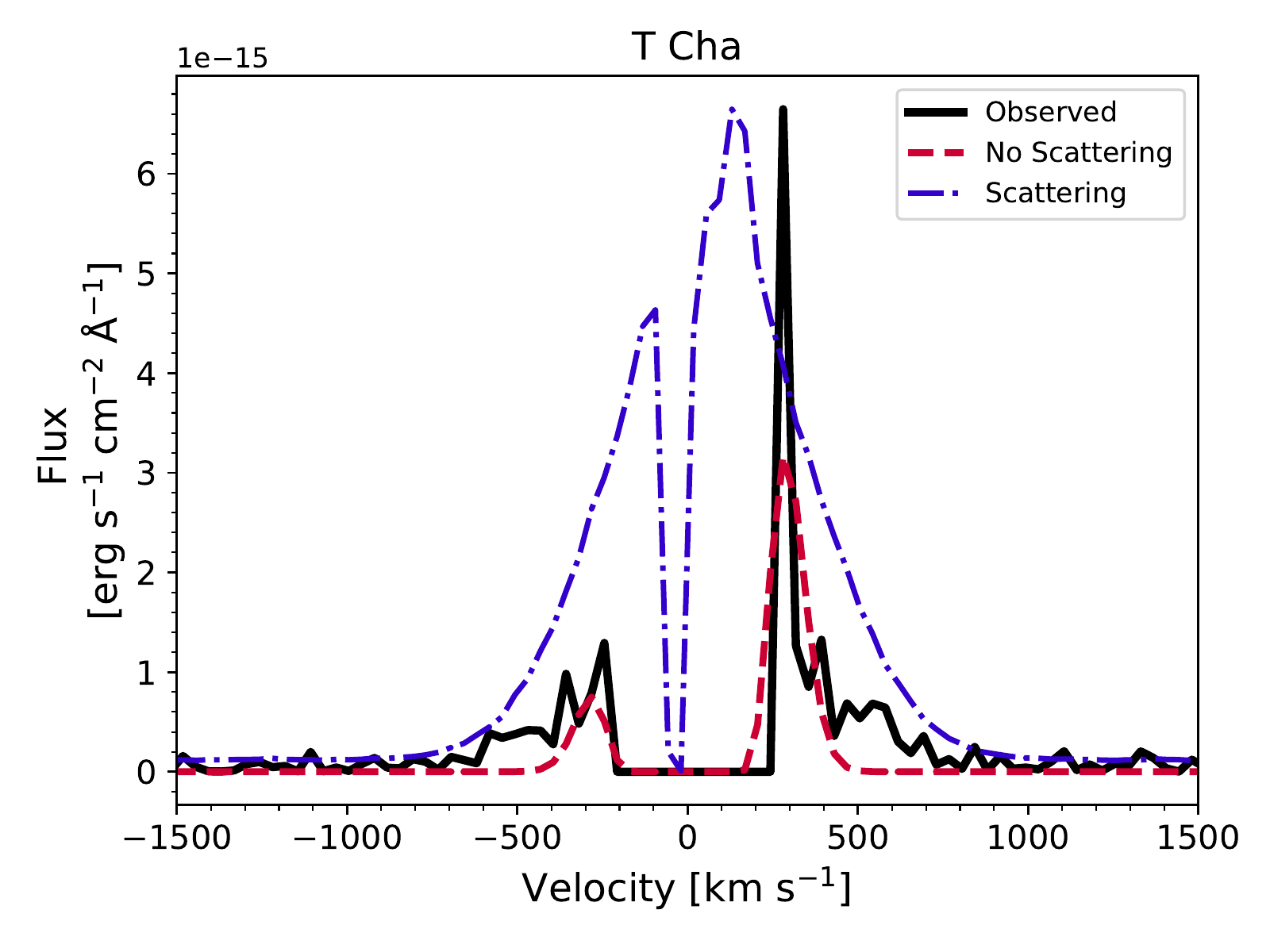}{0.5\textwidth}{(a) T Cha}
\fig{ 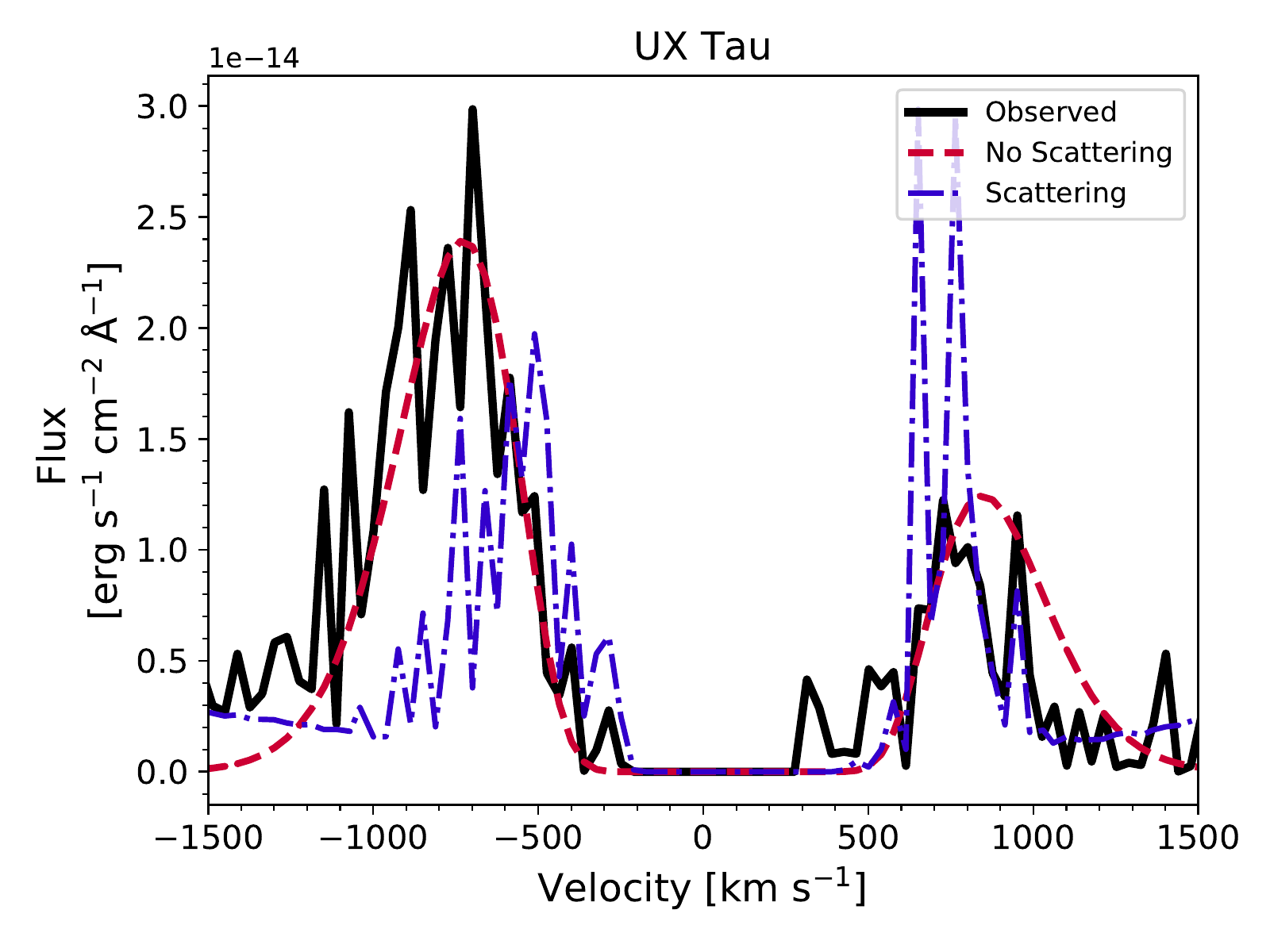}{0.5\textwidth}{(b) UX Tau A}}
\gridline{\fig{ LyA_Scattering_arxiv/Figures/V4046Sgr_LyA_modelcomp.pdf}{0.5\textwidth}{(c) V4046 Sgr}
\fig{ 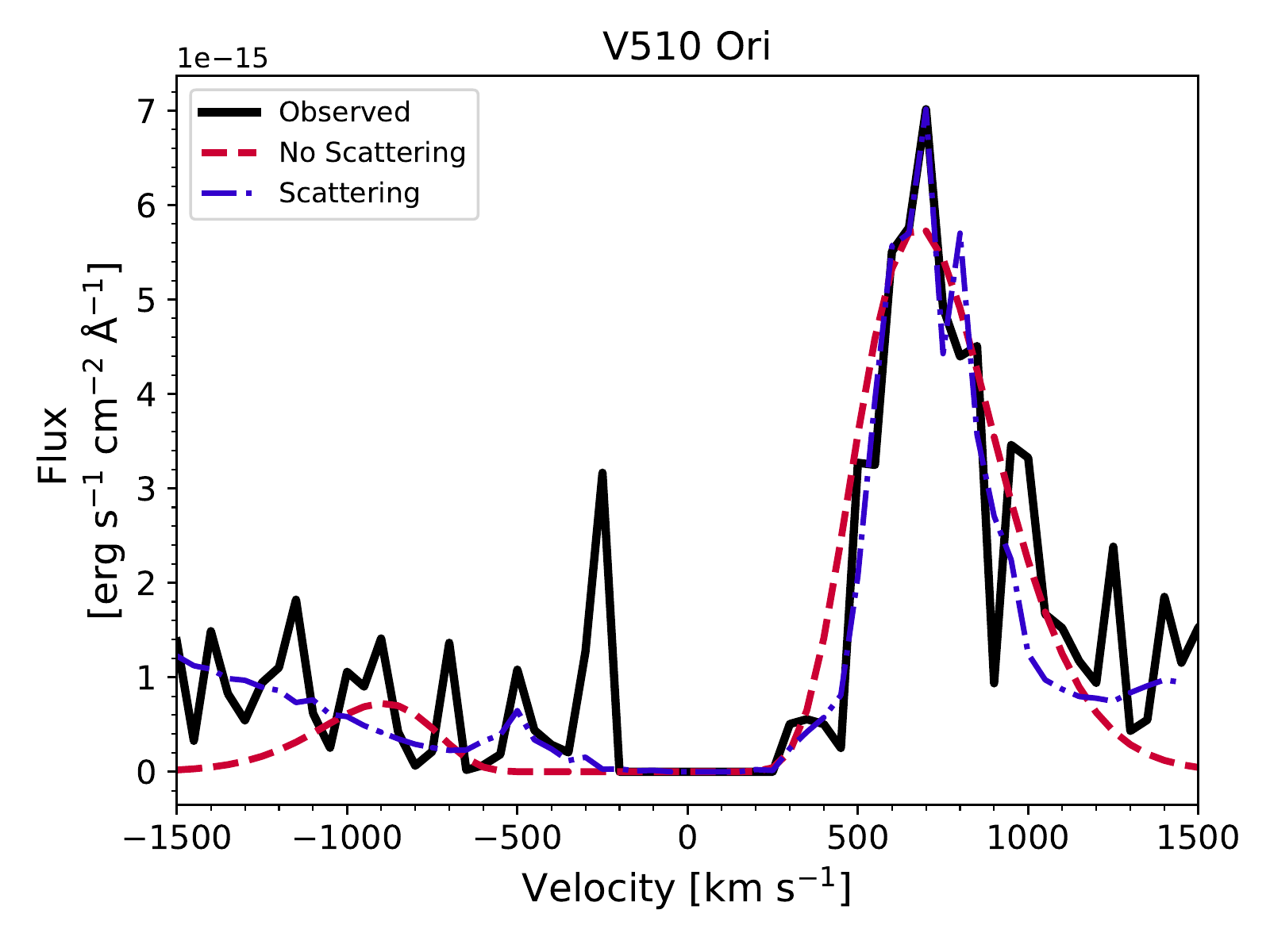}{0.5\textwidth}{(d) V510 Ori}}
\gridline{\fig{ 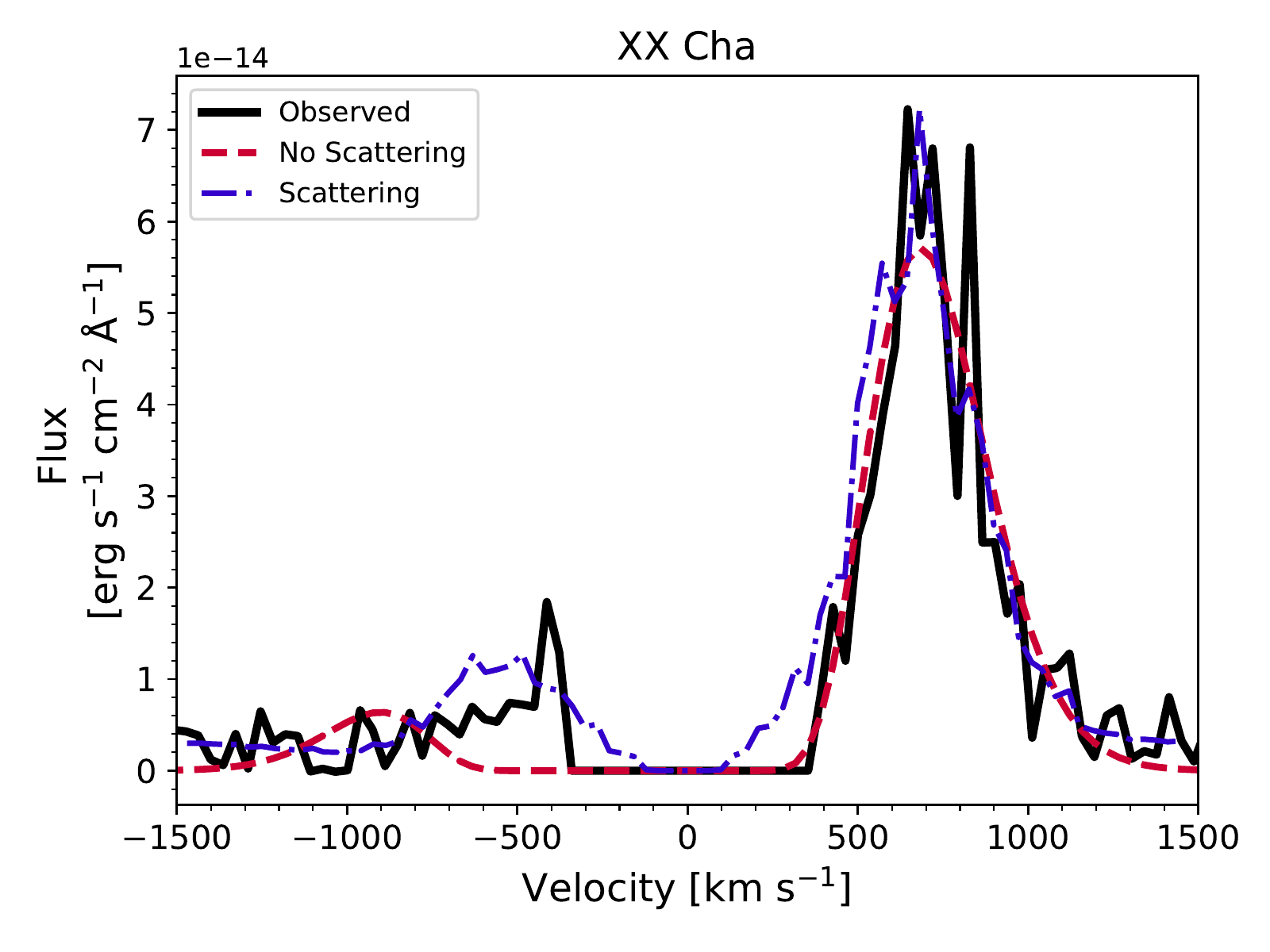}{0.5\textwidth}{(e) XX Cha}}
\caption{Observed Ly$\alpha$ emission line (black, solid) and best-fit models with scattering (blue, dash-dotted) and without scattering (red, dashed) for all 44 targets in our sample. Scattering models are not shown for targets with low S/N Ly$\alpha$ spectra, for which the MCMC chains did not converge.}
\end{figure*}

\begin{figure*}
\centering
\gridline{\fig{ 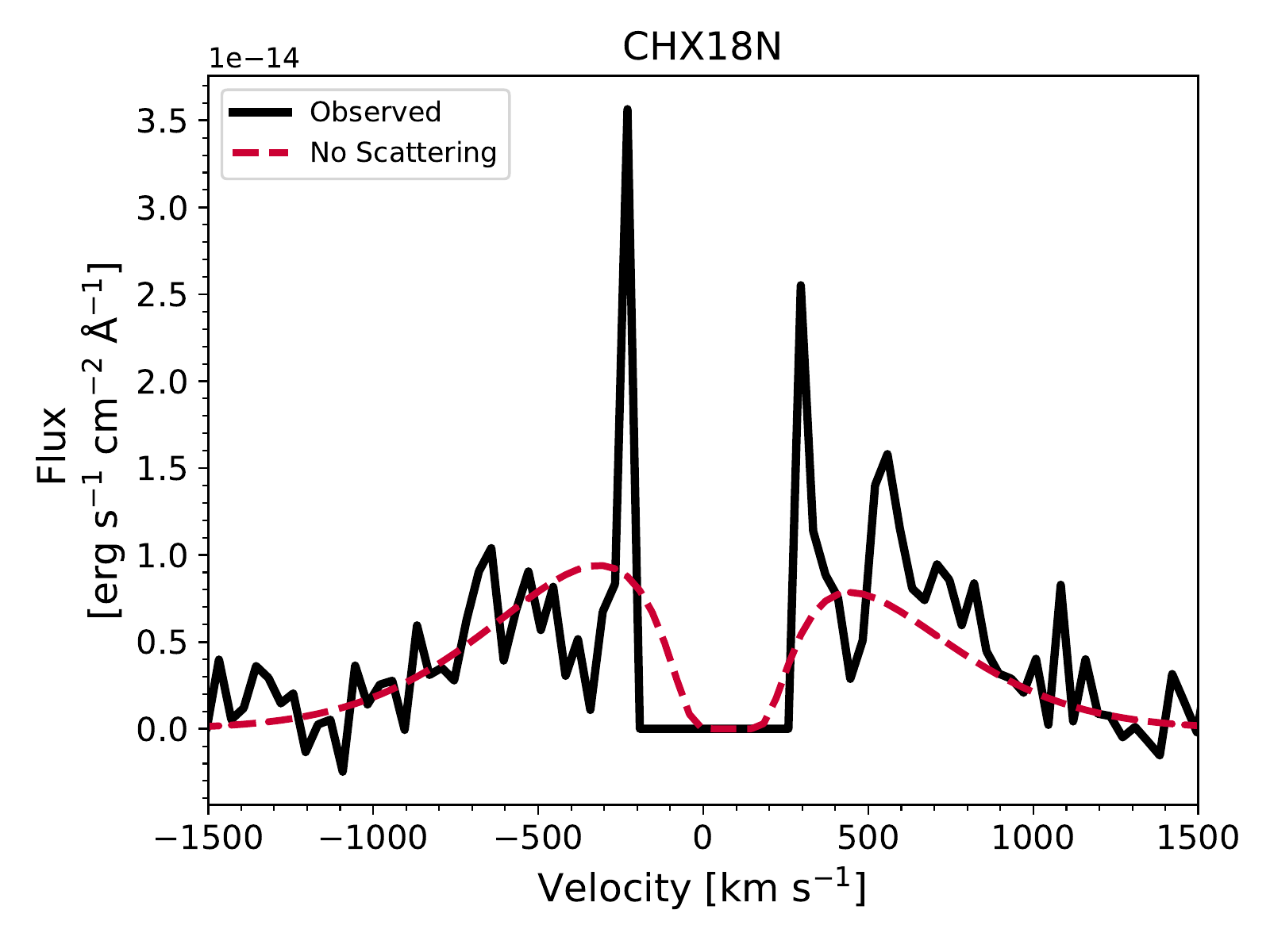}{0.3\textwidth}{(a)}
          \fig{ 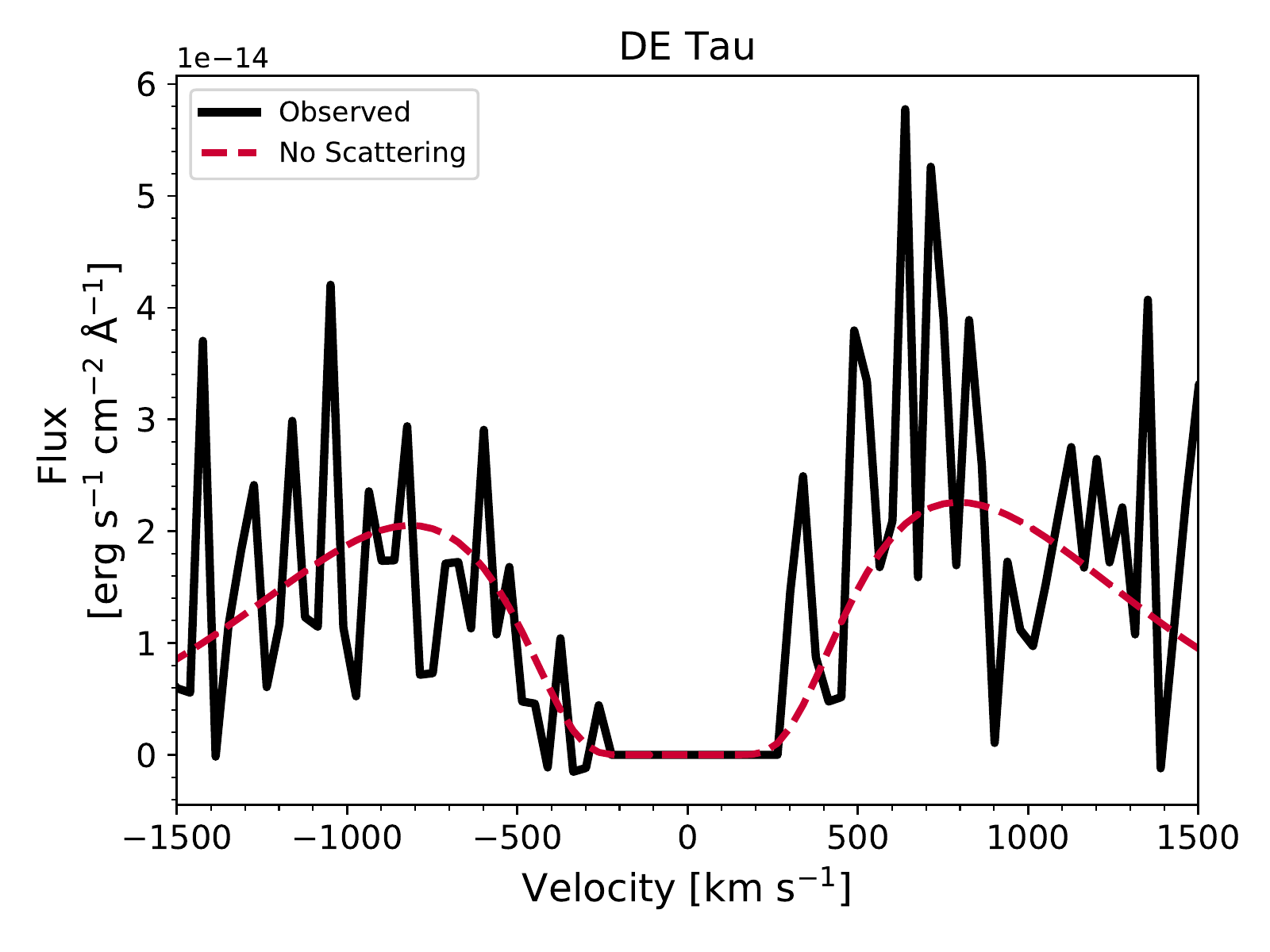}{0.3\textwidth}{(b)}
          \fig{ 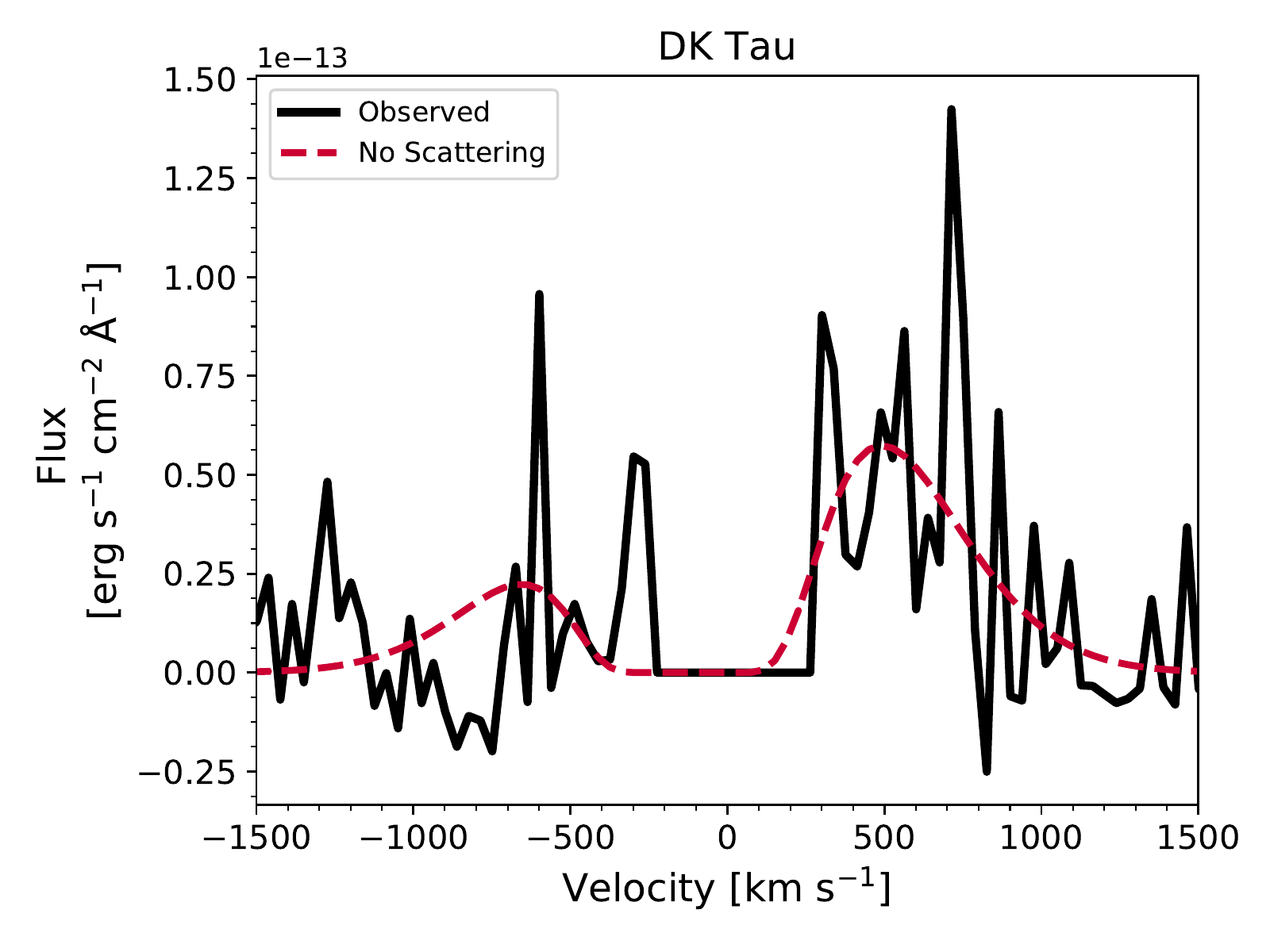}{0.3\textwidth}{(c)}}
\gridline{\fig{ 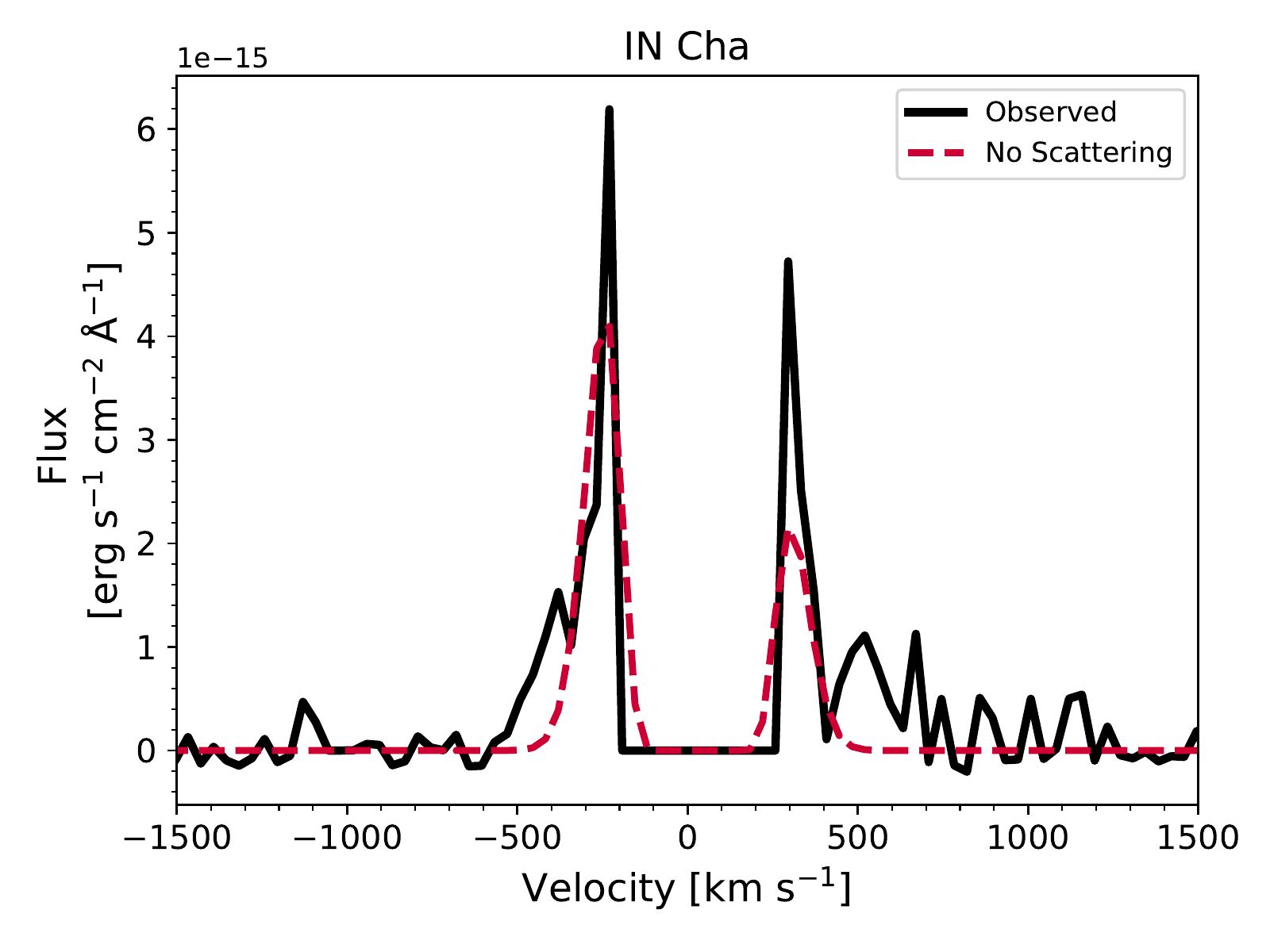}{0.3\textwidth}{(d)}
          \fig{ 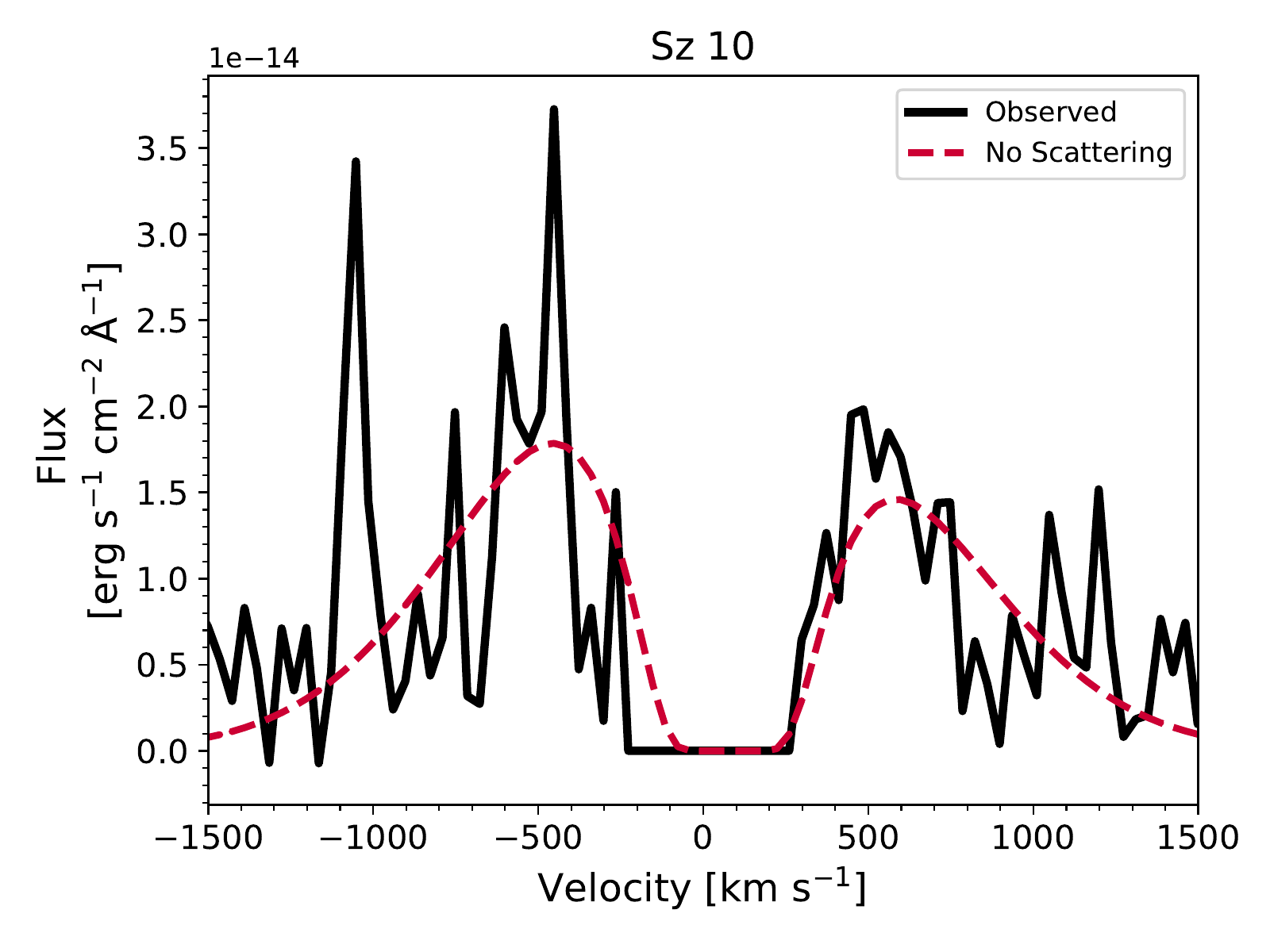}{0.3\textwidth}{(e)}
          \fig{ 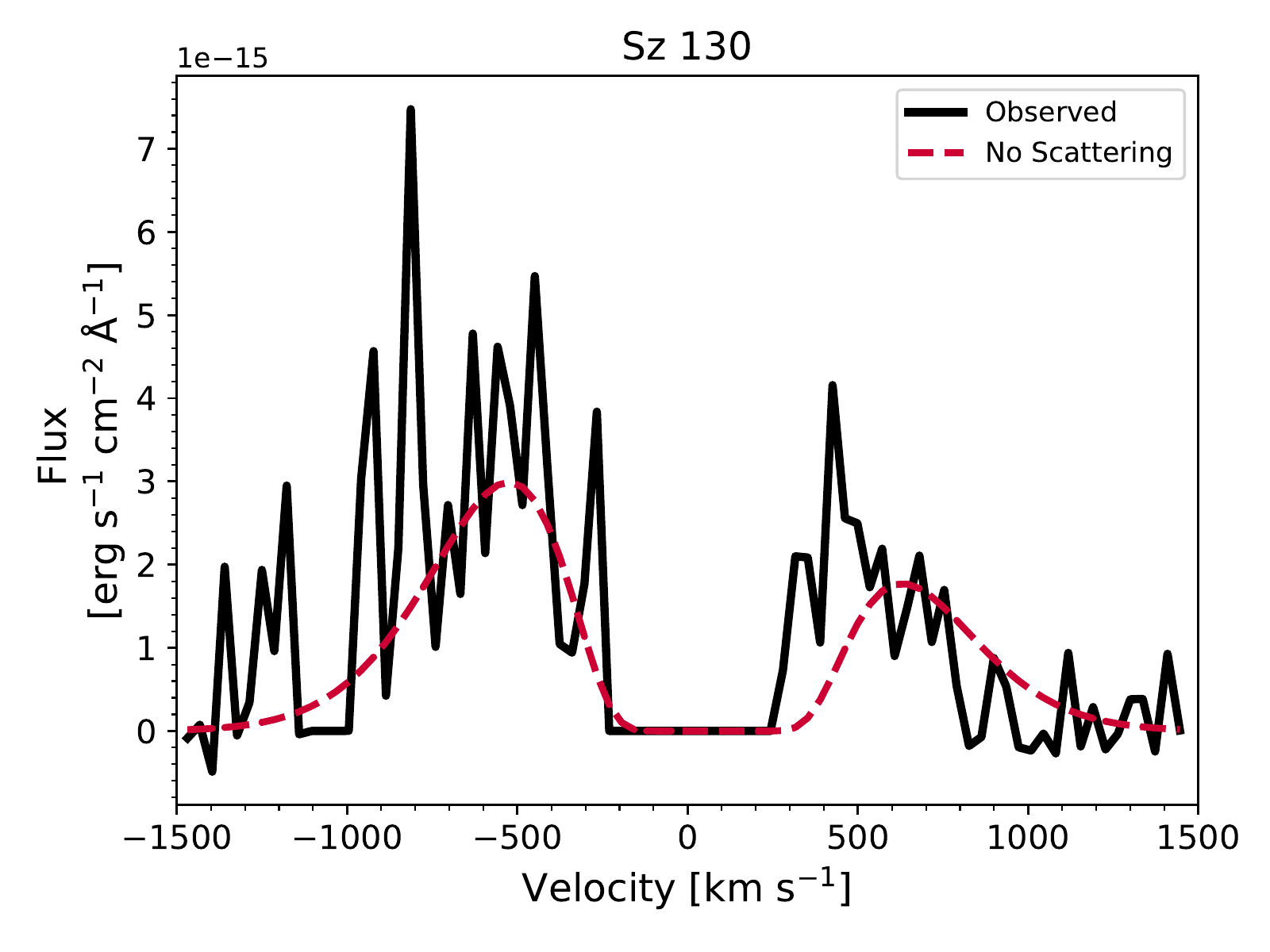}{0.3\textwidth}{(f)}}
\gridline{\fig{ 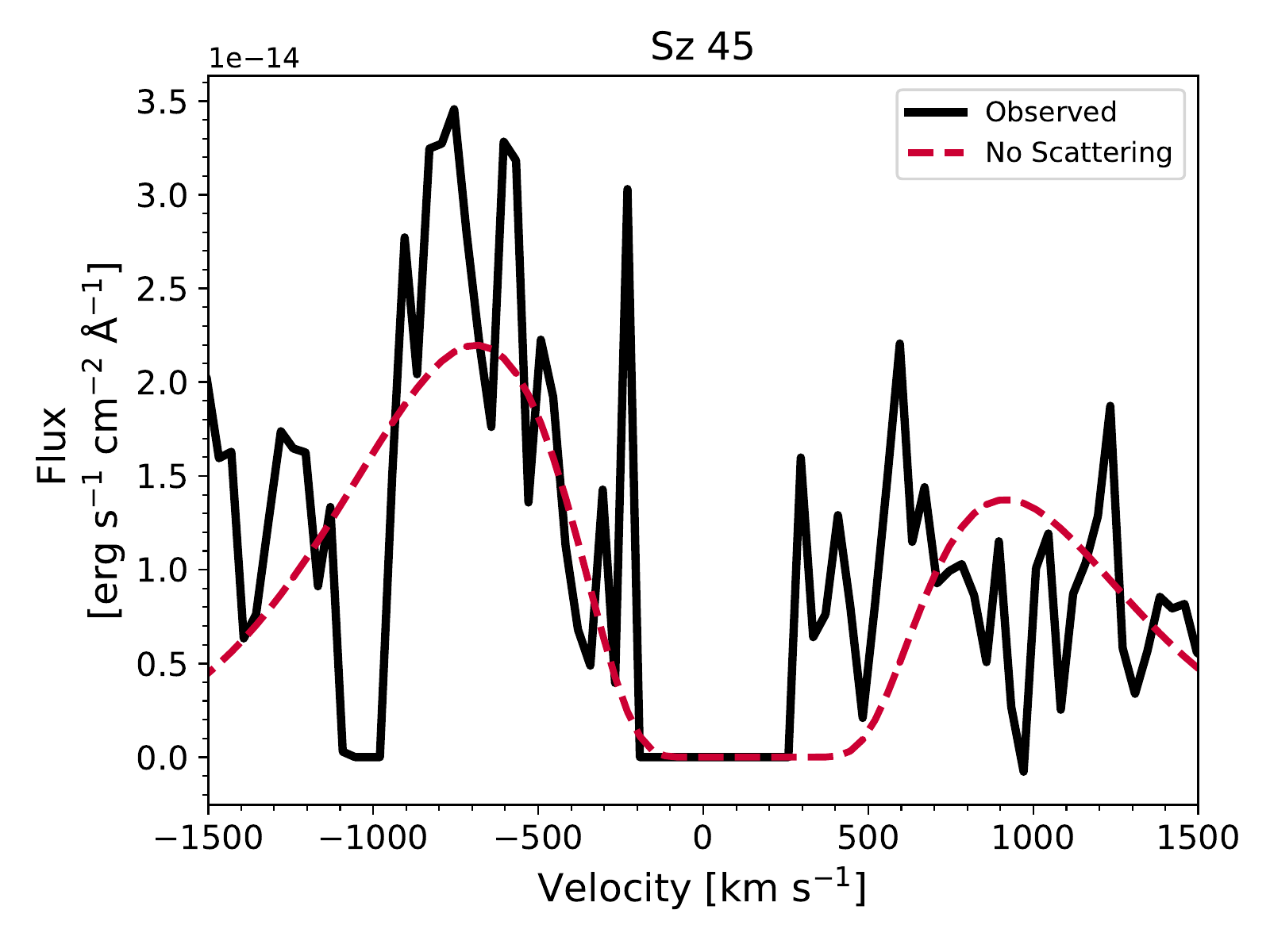}{0.3\textwidth}{(g)}
          \fig{ 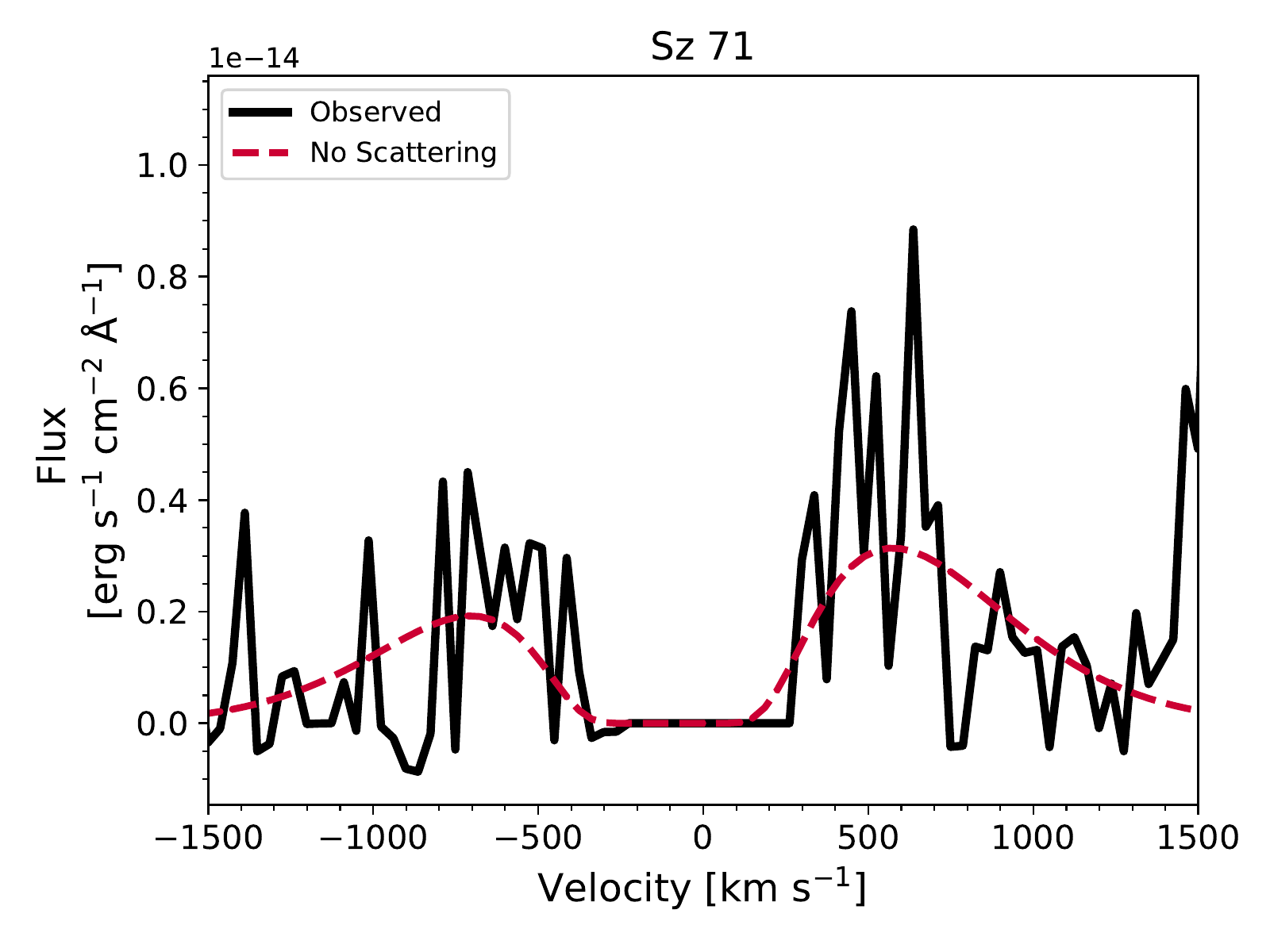}{0.3\textwidth}{(h)}
          \fig{ 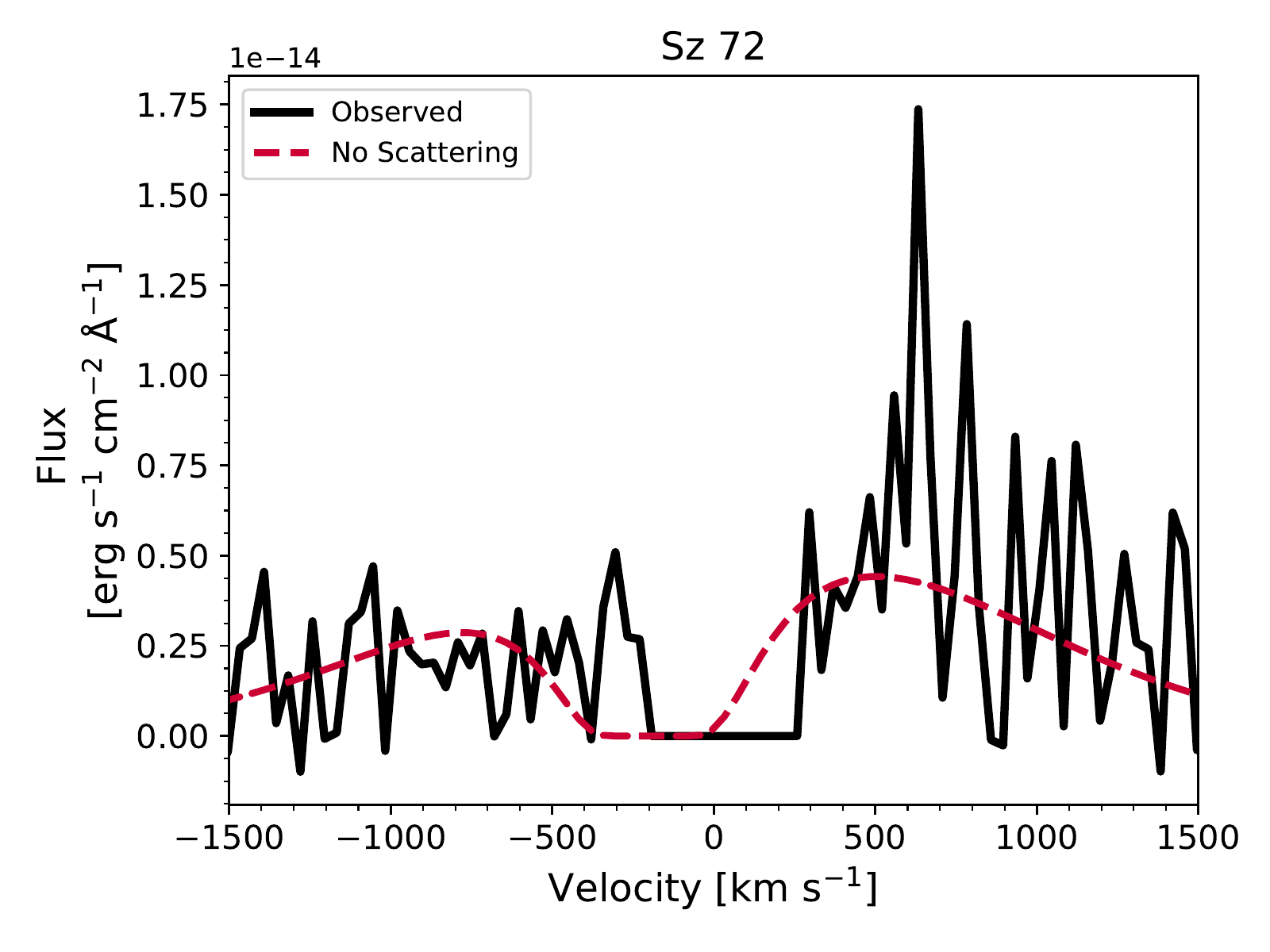}{0.3\textwidth}{(i)}}
\gridline{\fig{ 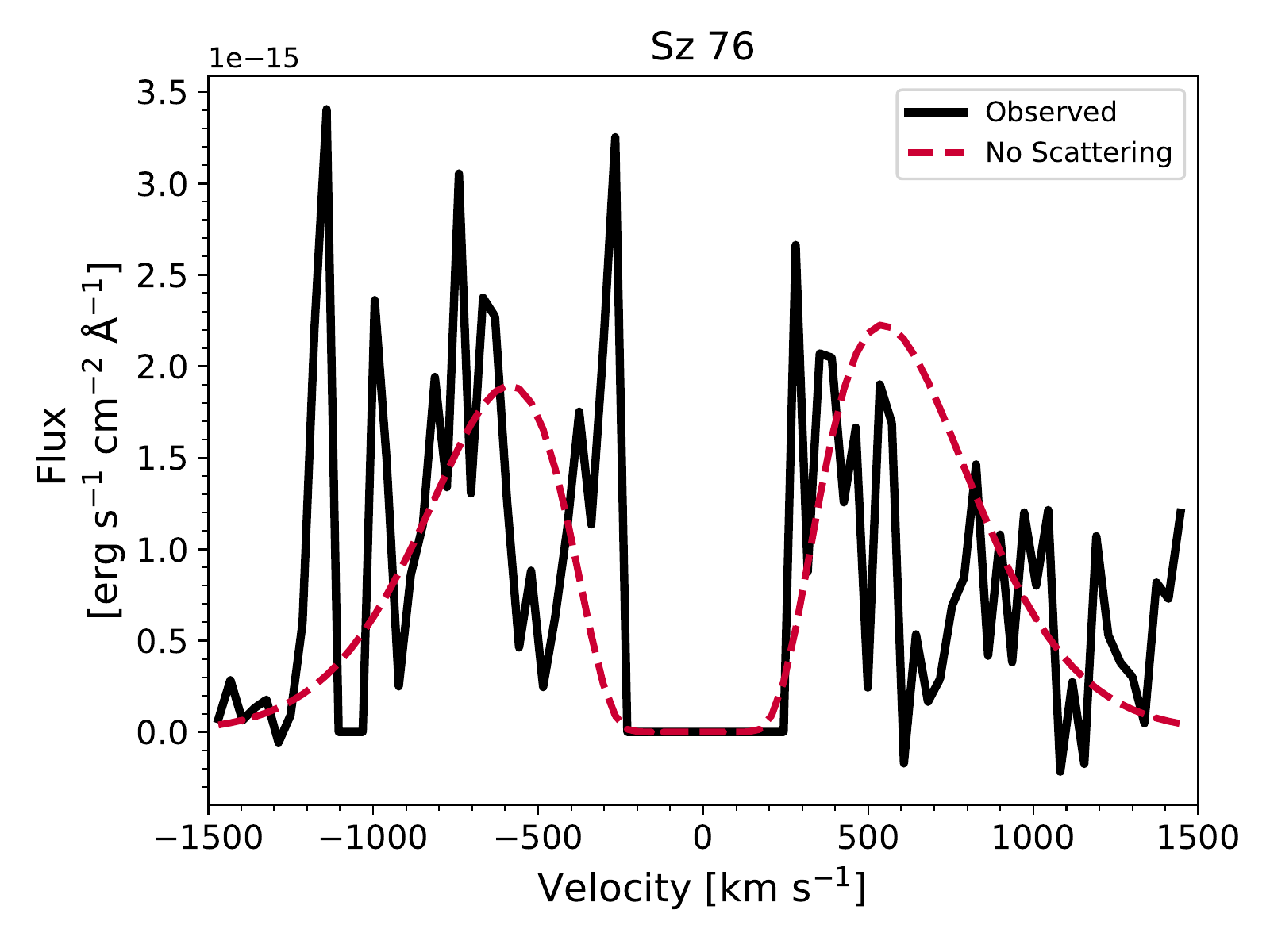}{0.3\textwidth}{(j)}
          \fig{ 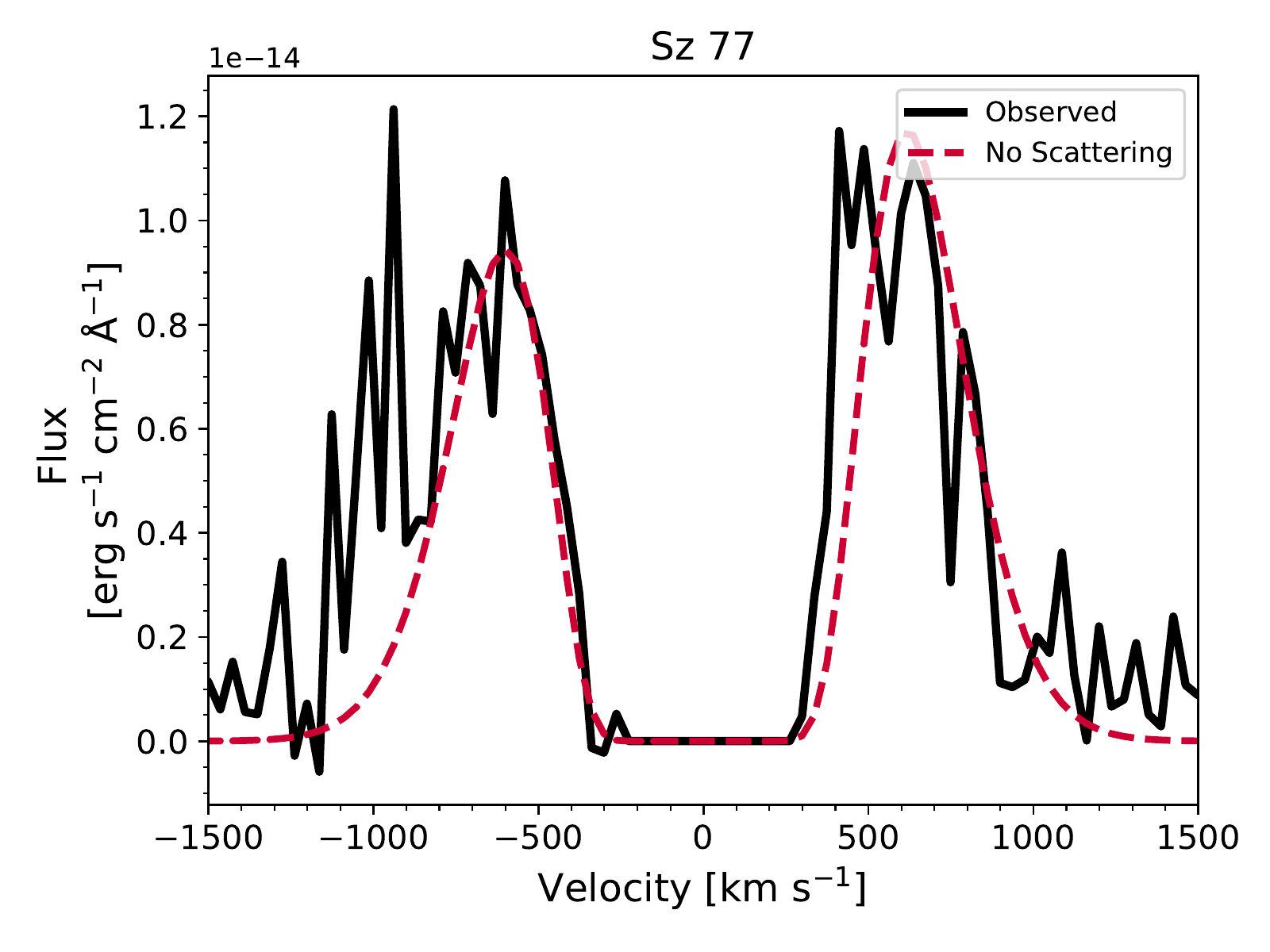}{0.3\textwidth}{(k)}}

\caption{Observed Ly$\alpha$ emission line (black, solid) and best-fit model without scattering (red, dashed) for targets with spectra that are too noisy to fit with the scattering model.}
\label{fig:noisy_LyA_modelcomp}
\end{figure*}

\begin{figure*}
\centering
\includegraphics[width=\textwidth]{ 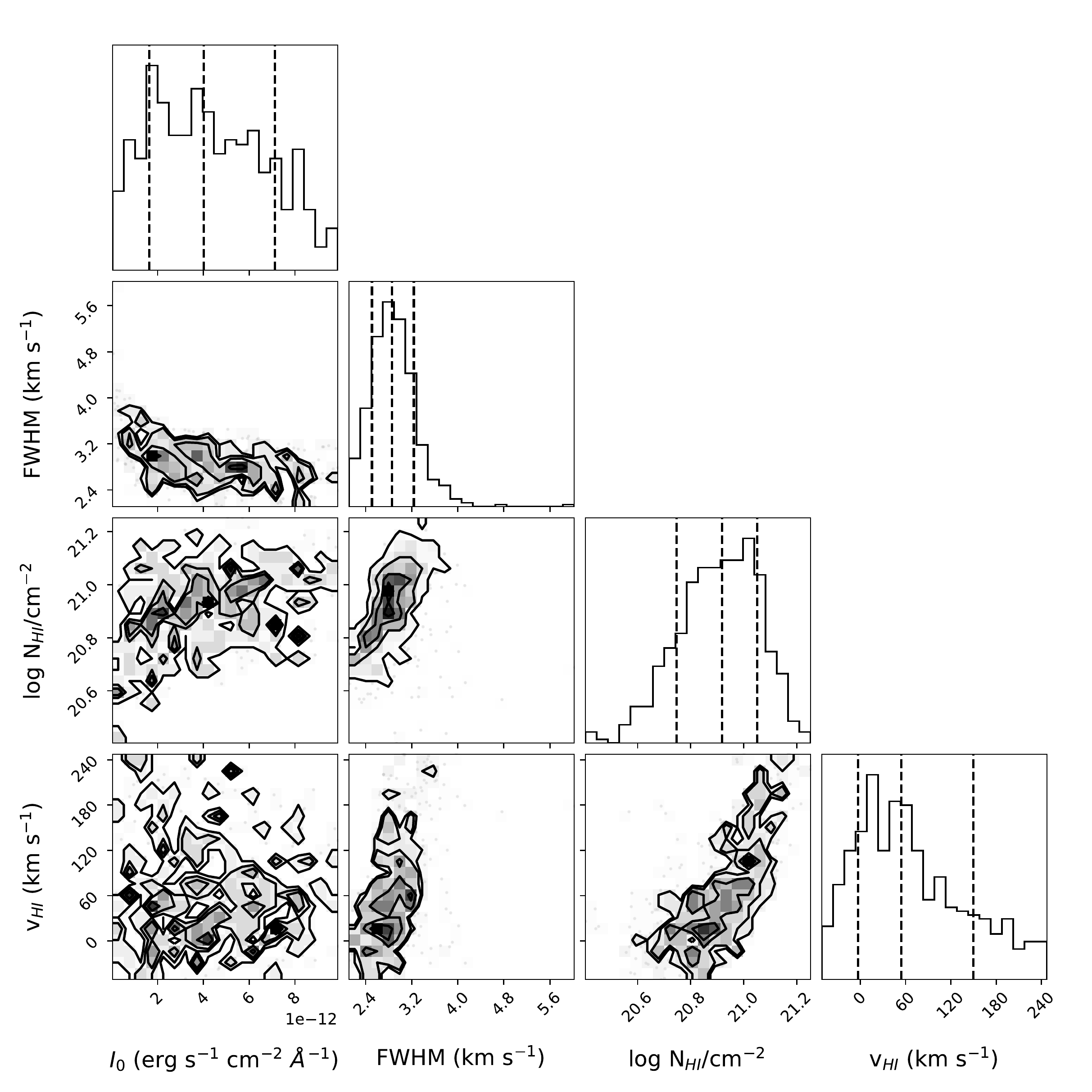}
\caption{Corner plot showing marginalized probability distributions for model parameters sampled for Sz 111, when Ly$\alpha$ scattering was not included. The Gaussian amplitude $\left(I_0 \right)$ is difficult to constrain from the emission line wings alone, but the samplers still converge within our criterion of $>$30\% acceptance fractions.}
\label{fig:Sz111_noscattering_corner}
\end{figure*}

\begin{figure*}
\centering
\includegraphics[width=\textwidth]{ 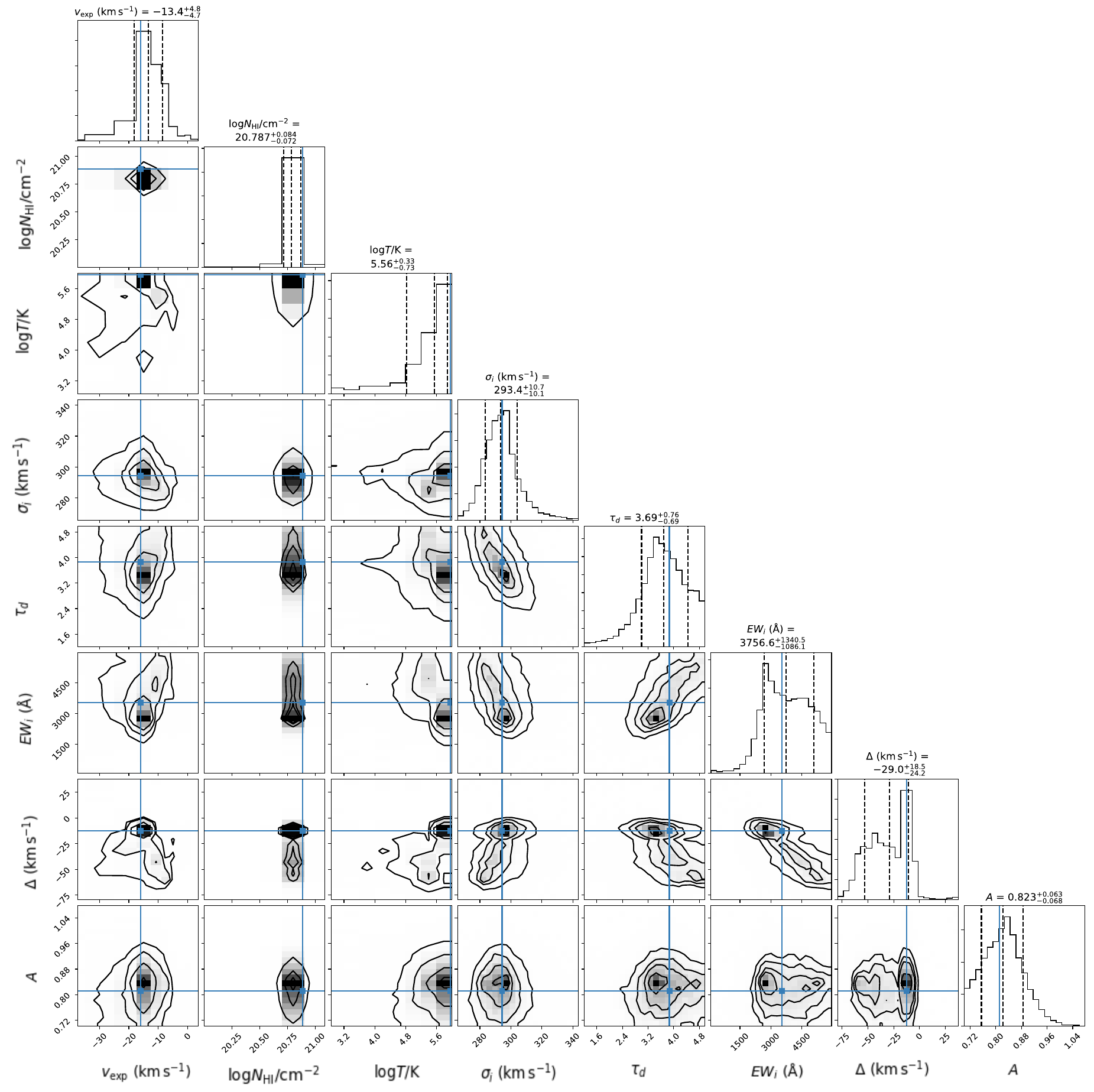}
\caption{Corner plot showing marginalized probability distributions for scattering model parameters sampled for Sz 111. Although the column density of H I (log $N_{\rm{H I}}$) is well constrained in this case, there are clear degeneracies between the intrinsic Ly$\alpha$ emission line width $\left(\sigma_i \right)$ and the effective temperature of the scattering medium (log $T/K$). This is consistent with the results of \citet{Li2022} and is likely why the H I column densities reported here do not change when $\sigma_i$ is held fixed.}
\label{fig:Sz111_scattering_corner}
\end{figure*}


\end{document}